\documentclass[12pt,a4paper]{article}
\usepackage[applemac]{inputenc}
\usepackage[T1]{fontenc}
\usepackage{amsmath,amsfonts,amssymb}
\usepackage[breaklinks=true]{hyperref}
\usepackage{mathtools}
\usepackage{graphicx,epsfig,float}
\usepackage{color,xcolor}
\usepackage{textcomp}
\usepackage{indentfirst}
\usepackage{multirow}
\usepackage{multicol}
\usepackage{enumerate}
\usepackage[font=small]{caption}
\usepackage{adjustbox}
\usepackage{amsthm}
\usepackage{graphics}
\usepackage{pstricks,pstricks-add,pst-math,pst-xkey}
\usepackage{bm}
\usepackage{nccmath}
\usepackage{url}
\usepackage{bbm}
\usepackage{setspace}
\usepackage[left=1in,right=1in,bottom=1.3in,top=1.3in]{geometry}
\usepackage{parskip}
\usepackage{breakcites}
\usepackage{algorithm}
\usepackage{algpseudocode}
\usepackage{tikz}
\usetikzlibrary{shapes,snakes}
\usetikzlibrary{automata, positioning}
\usetikzlibrary{matrix,arrows,decorations.pathmorphing,positioning,graphs,calc}
\usepackage{natbib}
\usepackage{subcaption}
\usepackage{subfiles}
\usepackage[title]{appendix}
\usepackage{titling}
\usepackage{mathtools}
\mathtoolsset{showonlyrefs}
\usepackage{placeins}
\usepackage{rotating}
\usepackage{booktabs}

\newcommand{\R}{{\mathbb{R}}}
\newcommand{\Lcal}{{\mathcal{L}}}
\newcommand{\ind}{\perp\!\!\!\perp} 
\theoremstyle{definition}

\theoremstyle{plain}

\theoremstyle{definition}

\DeclareMathOperator*{\argmax}{arg\,max}
\DeclareMathOperator*{\argmin}{arg\,min}
\DeclareMathOperator*{\U}{\text{Unif}}
\DeclareMathOperator*{\Betad}{\text{Beta}}
\DeclareMathOperator*{\Unifd}{\text{Unif}}
\DeclareMathOperator*{\Weibulld}{\text{Weibull}}
\DeclareMathOperator*{\Multinomiald}{\text{Multinomial}}
\DeclareMathOperator*{\Bernoullid}{\text{Bernoulli}}
\DeclareMathOperator*{\GPDd}{\text{GPD}}
\DeclareMathOperator*{\E}{\text{E}}
\DeclareMathOperator*{\ift}{\text{if }}

\definecolor{antiquefuchsia}{rgb}{0.57, 0.36, 0.51}

\providecommand{\keywords}[1]{{\small{\it Keywords:} #1}}
\def\LA#1{{\textcolor{magenta}{[LA: #1]}}}
\def\corr#1{{\textcolor{blue}{#1}}}
\def\todo#1{{\textcolor{red}{[{\bf To do:} #1]}}}
\def\writing#1{{\textcolor{purple}{#1}}}
\def\dint#1{\text{d}#1}

\def\JW#1{{\textcolor{blue}{[JW: #1]}}}
\def\RH#1{{\textcolor{orange}{[RH: #1]}}}

\allowdisplaybreaks

\begin{document}
\title{Neural Bayes estimation and selection of complex bivariate extremal dependence models}
\author{L. M. Andr\'e$^{1*}$, J. L. Wadsworth$^{2}$, R. Huser$^{3}$\\
\small $^{1}$ Namur Institute for Complex Systems, University of Namur, Rue Graf\'e 2, Namur 5000, Belgium\\
\small $^{2}$ School of Mathematical Sciences, Lancaster University, LA1 4YF, United Kingdom \\
\small $^{3}$ Statistics Program, Computer, Electrical and Mathematical Sciences and Engineering Division, \\
\small King Abdullah University of Science and Technology (KAUST), Saudi Arabia \\
\small $^*$ Correspondence to: \href{mailto:lidiamandre@gmail.com}{lidiamandre@gmail.com}}
\date{\today}

\maketitle
\pagenumbering{arabic}

\begin{abstract}
    Likelihood-free approaches are appealing for performing inference on complex dependence models, either because it is not possible to formulate a likelihood function, or its evaluation is very computationally costly. This is the case for several models available in the multivariate extremes literature, particularly for the most flexible tail models, including those that interpolate between the two key dependence classes of `asymptotic dependence' and `asymptotic independence'. We focus on approaches that leverage neural networks to approximate Bayes estimators. In particular, we explore the properties of neural Bayes estimators for parameter inference for several flexible but computationally expensive models to fit, with a view to aiding their routine implementation. Owing to the absence of likelihood evaluation in the inference procedure, classical information criteria such as the Bayesian information criterion cannot be used to select the most appropriate model. Instead, we propose using neural networks as neural Bayes classifiers for model selection. Our goal is to provide a toolbox for simple, fast fitting and comparison of complex extreme-value dependence models, where the best model is selected for a given data set and its parameters subsequently estimated using neural Bayes estimation. We apply our classifiers and estimators to analyse the pairwise extremal behaviour of changes in horizontal geomagnetic field fluctuations at three different locations.
\end{abstract}


\newpage
\onehalfspacing

\section{Introduction} \label{section:introduction}

\subsection{Motivation} \label{subsec:motivation}

Recent developments in bivariate extreme value modelling have produced new classes of models that allow for interpolation between the two key tail dependence regimes of asymptotic dependence and asymptotic independence; see for instance \citet{Engelkeetal2019}. These models simplify the approach to bivariate extremal modelling, by eliminating the need to pre-determine a dependence regime using unreliable empirical diagnostics. However, evaluating their likelihood functions often relies on numerical integration and inversion of functions, as well as censoring of non-extreme values, which makes likelihood evaluation burdensome. In other situations, the likelihood function might not be available at all. Nonetheless, despite the likelihood function being intractable or unavailable, it is often possible to simulate data from the model; this allows for the use of simulation-based likelihood-free algorithms to estimate model parameters. 

\subsection{Traditional likelihood-free inference approaches} \label{subsec:likelihoodfree}

One simulation-based approach is the pseudo-marginal Markov chain Monte Carlo \linebreak (MCMC) sampler proposed first by \citet{Beaumont2003} and later formalised by \citet{AndrieuRoberts2009}. When the target distribution is intractable, usually due to the likelihood function, pseudo-marginal MCMC is able to approximate the target function using an unbiased estimator obtained through importance sampling. The expected value of such an estimator corresponds to the true target distribution, enabling the algorithm to correctly sample from it. Pseudo-marginal MCMC methods, however, perform poorly when dealing with high-dimensional parameter vectors \citep{Alenlovetal2021} or when the distribution of the proposed estimator is, for example, heavy tailed, resulting in difficulties converging to the target distribution \citep{MurrayGraham2016}. Another commonly used likelihood-free procedure is approximate Bayesian computation (ABC; see, e.g., \citealp{Lintusaarietal2017} and \citealp{Sissonetal2018}). Specifically, this can be seen as a rejection sampling algorithm, where the model parameters are generated from a prior distribution and subsequently accepted or not based on the distance between a data sample simulated from the model, given the simulated parameters, and the original data sample, often evaluated based on informative summary statistics. Choosing a suitable prior distribution, and defining how similar the data samples are, constitute major drawbacks of using ABC to perform inference. An inadequate prior distribution might lead to misleading posterior estimates, particularly in situations where the selected summary statistics are not very informative. Conversely, a prior distribution which is too informative can result in a posterior that is skewed or biased towards the prior distribution, even if it leads to less variable estimates. \citet{Wood2010} propose the `synthetic likelihood' method, which constructs an approximate likelihood function by assuming that user-defined summary statistics follow a multivariate normal distribution. This approach is usually easier to tune than ABC and computationally more efficient, especially with higher dimensional data sets \citep{Priceetal2018}, but its usual underlying Gaussian assumption makes it inflexible in some cases, which may lead to sub-optimal inferences.

\subsection{Neural Bayes inference for extremes} \label{subsec:neuralBayesExtremes}

More recently, there has been a growing interest in likelihood-free estimation methods which use neural networks; see \citet{Zammitetal2024} for an in-depth review. The extremes literature has also started to be impacted by this new inference paradigm, mostly in the spatial setting; see for instance  \citet{DellOroGaetan2024}, \citet{Lenzietal2023}, \citet{MajumderReich2023},  \citet{Majumderetal2024}, \citet{Raietal2024}, \citet{Richardsetal2023, Richardsetal2024}, \citet{SainsburyDaleetal2024, SainsburyDaleetal2023} and \citet{Walchessenetal2024}. One approach for leveraging neural networks for likelihood-free inference is to obtain a point estimate of the vector of model parameters through a neural Bayes estimator \citep{SainsburyDaleetal2024}. Training the neural network to build such an estimator can be computationally expensive; however, this step only needs to be done once, with estimates subsequently obtained in milliseconds with new data, using a single graphics processing unit (GPU). This means that neural Bayes estimators are amortised \citep{Zammitetal2024}, which allows for their repeated use at almost no extra computational cost (see, e.g. \citealp{Richardsetal2023} for a compelling data illustration). Therefore, neural Bayes estimators are an appealing, and much faster, avenue to performing inference compared to state-of-the-art likelihood-based methods. Moreover, unlike other likelihood-free methods, such as ABC, this approach automatically learns the relevant summary statistics for the inference problem at hand. Given the computational complexity of the models of interest, this is the approach we take in this paper. Whilst the methodology developed for neural Bayes estimation has been mostly applied in the spatial and temporal contexts, we are interested in exploring its applicability in a simple bivariate setting, while allowing for censored data inputs. In order to achieve this, an appropriate neural network architecture will need to be designed, along with suitable prior choices. While the number of parameters to estimate may be similar to typical spatial models, the inference procedure may be more challenging in a bivariate setting compared to a spatial context, since there are fewer distinctive features (e.g., location, distance) in the data that the neural Bayes estimator can learn from.

When neural Bayes estimators are adopted for inference, typical model selection techniques, such as the Akaike information criterion (AIC) or the Bayesian information criterion (BIC), are often not available as they require knowledge of the likelihood function; see, e.g., \citet{ClaeskensHjort2008} for details on model selection criteria. Therefore, having a likelihood-free way of selecting the best model, for a given set of candidate models fitted to a certain data set, is desirable. \citet{Radevetal2023} show how the marginal likelihood (and hence, Bayes factors useful for model selection) may be approximated using a neural network-based approach. Similarly, neural methods targeting the likelihood-to-evidence ratio (see, e.g., \citealp{Cranmeretal2016}, \citealp{Hermansetal2020}) could be used for model selection using likelihood ratios. These methodologies, however, rely on full posterior and/or full likelihood neural approximations. \citet{Ahmedetal2022} and \citet{WixsonCooley2024}, on the other hand, propose using neural networks as classification tools for testing the extremal dependence type of a data set, however not in the context of model selection. Whilst the former only focus on spatial processes, the latter apply the classifier to bivariate data sets. More recently, \citet{Ahmedetal2024} propose using neural networks for selecting spatial max-stable processes. To do so, the extremal concurrence probability \citep{Dombryetal2018} is used as a summary statistic to extract the spatial features of a data set. However, if such summary statistic is not sufficient, its use as input to the neural network might result in a weaker ability to distinguish between key features. Here, we make a step further and instead propose a model selection procedure based on a classification problem, whereby a neural network architecture (analogous to that used for parameter estimation) automatically learns relevant summary statistics for the model selection problem at hand. More specifically, a neural Bayes classifier, designed to learn the distinguishing features of each model, is used (once trained) to efficiently estimate the posterior probability of a data set arising from a certain model. The obtained posterior probabilities can then be used as weights in Bayesian model averaging (e.g., \citealp{Hoetingetal1999}) type procedures, if desired.

Beyond parameter estimation, model selection and hypothesis testing, work at the interface between statistics of extremes and deep learning is not new and has been of growing interest recently; see, e.g., \citet{Cannon2010}, \cite{Cisnerosetal2024}, \citet{PascheEngelke2022}, \citet{RichardsHuser2022} for univariate extreme regression modelling, and \citet{MurphyBarltropetal2024} for extremal dependence modelling. 

\subsection{Main contributions and paper outline} \label{subsec:contribution}

In this paper, we contribute to the neural inference literature on extremes with the aim to provide a supervised learning toolbox for simple fitting and selection of complex bivariate extremal dependence models, which avoids the subjective, and often awkward, selection of summary statistics. We start by exploring the utility of neural Bayes estimators in this specific setting; this is done through assessment of the estimation accuracy of the model parameters and key dependence measures. The uncertainty of the neural Bayes estimators is assessed and compared through two different approaches: a non-parametric bootstrap approach, and an estimator trained to target posterior quantiles. We then outline the neural Bayes classifier, and examine its success for model selection. When available, we compare its performance to a likelihood-based information criterion. The end goal is to make the entire statistical pipeline amortised. First, the best model for a given data set is selected through the neural Bayes classifier, and then estimates of the model parameters are obtained through a neural Bayes estimator.

This paper is organised as follows: in Section~\ref{section:methodology}, we introduce the methodology used for parameter estimation using neural networks for both uncensored and censored data, and describe our model selection procedure based on a classification task. Section~\ref{section:models} presents an overview of bivariate extreme value modelling and introduces the models of interest for which likelihood-based inference is burdensome. Simulation studies assessing our proposed inference and model selection frameworks are discussed in Section~\ref{section:simulation}. We then apply the proposed toolbox to study the pairwise extremal behaviour of the changes in horizontal geomagnetic field fluctuations between three locations in Section~\ref{section:application}, followed by a conclusion in Section~\ref{section:conclusion}.
\section{Inference methodology} \label{section:methodology}

In this section, we review background on neural point estimation using the methodology developed by \citet{SainsburyDaleetal2024} and \citet{Richardsetal2023} and describe our neural approach to model selection. In Section~\ref{subsec:nbe}, we review neural Bayes estimators, firstly outlining the general framework when the sample size is fixed and known, and data are fully observed (i.e., uncensored). We then explain how to adapt the estimation procedure to account for variable sample size and censored data, respectively. In Section~\ref{subsec:modelselection}, we present our neural Bayes classifier for model selection and describe how to construct it. Implementation details are given in Section~\ref{subsec:compimpl}. Although we later focus on bivariate models, we use data dimension $d$ in this section for generality.

\subsection{Neural Bayes estimators}\label{subsec:nbe}

\subsubsection{General framework} \label{subsec:NBEgeneral}

Let $\bm Z_1, \ldots, \bm Z_n\in\mathcal{S}\subseteq \mathbb{R}^d$ be $n$ independent and identically distributed random vectors, and let their collection be represented by $\bm Z = (\bm Z_1', \ldots, \bm Z_n')'\in\mathcal{S}^n.$ We assume that $\bm Z$ admits the parametric joint density $f(\bm z \mid \bm \theta)=\prod_{i=1}^n f(\bm z_i\mid\bm \theta),$ where $\bm\theta \in \mathbb{R}^p$ is the vector of parameters. A point estimator $\bm{\hat \theta}(\cdot)$ maps data $\bm Z$ to parameter estimates from the parameter space $\Theta,$ i.e., $\bm{\hat\theta}: \mathcal{S}^n \to \Theta.$ Given a non-negative loss function $L(\bm \theta, \bm{\hat\theta}(\cdot)),$ a Bayes estimator minimises a weighted average of the risk at $\bm \theta,$ $R(\bm \theta, \bm{\hat\theta}(\cdot))=\E[L(\bm \theta, \bm{\hat\theta}(\bm Z))],$ which may be expressed as
\begin{equation}\label{eq:bayesrisk}
    r_\Omega(\bm{\hat\theta}(\cdot))=\int_\Theta R(\bm \theta, \bm{\hat\theta}(\cdot))\dint\Omega(\bm\theta) = \int_\Theta\int_{\mathcal{S}^n} L(\bm \theta, \bm{\hat\theta}(\bm z))f(\bm z\mid \bm \theta)\dint\bm z\,\dint\Omega(\bm\theta),
\end{equation}
where $\Omega(\cdot)$ is a prior measure for $\bm \theta.$ Equation \eqref{eq:bayesrisk} is known as the Bayes risk. It is key to realise that Bayes estimators are functionals of the posterior distribution induced by the prior $\Omega(\cdot)$. In particular, if $L(\cdot,\cdot)$ is the squared error loss, $\bm{\hat\theta}(\bm Z)$ is the posterior mean of $\bm\theta$. If $L(\cdot,\cdot)$ is instead the absolute error loss, $\bm{\hat\theta}(\bm Z)$ is the posterior median, and alternative posterior quantiles can be estimated using the quantile loss function; see Section~\ref{subsec:parestimationsim} for further details. Under suitable regularity conditions and the squared error loss, these estimators are consistent and asymptotically efficient; see for instance \citet[Ch.~5 and 6]{LehmanCasella1998}.

In practice, however, Bayes estimators are rarely available in closed form, and the Bayes risk in equation \eqref{eq:bayesrisk} is difficult to evaluate. This can be overcome by approximating these estimators using a neural network, since these are universal function approximators \citep{Horniketal1989}. In this context, a neural point estimator $\bm{\hat \theta}(\cdot; \bm \gamma)$ is constructed as a neural network, with parameters $\bm \gamma,$ that returns a point estimate from data input $\bm Z.$ Bayes estimators may thus be approximated with $\bm{\hat \theta}(\cdot; \bm \gamma^*)$ where \linebreak $\bm\gamma^* = \argmin_{\bm \gamma} \hat r_\Omega(\bm{\hat \theta}(\cdot ; \bm \gamma)),$ and
\begin{equation}\label{eq:bayesriskmcmc}
    \hat r_\Omega(\bm{\hat \theta}(\cdot ; \bm \gamma)) = \frac{1}{n_{\mathcal{T}}}\sum_{\{\bm \theta, \bm z\} \in \mathcal{T}_{\bm \theta, \bm z}} L(\bm \theta,\bm{\hat\theta}(\bm z; \bm \gamma)) \approx r_\Omega(\bm{\hat \theta}(\cdot ; \bm \gamma)).
\end{equation}
In equation~\eqref{eq:bayesriskmcmc}, $\mathcal{T}_{\bm \theta, \bm z}$ is a training set of size $n_{\mathcal{T}}$ consisting of draws $\bm\theta$ from the the prior $\Omega(\cdot)$ and corresponding data set realisations $\bm z=(\bm z_1',\ldots,\bm z_n')'$ sampled from the model given $\bm \theta$, i.e., from $\prod_{i=1}^n f(\bm z_i\mid\bm \theta).$ The neural point estimator $\bm{\hat \theta}(\cdot ; \bm \gamma^*)$ is called a neural Bayes estimator (NBE) as it minimises a Monte Carlo approximation \eqref{eq:bayesriskmcmc} of the Bayes risk \eqref{eq:bayesrisk}; see  \citet{SainsburyDaleetal2024} for more details. 

The discrepancy between the neural Bayes estimator and the true Bayes estimator will depend on a few factors, one of which is the neural network architecture. Through a judicious choice of architecture, NBEs can be enforced to satisfy the fundamental property that $\bm{\hat\theta}(\bm Z;\bm \gamma)=\bm{\hat\theta}(\tilde{\bm Z};\bm \gamma)$ for any vector $\tilde{\bm Z}$ whose elements are permutations of the independent replicates in $\bm Z;$ this can be achieved by exploiting a neural network architecture known as DeepSets \citep{Zaheeretal2017, SainsburyDaleetal2024}. Let $\bm\psi:\mathbb{R}^d \to \mathbb{R}^q$ and $\bm \phi: \mathbb{R}^q\to\mathbb{R}^p$ be two multilayer neural networks parameterised by $\bm \gamma_{\bm \psi}$ and $\bm \gamma_{\bm \phi},$ respectively. Following \citet{SainsburyDaleetal2024}, the NBE is then represented~as
\begin{equation}\label{eq:nbedeepsets}
    \bm{\hat \theta}(\bm Z ; \bm \gamma) = \bm \phi\left(\bm S(\bm Z; \bm \gamma_{\bm \psi}); \bm \gamma_{\bm \phi}\right) \quad \text{with} \quad \bm S(\bm Z; \bm \gamma_{\bm \psi})=\frac{1}{n}\sum_{i=1}^n \bm \psi(\bm Z_i;\bm \gamma_{\bm \psi}),
\end{equation}
where the average is computed componentwise, $\bm S$ denotes a vector of learnt summary statistics, and $\bm \gamma = (\bm \gamma_{\bm \psi}', \bm \gamma_{\bm \phi}')'$ are the parameters of the neural networks $\bm \psi$ and $\bm \phi$. In the multivariate unstructured setting, a multilayer perceptron (MLP) neural network may be used for both $\bm \psi$ and $\bm \phi;$ see Section~\ref{subsec:compimpl} for further details. A schematic of the DeepSets architecture is shown in Section~\ref{supsec:deepsets} of the Supplementary Material.

More recently, \citet{Rodderetal2025} have provided theoretical and asymptotic guarantees for NBEs. In particular, the authors show that, under the squared error loss function, an MLP with ReLU activation function (see Section~\ref{subsec:compimpl}), and other regularity assumptions, these estimators are fully efficient compared to the corresponding analytical Bayes estimators, and are consistent in a Bayesian sense. They do so by decomposing the empirical risk~\eqref{eq:bayesriskmcmc} into the analytical Bayes risk, an approximation error term, a generalisation error term, and an algorithm error term, which all converge to zero under an appropriate asymptotic regime.

The performance of the NBE is inherently influenced by the choice of the prior distributions. While an informative prior might reduce the volume of the parameter space, and allow lower values for $n_\mathcal{T}$ in equation~\eqref{eq:bayesriskmcmc}, this restricts the use of the trained estimators across several applications; when the latter is desirable, a vague prior is advisable \citep{SainsburyDaleetal2024}. When the likelihood is available, and feasible, informative prior distributions can be constructed using likelihood-based estimates \citep{Lenzietal2023}. Alternatively, informative priors can be obtained by sequentially incorporating the knowledge from previously estimated posterior distributions; however, this is no longer an amortised approach, as the neural network must be retrained whenever the prior changes. 

\subsubsection{Variable sample size} \label{subsec:varsample}

When training a neural Bayes estimator on a data set with a fixed number, $n,$ of replicates, this estimator will generally not be Bayes for data sets with a different sample size $\tilde{n}\neq n.$ Therefore, in order to ensure that the trained NBE approximately minimises the Bayes risk $r_\Omega(\bm{\hat\theta} (\cdot;\bm \gamma))$ for varying sample sizes $n,$ \citet{SainsburyDaleetal2024} propose two approaches: either obtaining a piecewise neural Bayes estimator by pre-training the estimator for specific fixed sample sizes \citep{Goodfellowetal2016}, or treating the sample size as a random variable $N;$ we adopt the latter in our work.

Let us assume that the sample size $N$ follows a discrete uniform distribution, that is $N\sim \Unifd(\{n_1, n_1+1,\ldots, n_2\})$ where $\Pr(N=n)=1\slash (n_2 - n_1 +1)$ for $n\in \{n_1,\ldots, n_2\}$ and $n_1 < n_2\in\mathbb{N}.$ Further, the sample size $N$ is assumed independent of the model parameters $\bm \theta.$ This extra random variable modifies the Bayes risk function \eqref{eq:bayesrisk}, which can now be approximated by
\begin{equation}\label{eq:bayesriskmcmcM}
    \hat r_\Omega(\bm{\hat \theta}(\cdot ; \bm \gamma)) = \frac{1}{n_{\mathcal{T}}}\sum_{\{n,\bm \theta,\bm z\} \in \mathcal{T}_{n, \bm \theta, \bm z}}L(\bm \theta,\bm{\hat\theta}(\bm z; \bm \gamma)),
\end{equation}
where $\mathcal{T}_{n, \bm \theta, \bm z}$ is a training set of size $n_{\mathcal{T}}$ consisting of sample sizes $n$ drawn from the $\Unifd(\{n_1, n_1+1,\ldots, n_2\})$ distribution, parameters $\bm\theta$ drawn from the prior $\Omega(\cdot),$ and corresponding data set realisations $\bm z$ drawn from the model given $n$ and $\bm \theta$. Thus, during training, it is now necessary to simulate the sample size along with model parameters from the prior and replicated data from the model; however, the general inference method remains the same.

\subsubsection{Censored data} \label{subsec:censdata}

The use of neural Bayes estimators for censored data was later considered by \citet{Richardsetal2023}. This is essential in multivariate models aimed at capturing the extremal dependence structure and common practice in multivariate extremes; see \citet{LedfordTawn1996} and \cite{Smithetal1997} for early examples of this approach. In these models, low observations are often censored to prevent these non-extreme values from affecting the estimation of the tail dependence. As shown by \citet{Huseretal2016}, treating the censored data as observed leads to more biased estimators.

Consider the random vector $\bm Z_i=(Z_{i1}, \ldots, Z_{id})',$ $i=1, \ldots, n,$ and let $F_{j}^{-1}$ be the inverse cumulative distribution function (cdf) of variable $Z_{ij},$ $j=1, \ldots, d.$ Various censoring schemes can be adopted. One possibility, used by \citet{Richardsetal2023}, is to censor the observations that fall below a high marginal $\tau$-quantile, i.e., if $Z_{ij} < F_j^{-1}(\tau; \bm \theta),$ then $Z_{ij}$ is treated as censored. Instead, in this paper, we censor the observations only if all the components are below their respective $\tau$-marginal quantile, i.e., if $\max_{j=1,\ldots, d}Z_{ij}\slash F_j^{-1}(\tau; \bm \theta)<1,$ then the entire vector $\bm Z_{i}$ is treated as fully censored, otherwise, if at least one component of $\bm Z_i$ has a value above its marginal $\tau$-quantile, the entire vector is treated as uncensored. We do so for consistency in later comparisons between models (see Sections~\ref{subsec:flexiblemodels} and \ref{subsec:modelselsim}), as this is the only censoring scheme that is implemented for likelihood-based inference in some of the models we consider. In addition, this censoring scheme can be less computationally demanding for likelihood-based estimation.

In order for the neural Bayes estimator to account for censored data, \citet{Richardsetal2023} propose standardising data $\bm Z_i$ $(i = 1, \ldots, n)$ to a common marginal scale and setting the censored observations to some constant $c\in \mathbb{R}$, yielding $\bm Z^*_i\in \mathbb{R}^d$ $(i=1, \ldots, n).$ To improve performance of the NBE, this constant $c$ is set to a value outside of the support of the data. In order for information on the censored observations to be passed to the NBE, \citet{Richardsetal2023} propose additionally creating a one-hot encoded vector $ \bm I_i = (I_{i1}, \ldots, I_{id})'$ that identifies which indices of $\bm Z^*_i$ are censored (taking the value $I_{ij}=1$), and which are not (taking the value $I_{ij}=0$). Then, the neural Bayes estimator is trained using an augmented data set $\bm A$ containing the data $\bm Z^* = ((\bm Z^*_1)', \ldots, (\bm Z^*_n)')'$ and the indicator vector $\bm I  = (\bm I_1', \ldots \bm I_n')',$ that is $\bm A = \left((\bm Z^*)', \bm I'\right)'.$ Passing $\bm A$ as the input to the neural network in place of $\bm Z$ in equation \eqref{eq:nbedeepsets} is sufficient to ensure all information about the censoring scheme is given to the NBE. In contrast to the approach of \citet{Richardsetal2023}, under the adopted censoring scheme, each vector $\bm I_i$ is either a $d$-dimensional vector of zeros or a $d$-dimensional vector of ones. Nevertheless, we believe that incorporating $\bm I$ as an input to the neural network remains beneficial, as it carries information about the number of exceedances during training. Moreover, we propose handling the augmented data set $\bm A$ with a bilinear layer (see Section~\ref{subsec:compimpl}) as the input layer of the neural network $\bm\psi(\cdot)$ instead. 

Similarly to the sample size, these NBEs are only (approximately) optimal when applied to data sets where the censoring level is kept the same as the one used for training. However, having a single neural estimator which performs well for any valid $\tau \in (0,1)$---or a subset thereof---is often desirable. This can be achieved by feeding $\tau$ as an extra input to the outer neural network $\bm \phi(\cdot)$ as
\begin{equation}\label{eq:nbedeepsetscensored}
    \bm{\hat \theta}(\bm A,\tau ; \bm \gamma) = \bm \phi\left(\bm S(\bm A,\tau; \bm \gamma_{\bm \psi}); \bm \gamma_{\bm \phi}\right) \quad \text{with} \quad \bm S(\bm A,\tau; \bm \gamma_{\bm \psi})= \left(\bm S(\bm A; \bm \gamma_{\bm \psi})', \tau\right)', 
\end{equation}
where $\bm S(\bm A; \bm \gamma_{\bm \psi}) = \frac{1}{n}\sum_{i=1}^n \bm \psi(\bm A_i;\bm \gamma_{\bm \psi})$ as in \eqref{eq:nbedeepsets} with $\bm A_i$ in place of $\bm Z_i$.

In this context, the censoring level must be treated as a random variable $T,$ which requires an additional prior. As we are not interested in censoring too low, a prior $T \sim \Unifd(\tau_1,\tau_2)$ concentrated around high quantiles (i.e., with probability values $0<\tau_1 <\tau_2 < 1$ close to 1) is a natural choice. Thus, each data set now has a censoring level associated with it, making it possible to have different censoring levels for the different training parameter vectors and corresponding data samples. The Monte Carlo approximation of the Bayes risk in this case takes now the form
\begin{equation}\label{eq:bayesriskmcmccen}
    \hat r_\Omega(\bm{\hat \theta}(\cdot,\cdot;\bm \gamma)) = \frac{1}{n_{\mathcal{T}}}\sum_{(\tau,n,\bm \theta,\bm a) \in \mathcal{T}_{\tau,n,\bm\theta,\bm a}} L(\bm \theta,\bm{\hat\theta}(\bm a,\tau; \bm \gamma)),
\end{equation}
where $\mathcal{T}_{\tau,n,\bm\theta,\bm a}$ is a training set of size $n_{\mathcal{T}}$ consisting of censoring levels $\tau$ drawn from the continuous $\Unifd(\tau_1,\tau_2)$ distribution with density $1\slash (\tau_2-\tau_1)$ over $[\tau_1,\tau_2],$ sample sizes $n$ drawn from the discrete $\Unifd(\{n_1, n_1+1,\ldots, n_2\})$ distribution as above, parameters $\bm\theta$ drawn from the prior $\Omega(\cdot),$ and corresponding data set realisations $\bm a$, drawn from the model given $n$ and $\bm \theta,$ that are masked at the simulated censoring level $\tau$ and augmented with the threshold exceedances indicator vector as described above.

\subsection{Model selection with neural Bayes classifiers} \label{subsec:modelselection} 

From a Bayesian perspective, model selection is naturally performed by treating the model type as a random variable, $M$, which is inferred along with model parameters, based on data set $\bm Z=(\bm Z_1',\ldots,\bm Z_n')'\in\mathcal{S}^n$. Consider a finite collection of $K\geq2$ candidate parametric models, such that $M$ takes values in $\{1, \ldots, K\}$, and assume that a priori $p_m:=\Pr(M=m)>0$ $(m=1,\ldots,K)$ with $\sum_{m=1}^K p_m=1$. In our experiments, we specify a uniform prior, i.e., $p_m=1\slash K$ for all $m=1,\ldots,K$, indicating that the data $\bm Z$ have an equal probability of being drawn from each model a priori; however, other choices are possible. Our interest thus lies in approximating the posterior distribution of $(\bm\theta',M)'$, i.e., the distribution of ${(\bm\theta',M)'\mid \bm Z}$---or functionals thereof---where $\bm\theta=(\bm\theta_1',\ldots,\bm\theta_K')'$ is the combined vector of model parameters for all candidate models, and $\bm\theta_m$ is the parameter vector for the $m$-th model. This joint inference task may be drastically simplified (without loss) by noting that the conditional distribution of ${(\bm\theta',M)'\mid \bm Z}$ can be written as the product between the distribution of ${\bm\theta\mid (\bm Z',M)'}$ and that of ${M\mid\bm Z}$, both of which can be inferred separately. Inferring ${\bm\theta\mid (\bm Z',M)'}$ can be further simplified by splitting the problem into model-specific inference tasks, i.e., by separately inferring ${\bm\theta_m\mid (\bm Z',M=m)'}$ for each $m=1,\ldots,K$. This can be achieved efficiently by training different NBEs for each specific candidate model $m=1,\ldots,K$; recall Section~\ref{subsec:nbe}. Therefore, to perform amortised model selection, the only missing piece is a fast inference technique to compute, or rather approximate, the distribution of ${M\mid\bm Z}$. Bayes' theorem implies that 
\begin{equation}
     \Pr(M=m\mid \bm Z=\bm z)=\frac{p_m\int_{\Theta_m}f_m(\bm z\mid\bm\theta_m)\dint\Omega_m(\bm \theta_m)}{\sum_{m=1}^K p_m\int_{\Theta_m}f_m(\bm z\mid\bm\theta_m)\dint\Omega_m(\bm \theta_m)},\label{eq:MgivenZ}
\end{equation}
where $\Theta_m$ is the parameter space of $\bm\theta_m,$ $\Omega_m(\cdot)$ is a prior measure for $\bm\theta_m,$ and ${f_m(\bm z\mid\bm\theta_m)}$ denotes the density of $\bm Z$ (evaluated at the value $\bm z$) given $M=m$ and $\bm\theta_m.$ In other words, $f_m(\bm z\mid\bm\theta_m)$ is the parametric joint density of the $m$-th candidate model (for $n$ independent replicates) for fixed parameter vector $\bm\theta_m.$ Equation~\eqref{eq:MgivenZ} thus shows that the distribution of ${M\mid\bm Z}$ is unavailable when the individual model densities are intractable.

Nevertheless, similarly to amortised parameter estimation, it is possible to construct a neural network that approximates the true discrete distribution of ${M\mid\bm Z=\bm z}$ for \emph{any} data input $\bm Z=\bm z$. Let $\bm p(\bm z)=(p_1(\bm z),\ldots,p_K(\bm z))'$ be the true (target) posterior probabilities of ${M\mid \bm Z=\bm z}$, and let $\hat{\bm p}(\bm z;\bm\gamma)=(\hat p_1(\bm z;\bm\gamma),\ldots,\hat p_K(\bm z;\bm\gamma))'$ denote the corresponding approximate posterior probabilities (appropriately constrained to be positive and sum to one), obtained as the output of a neural network parameterised by the vector $\bm\gamma$. We further write $f(\bm z)$, $f_m(\bm z)$ and $f_m(\bm z\mid\bm\theta_m)$ for the densities of $\bm Z$ (marginalised over both $\bm\theta_m$ and $M$), $\bm Z\mid M=m$ (marginalised over $\bm\theta_m$ only), and $\bm Z\mid(\bm\theta_m,M=m)$, respectively, and let $\mathcal{S}^n$ and $\mathcal{S}_m^n$ be the sample spaces of $\bm Z$ and $\bm Z\mid M=m$, respectively. Following the arguments in \citet{Zammitetal2024}, the required inference task reduces to finding the optimal $\bm\gamma$ that minimises the \emph{expected} forward Kullback--Leibler divergence between $\bm p(\bm z)$ and $\hat{\bm p}(\bm z; \bm{\gamma})$, that~is
\begin{align}\label{eq:cross-entropy.loss}
    \bm\gamma^\star&=\argmin_{\bm \gamma}{\E}_{\bm Z}\left[\sum_{m=1}^K\log\left\{\frac{p_m(\bm Z)}{\hat p_m(\bm Z;\bm\gamma)}\right\}p_m(\bm Z)\right]\nonumber\\
    &=\argmin_{\bm \gamma}{\E}_{\bm Z}\left[-\sum_{m=1}^K\log\{\hat p_m(\bm Z;\bm\gamma)\}p_m(\bm Z)\right]\nonumber\\
    &=\argmin_{\bm \gamma}-\sum_{m=1}^K\int_{\mathcal{S}^n}\log\{\hat p_m(\bm z;\bm\gamma)\}p_m(\bm z)f(\bm z)\dint\bm z\nonumber\\
    &=\argmin_{\bm \gamma}-\sum_{m=1}^K\int_{\mathcal{S}_m^n}\log\{\hat p_m(\bm z;\bm\gamma)\}p_mf_m(\bm z)\dint\bm z\nonumber\\
    &=\argmin_{\bm \gamma}-\sum_{m=1}^Kp_m\int_{\Omega_m}\int_{\mathcal{S}_m^n}\log\{\hat p_m(\bm z;\bm\gamma)\}f_m(\bm z\mid\bm\theta_m)\dint\bm z\,\dint\Omega_m(\bm\theta_m)\nonumber\\
    &\approx\argmin_{\bm \gamma}-\frac{1}{n_{\mathcal{T}}}\sum_{\{m,\bm\theta_m,\bm z\}\in\mathcal{T}_{m,\bm\theta_m,\bm z}}\log\left\{\hat{p}_m(\bm z;\bm\gamma)\right\},
\end{align}
where $\mathcal{T}_{m,\bm\theta_m,\bm z}$ is a training set of size $n_{\mathcal{T}}$ consisting of models $m$ drawn from the prior for $M$ with probabilities $\bm p=(p_1,\ldots,p_K)'$, model parameters $\bm\theta_m$ drawn from the prior $\Omega_m(\cdot)$ (i.e., given the model $m$), and corresponding data set realisations $\bm z$ drawn from the fixed model $m$ given parameters $\bm\theta_m$. Note that the fourth equality above is justified by a simple application of Bayes' theorem. With data $\bm Z$, amortised model selection can thus proceed based on $\hat{\bm p}(\bm Z;\bm\gamma^\star)$ by selecting the model $m\in\{1,\ldots,K\}$ that maximises the inferred posterior probability $\hat{p}_m(\bm Z;\bm\gamma^\star).$ This is akin to a classification problem, where each model type is treated as a different `class' (among $K$ classes), and where the data $\bm Z$ are labelled according to the class that is the most likely a posteriori. The optimisation in equation~\eqref{eq:cross-entropy.loss} is indeed identical to minimising the categorical cross-entropy loss function, which is a popular and natural choice in multiclass classification problems. Thus, by analogy with NBEs, we call our amortised model selection tool the neural Bayes classifier (NBC). If suitably trained, the NBC is expected to perform well for any data $\bm Z$ drawn from one of the $K$ models, provided that $n_{\mathcal{T}}$ is sufficiently large in equation~\eqref{eq:cross-entropy.loss}. However, as with NBEs, the architecture of NBCs should depend on the structure of the data at hand and the desired level of amortisation. In particular, with replicated data (i.e., with $n\geq2$), the DeepSets architecture introduced in Section~\ref{subsec:nbe} can be applied, and similar techniques to those used for parameter estimation can be employed to ensure the NBC remains suitable for censored data, variable sample sizes, and/or variable censoring levels. We note, however, that there is no restriction in using a different neural network architecture to that used for the NBEs.

\subsection{Implementation details} \label{subsec:compimpl}

In this work, a multilayer perceptron (MLP) neural network is used for both $\bm\psi$ and $\bm\phi$ in the DeepSets construction~\eqref{eq:nbedeepsets}, for both parameter and model selection procedures. Two key components of a standard MLP are its activation functions and vector of parameters $\bm \gamma$, with the latter containing what is often referred to as weights and biases. The activation functions introduce non-linearity into the MLP construction, allowing the network to learn and represent the complexities of the underlying data. The parameters, on the other hand, determine the strength (weights) of the connection between two neurons (i.e., nodes of the neural network), and shift the input of the activation functions (biases). Consider a generic MLP with $L$ layers, and let $\bm W_l$ and $\bm b_l$ represent the matrix of weights and vector of biases (of appropriate dimensions) of layer $l=1, \ldots, L,$ respectively. The MLP is constructed by taking the input $\bm X$ and transforming it into the output $\bm Y(\bm X)$ as
\begin{align*}
    \bm h_1 &= \sigma_1(\bm W_1 \bm X + \bm b_1), \\
    \bm h_l &= \sigma_l(\bm W_l \bm h_{l-1} + \bm b_l), \quad l = 2, \ldots, L-1, \\
    \bm Y(\bm X) & = \sigma_L(\bm W_L \bm h_{L-1} + \bm b_L),
\end{align*}
where $\bm h_l$ are often denoted as the hidden states \citep{Zammitetal2024}, and $\sigma_l(\cdot)$ is the activation function of the $l$-th layer ($l=1,\ldots,L$). As mentioned in Section~\ref{subsec:censdata}, a bilinear layer is used as the input layer $l = 1$ of the neural network $\bm \psi$ when handling censored data. As opposed to the layer $\bm h_1$ defined above, this type of layer allows for two separate inputs, that may be of different dimension, to be combined into a single input. In particular, given the augmented data set $\bm A_i = ((\bm Z_i^*)', \bm I'_i)'$ for $i=1, \ldots, n,$ the $k$-th element of the first hidden state $\bm h_1$ for each data replicate $i$ takes the form $h_1^{(k)} = \sigma_1\left((\bm Z_i^*)'\bm W^{*(k)}_1\bm I_i + b_1^{(k)}\right)$ with $\bm W^{*(k)}_1\in\mathbb{R}^{d\times d}$, for $k = 1, \ldots, d_{\bm h_1}$ where $d_{\bm h_1}$ is the output dimension of $\bm h_1.$ This corresponds to a sum of elementwise interactions between the standardised data input $\bm Z_i^*\in \mathbb{R}^d$ and the indicator vector $\bm I_i\in \mathbb{R}^d,$ $i=1, \ldots, n.$ As before, the $n$ data replicates are combined using the DeepSets framework (recall Sections~\ref{subsec:NBEgeneral} and \ref{subsec:censdata}). We note, however, that with the adopted censoring scheme (i.e., with $\bm I_i$ either being a $d$-dimensional vector of zeros or ones), identical results are obtained when considering the original data input $\bm Z$ instead.  

Unless stated otherwise, the Rectified Linear Unit (ReLU) activation function is used when training the neural Bayes estimator and classifier, except in the final layer $L$. Writing $\bm x_l = \bm W_l \bm h_{l-1}+\bm b_l$ for $l \geq 2,$ (and $\bm x_1 = \bm W_1 \bm X + \bm b_1$ for $l =1$), the ReLU function returns the same value if the corresponding element of $\bm x_l$ is positive, and 0 otherwise, i.e, $\sigma_l(\bm x_l) = \max\{\bm x_l, \bm 0\}$ applied elementwise. For the final layer $L$ of the $\bm\phi$ network (recall the DeepSets framework in~\eqref{eq:nbedeepsets}), different activation functions are used so that constraints on the support of the parameters are satisfied. In particular, when training the NBE, if the $j$-th parameter is strictly positive, the softplus activation function is applied, i.e., $\sigma_L(x_{L;j})=\log(1+\exp(x_{L;j})),$ where $x_{L;j}$ denotes the $j$-th element of the vector $\bm x_L$. If the parameter is bounded in the interval $[a,b],$ a layer compression is used instead; such a layer uses a logistic function $\sigma_L(x_{L;j}) = {a + (b-a)\slash (1+\exp(-x_{L;j}))},$ restricting the $j$-th parameter to be within $[a,b].$ The identity function, $\sigma(x_{L;j})=x_{L;j},$ is used for parameters whose support is the real line. When training the NBC, the softmax activation function is used in the final layer $L$ so that the output $\hat{\bm p}(\bm Z;\bm\gamma)$ is a valid vector of probabilities, i.e., all elements of $\hat{\bm p}(\bm Z;\bm\gamma)$ are positive and $\sum_{m=1}^K\hat{p}_m(\bm Z;\bm\gamma)=1$. More specifically, writing $\bm x_L=(x_{L;1}, \ldots, x_{L;K})'\in \mathbb{R}^{K}$, the softmax activation function takes the form $\sigma_L(\bm x_L) = \exp(\bm x_L)\slash \sum_{m=1}^K\exp(x_{L;m}),$ with componentwise operations again. 

As mentioned in Sections~\ref{subsec:nbe} and \ref{subsec:modelselection}, neural Bayes estimators and classifiers are trained by minimising a specific objective function defined either as the empirical Bayes risk (for parameter estimation) or as the expected Kullback--Leibler divergence (for model selection), with respect to the neural network parameters $\bm \gamma.$ More specifically, the training process involves learning the optimal parameters $\bm \gamma$ that map data inputs to parameter estimates or posterior probabilities. This optimisation is done via back-propagation and the stochastic gradient descent (SGD) algorithm, where parameters are iteratively updated to minimise the selected objective function. Moreover, during training, two types of data sets are used: the \emph{training} and \emph{validation} sets, denoted respectively by $\mathcal{T}$ and $\mathcal{V}$ from now on. Both data sets are passed through the network and are refreshed after every epoch; here an epoch is defined as one full cycle through all the input data in the training set during the SGD process. Precisely, the training set $\mathcal{T}$ contains data used to train the model, which are used to update the parameters of the network by minimising the objective function using SGD. On the other hand, the validation set $\mathcal{V}$ does not contribute to the parameter updates; instead, it is used to assess the ability of the model to generalise to new data by evaluating the same objective function on the validation set, which helps to avoid overfitting, and it is used to define an early-stopping criterion for the algorithm (see, e.g., \citealp{Prechelt1998}). Finally, the performance of the trained estimator to model new data is objectively assessed by evaluating the same objective function on a \emph{test} set, which is not used during training (nor validation) at all. All the computations are performed using the {\it NeuralEstimators} \citep{SainsburysDaleRpackage} and {\it Flux} \citep{Innes2018} packages in \textsf{Julia}; see \url{https://msainsburydale.github.io/NeuralEstimators.jl/dev/} and \url{https://fluxml.ai/Flux.jl/stable/}, respectively, for the full documentation.
\section{Bivariate models of interest} \label{section:models}

From now on, we work in the bivariate setting. Specifically, we focus on the modelling of the joint tail behaviour of the (single) random vector $\bm Z = (Z_{1}, Z_{2})'.$ We note, however, that the inference methodology can be applied to higher dimensions. We are particularly interested in bivariate models that are suitable for the modelling of both types of extremal dependence structures. More specifically, we focus on flexible models that allow interpolation between asymptotic dependence and independence, as well as on the weighted copula model (WCM) proposed by \citet{Andreetal2024}, which is designed to represent both the body and tail of a data set. We aim to provide a tool for fast inference for a variety of bivariate models exhibiting complex dependence structures. Evaluation of their likelihood functions relies heavily on numerical integration and inversion of functions; this results in computationally costly likelihood-based inference procedures and may otherwise limit the use of these models in practice. Furthermore, likelihood-based inference for the WCM, which is a dynamic mixture model, is currently infeasible when one of its components is taken as one of the models able to interpolate between the two classes of extremal dependence. Section \ref{subsec:copulas} reviews basics of copula modelling, while some background on extremal dependence measures is given in Section \ref{subsec:measures}. The models of interest are introduced in Sections \ref{subsec:flexiblemodels} and \ref{subsec:wcm}. 

\subsection{Copula modelling} \label{subsec:copulas}

The dependence between variables $Z_1$ and $Z_2$ can be characterised by means of copulas. Let $Z_1\sim F_{Z_1}$ and $Z_2 \sim F_{Z_2}$ be the margins and let $F_{Z_1,Z_2}$ denote their joint distribution function. According to Sklar's theorem \citep{Sklar1959}, the underlying copula $C:[0,1]^2\to[0,1]$ of $\bm Z = (Z_1, Z_2)'$ is obtained as
\begin{equation}\label{eq:sklar}
    C(u_1, u_2)=F_{Z_1,Z_2}\left(F_{Z_1}^{-1}(u_1), F_{Z_2}^{-1}(u_2)\right), \quad (u_1, u_2)'\in [0,1]^2.
\end{equation}
When $Z_1$ and $Z_2$ are continuous variables, the copula $C$ is unique and represents the joint distribution function of $\bm U=(U_1, U_2)',$ where $U_1=F_{Z_1}(Z_1)$ and $U_2=F_{Z_2}(Z_2)$ are $\Unifd(0,1)$ random variables. This result is useful since it is sufficient to marginally transform the data to a uniform scale through the probability integral transform and represent their joint behaviour via a copula model $C.$ When it exists, the copula density $c(u_1, u_2)$ can be obtained by taking the second derivative of $C$ with respect to $u_1$ and $u_2$ as
\begin{equation}
    c(u_1, u_2) = \frac{f_{Z_1, Z_2}\left(F_{Z_1}^{-1}(u_1), F_{Z_2}^{-1}(u_2)\right)}{f_{Z_1}\left(F_{Z_1}^{-1}(u_1)\right)f_{Z_2}\left(F_{Z_2}^{-1}(u_2)\right)}, \quad (u_1, u_2)'\in [0,1]^2,
\end{equation}
where $f_{Z_i}$ is the probability density function (pdf) of variable $Z_i$ for $i=1, 2,$ and $f_{Z_1, Z_2}$ is the joint pdf of $Z_1$ and $Z_2.$ When $F_{Z_i}$ does not have an explicit form, it often needs to be computed through numerical integration; this might also be necessary for $f_{Z_i}.$ Additionally, inversion techniques are often required to compute $F_{Z_i}^{-1}.$ All of these can require substantial computational resources.

\subsection{Bivariate extremal dependence measures} \label{subsec:measures}

When interest lies in the joint extremes of a bivariate random vector, a key element is to correctly identify its extremal dependence behaviour, i.e., whether large values in different components of this vector are likely to occur simultaneously or not. Misidentifying the extremal dependence structure may indeed lead to inaccurate representations of the extremes, and incorrect extrapolations (see, e.g., \citealp{LedfordTawn1996, LedfordTawn1997}, \citealp{HeffernanTawn2004}, \citealp{HuserGenton2016}). Intuitively speaking, asymptotic dependence (AD) is present if the most extreme values of the components of the random vector $(Z_1, Z_2)'$ can occur together, and asymptotic independence (AI) is present otherwise. This extremal behaviour is often quantified through the tail dependence coefficient $\chi \in [0,1]$ (see, e.g., \citealp{Joe1997}) and/or through the residual tail dependence coefficient $\eta\in(0,1]$ \citep{LedfordTawn1996}. The coefficient $\chi$ can be obtained as $\chi=\lim_{u\to 1}\chi(u),$ when it exists, with
\begin{equation}\label{eq:chi}
    \chi(u)=\Pr(F_{Z_2}(Z_2)>u \mid F_{Z_1}(Z_1)>u)=\frac{\Pr(F_{Z_1}(Z_1)>u, F_{Z_2}(Z_2)>u)}{1-u}, \; u\in(0,1).
\end{equation} 
The vector $(Z_1,Z_2)'$ is asymptotically independent if $\chi=0,$ and asymptotically dependent if $\chi>0.$ Given a function $\mathcal{L}$ that is slowly-varying at zero (i.e, $\mathcal{L}(cx)\slash\mathcal{L}(x) \to 1$ as $x\to 0$ for any $c>0$), \citet{LedfordTawn1996} assume the joint tail may be expressed as
\begin{equation}\label{eq:ledtawn}
    \Pr(F_{Z_2}(Z_2)>u \mid F_{Z_1}(Z_1)>u) = \mathcal{L}(1-u)(1-u)^{1\slash \eta -1}, \quad \eta\in(0, 1],\, u\to 1.
\end{equation}
If $\eta=1$ and $\mathcal{L}(1-u) \not\to 0$ as $u\to1,$ then $(Z_1,Z_2)'$ is asymptotically dependent with $\chi = \lim_{u\to 1}\mathcal{L}(1-u),$ otherwise it is asymptotically independent, with larger values of $\eta\in(0,1]$ indicating stronger dependence. Similarly to $\chi(u)$ given in equation~\eqref{eq:chi}, a sub-asymptotic version of $\eta$ can be obtained from equation~\eqref{eq:ledtawn}.

Taken together, $\chi>0$ provides a summary measure of dependence within the AD class (with $\eta=1$), while $\eta\leq 1$ provides a summary within the AI class (with $\chi = 0$). Available models in the multivariate extremes literature are often only suitable for one extremal dependence class. That is, aside from boundary points of the parameter space, traditional models either yield $\{\chi > 0, \eta = 1\}$ or $\{\chi = 0, \eta\leq 1\},$ but cannot span across both classes. However, in recent years, more flexible methods which are able to capture both types of dependence have been proposed; see for instance \citet{Wadsworthetal2017}, \citet{HuserWadsworth2019} and \citet{Engelkeetal2019}. Each of these can be written as a random scale construction, which we introduce in the following section.

\subsection{Random scale construction models} \label{subsec:flexiblemodels}

A range of bivariate dependence models can be constructed using a random scale representation based on `radial' and `angular' coordinates. More specifically, a random scale mixture vector $(Z_1,Z_2)'$ is constructed as follows
\begin{equation}\label{eq:randomconstruction}
    (Z_1, Z_2)' = R(V_1, V_2)', \quad R\ind (V_1, V_2)',
\end{equation}
where $R>0$ is the `radial' variable and is assumed to follow a non-degenerate distribution, and $(V_1, V_2)' \subseteq \mathbb{R}^2$ is the vector of `angular' components. Different models can be obtained by varying the distributions and constructions of $R$ and $(V_1,V_2)'.$ Depending on the precise specification, these may be able to interpolate between the two regimes of extremal dependence; see \citet{Engelkeetal2019} for a detailed overview of the dependence properties arising from this construction. Interest lies in exploiting the copula $C$ of these flexible models. Moreover, since these aim at capturing the extremal dependence of the vector $(Z_1,Z_2)',$  non-extreme values are often censored to prevent their influence on the joint tail. We now present four particularly interesting bivariate models with construction~\eqref{eq:randomconstruction}, each of which can yield both $\{\chi>0,\eta=1\}$ and $\{\chi=0,\eta\leq 1\},$ depending on their parameter vectors, with the transition between these two regimes occurring at interior points of the parameter space.

\paragraph{Model W.} Let $V\sim \Betad(\alpha, \alpha),$ and let $R$ follow a generalised Pareto distribution (GPD) with scale parameter 1 and shape parameter $\xi\in\mathbb{R},$ i.e, $R\sim \GPDd(1,\xi)$ with cdf $F_R(r)=1-(1+\xi r)^{-1\slash \xi},$ $r \geq 0.$ \citet{Wadsworthetal2017} propose a model with
\begin{equation}
    (V_1,V_2)'=\frac{(V,1-V)'}{\max(V, 1-V)} \in \Sigma=\{{\bf v} = (v_1, v_2)' \in \mathbb{R}^2_+: \max(v_1, v_2)=1\}. 
\end{equation}
Given the model construction, $C(u_1, u_2)$ is assumed to hold only when $\|(Z_1,Z_2)'\|_\infty=\max(Z_1, Z_2)$ is large. This model is able to smoothly interpolate between the two classes of asymptotic dependence through the parameter $\xi.$ For $\xi >0,$ $(Z_1,Z_2)'$ is asymptotically dependent with 
\begin{equation}\label{eq:chiadwads}
    \chi_{W} = \E\left(\min\left\{\frac{V_1^{1\slash \xi}}{\E\left(V_1^{1\slash \xi}\right)}, \frac{V_2^{1\slash \xi}}{\E\left(V_2^{1\slash \xi}\right)}\right\}\right)>0 \quad \text{and} \quad \eta_{W}=1.
\end{equation}
If $\xi\leq 0,$ then asymptotic independence is present with $\eta_{W}=(1-\xi)^{-1} \leq 1$ and $\chi_{W} = 0.$ More details can be found in \citet{Wadsworthetal2017}. 

\paragraph{Model HW.} \citet{HuserWadsworth2019} detail a model construction providing flexible extremal dependence structures for spatial processes. However, they also propose a bivariate model able to transition between different types of dependence. In this model, both $R$ and the vector $(V_1,V_2)'$ are marginally Pareto distributed with different shape parameters. More specifically, the marginal cdfs of each variable are given by $F_R(r) = 1-r^{-1\slash \delta},$ $r\geq 1,$ and $F_{V_1}(v_1) = 1-v_1^{-1\slash (1-\delta)},$ $v_1\geq 1,$ with $F_{V_2}(v_2)$ defined analogously, for $\delta \in (0,1).$ Additionally, we assume here that $(V_1,V_2)'$ follows a bivariate Gaussian copula with correlation parameter $\omega \in (-1,1);$ see also \citet{GongHuser2022b, GongHuser2022}.

This model is able to interpolate between the two classes of extremal dependence through the parameter $\delta;$ when $\delta > 1\slash 2,$ the tail of $R$ is heavier than $(V_1,V_2)'$ and so $(Z_1,Z_2)'$ is asymptotically dependent with
\begin{equation}\label{eq:chihw}
    \chi_{HW} = \frac{\E(\min\{V_1,V_2\}^{1\slash\delta})}{\E(V_1^{1\slash\delta})}>0 \quad \text{and} \quad \eta_{HW}=1.
\end{equation} 

When $\delta \leq 1\slash 2,$ $(Z_1,Z_2)'$ is asymptotically independent with
\begin{equation}\label{eq:etahw}
    \eta_{HW} = \begin{cases}
                    1, & \ift \delta = 1\slash 2, \\
                    \delta \slash (1-\delta), & \ift \eta_V\slash(1+\eta_V) < \delta < 1\slash 2, \\
                    \eta_V, & \ift \delta \leq \eta_V\slash(1+\eta_V),
                \end{cases}
\end{equation}
where $\eta_V<1$ is the residual tail dependence coefficient \eqref{eq:ledtawn} of vector $(V_1,V_2)'.$ More details can be found in \citet{HuserWadsworth2019}.

The final two models were proposed by \citet{Engelkeetal2019}. Similarly to the model introduced by \citet{Wadsworthetal2017}, let $V\sim \Betad(\alpha,\alpha),$ $\alpha >0.$ 

\paragraph{Model E1.} For the first model, $R$ follows a Weibull distribution with distribution function $F_R(r)=1-\exp(-r^{\beta}),$ $r>0,\, \beta>0,$ and the angular components are constructed as follows
\begin{equation}
    (V_1,V_2)'=\frac{(V,1-V)'}{\nu(V,1-V)'},
\end{equation}
where $\nu(V,1-V)=\mu\max(V, 1-V)+(1-\mu)\min(V,1-V)$ and $\mu \geq 1\slash 2.$ For this model, the extremal dependence is controlled by $\mu;$ when $\mu \leq 1,$ $Z_1$ and $Z_2$ are asymptotically independent with $\chi_{E1} = 0$ and $\eta_{E1}= \mu^\beta.$ If $\mu >1,$ they are asymptotically dependent with
\begin{equation}
    \chi_{E1} = \frac{2(\mu-1)}{2\mu-1} \quad \text{and} \quad \eta_{E1}= 1.
\end{equation}

\paragraph{Model E2.} For the second model, $R\sim \GPDd(1,\xi)$ with $\xi \in \mathbb{R}$ as in \citet{Wadsworthetal2017}. The angular components $V_1$ and $V_2$ are now independent of each other and distributed as $V,$ that is $V_1, V_2\overset{\text{iid}}{\sim} \Betad(\alpha, \alpha).$ The extremal dependence of this model is determined by $\xi.$ In particular, if $\xi > 0,$
\begin{equation}
    \chi_{E2} = \frac{\E\left(\min\{V_1, V_2\}^{1\slash\xi}\right)}{\E\left(V^{1\slash\xi}\right)} \quad \text{and} \quad \eta_{E2}= 1,
\end{equation}
and $(Z_1,Z_2)'$ are asymptotically dependent. When $\xi \leq 0,$ the variables are asymptotically independent with
\begin{equation}
    \chi_{E2} = 0 \quad \text{and} \quad \eta_{E2}= \begin{cases}
        1, & \ift \xi =0, \\
        (1-\xi\alpha)\slash (1- 2 \xi \alpha), &\ift \xi <0.
    \end{cases}
\end{equation}

Although Models E1 and E2 were proposed in \citet{Engelkeetal2019}, to the best of our knowledge they have not been used for model inference or in concrete data applications in the literature previously.


\subsection{Weighted copula model} \label{subsec:wcm}

\citet{Andreetal2024} propose a model that is able to accurately represent both the body and tail regions of a data set, while ensuring a smooth transition between them. Let $c_t$ denote the density of a copula tailored to the tail, with parameter $\bm \lambda_t,$ and $c_b$ denote the density of a copula tailored to the body, with parameter $\bm \lambda_b.$ For $(x_1,x_2)'\in[0,1]^2$, the density of the proposed model is given by dynamically mixing $c_t$ with $c_b$ as follows
\begin{equation} \label{eq:wcm}
    h(x_1,x_2; \bm \theta) = \frac{\pi(x_1,x_2;\kappa)c_t(x_1,x_2;\bm \lambda_t) + [1-\pi(x_1,x_2;\kappa)]c_b(x_1,x_2;\bm\lambda_b)}{c(\bm\theta)},
\end{equation}
where $\pi(x_1,x_2;\kappa): [0,1]^2 \to [0,1]$ is a dynamic weighting function, $\bm \theta =(\bm\lambda_t', \bm\lambda_b', \kappa)'$ is the vector of model parameters, and $c(\bm\theta)$ a normalising constant. Note that model~\eqref{eq:wcm} does not have uniform margins in general, which makes likelihood-based inference difficult. The weighting function is dynamic and specified such that it is increasing in $x_1$ and $x_2$ for a fixed value of $\kappa;$ in particular, more weight is given to $c_b$ for small values of $(x_1,x_2)',$ and more weight is given to $c_t$ for large values of $(x_1,x_2)'.$ Similarly to the models introduced in Section \ref{subsec:flexiblemodels}, the interest lies in exploiting the copula \linebreak $C(u_1,u_2;\bm \theta) = H\left(H_{X_1}^{-1}(u_1),H_{X_2}^{-1}(u_2);\bm \theta\right)$ of density \eqref{eq:wcm}, where $H_{X_1}$ and $H_{X_2}$ are the marginal cdfs of the cdf $H$ corresponding to $h$ in~\eqref{eq:wcm}. As shown in \citet{Andreetal2024}, the model shows some interesting features regarding extremal dependence, with $c_b$ potentially having an influence on $\chi$ depending on the weighting function chosen. More details can be found in \citet{Andreetal2024}.
\section{Simulation studies} \label{section:simulation}

Several simulation studies are performed to assess the trained neural Bayes estimators in different scenarios. We present selected studies in this section, and give the remaining ones in the Supplementary Material. All general settings and priors are outlined in Section~\ref{subsec:generalsetting}. In Section~\ref{subsec:parestimationsim} we then study the performance of NBEs in estimating the model parameters of the models described in Sections~\ref{subsec:flexiblemodels} and \ref{subsec:wcm}. We assess the efficacy of the neural Bayes classifier for model selection in Section~\ref{subsec:modelselsim}, focusing on the four models from Section~\ref{subsec:flexiblemodels}. In Section~\ref{subsec:missscenarios} we investigate the performance of the NBCs and NBEs trained for model selection and parameter estimation, respectively, in misspecified scenarios. 

\subsection{General settings} \label{subsec:generalsetting}

In all simulations, the sample size $n$ and, when applicable, the censoring level $\tau$ are assumed random and independent realisations of variables distributed, respectively, as $N\sim\Unifd(\{100,\allowbreak 101,\ldots, 1500\})$ and $T \sim \Unifd(0.5, 0.99).$ Each parameter of the four models mentioned in Section~\ref{subsec:flexiblemodels} are also assumed independent a priori and uniformly distributed. In particular, for Model W \citep{Wadsworthetal2017} we take $\alpha \sim \Unifd(0.2, 15)$ for the parameter of the Beta distribution, and $\xi \sim \Unifd(-2,1)$ for the shape parameter of the GPD. For Model HW \citep{HuserWadsworth2019}, we take $\delta \sim \Unifd(0,1)$ and $\omega\in \Unifd(-1,1).$  For Model E1 \citep{Engelkeetal2019}, we take $\alpha \sim \Unifd(0.2, 15)$ for the parameter of the Beta distribution, $\beta\sim \Unifd(0, 15)$ for the Weibull parameter, and $\mu\sim \Unifd(0.5,4).$ For Model E2 \citep{Engelkeetal2019}, we assume that $\alpha \sim \Unifd(0.2,15)$ and $\xi\sim\Unifd(-2,1).$ For the weighted copula model from equation~\eqref{eq:wcm}, we take the weighting function to be $\pi(x_1, x_2;\kappa)=(x_1x_2)^{\exp(\kappa)},$ with $\kappa \sim \Unifd(-3.51, 1.95)$ as the prior for the weighting function parameter; based on previous analysis of this model, $\kappa\in(-3.51, 1.95)$ ensures that there is a good representation of both copula components, $c_b$ and $c_t,$ in each drawn sample.

In order to train the NBE for parameter estimation, training and validation sets are generated (recall Sections~\ref{subsec:nbe} and \ref{subsec:compimpl}). For the training set $\mathcal{T}$, we sample $100\,000$ parameter vectors $\bm\theta$ independently, and for each of them we sample five different data set realisations $\bm z$ from the model, each associated with a different (randomly sampled) sample size $n$ and censoring level $\tau$ (when applicable). Therefore, the training set $\mathcal{T}$ for the NBE is of size $n_{\mathcal{T}}=500\,000.$ For the validation set $\mathcal{V}$, we use the same sampling strategy, but we only sample $20\,000$ parameter vectors, thus resulting in a set of size $n_{\mathcal{V}}=100\,000.$ In order to train the NBC for model selection (recall Section~\ref{subsec:modelselection}), a similar approach is used, but the model is now also treated as random. For the training set $\mathcal{T}$, we start by generating $100\,000$ models $m$, but only sample a single parameter vector, sample size, censoring level (when applicable), and corresponding data set for each model. Therefore, the size of the training set for the NBC is $n_{\mathcal{T}}=100\,000.$ For the validation set, we use the same strategy but only generate $20\,000$ models, resulting in a validation set of size $n_{\mathcal{V}}=20\,000.$ In all simulations, the rank transform is used to transform the margins of $\bm Z$ to an approximate uniform scale, as required when fitting our copula models. 

The number of layers assumed for each neural network $\bm \phi$ and $\bm \psi$ (recall equation~\eqref{eq:nbedeepsets}) and its parameters are determined experimentally, and in order to reduce the computational intensity, we adopt the `simulation-on-the-fly' technique; see \citet{SainsburyDaleetal2024} for more details. We also adopt an early stopping rule whereby the training step finishes if the objective function (i.e., empirical Bayes risk for the NBE or expected Kullback--Leibler divergence for the NBC) computed based on the validation set has not decreased in 5 consecutive epochs for parameter estimation and 10 consecutive epochs for the model selection procedure. All the simulations are performed using a high-end computing cluster with a NVidia V100 32 GB GPU hardware with 192 GB of memory; see \url{https://lancaster-hec.readthedocs.io/en/latest/gpu.html} for more details (last accessed on 02/09/2025). For reproducibility and to make our methodology available to a broad readership, all the trained NBEs for parameter estimation and NBCs for model selection are available on \url{https://github.com/lidiamandre/NBE_classifier_depmodels}.

\subsection{Parameter estimation assessment} \label{subsec:parestimationsim}

We take the absolute error loss function in equation~\eqref{eq:bayesriskmcmc} and its analogues, thus targeting the marginal posterior median. We first demonstrate the performance of NBEs for uncensored data from the weighted copula model from Section \ref{subsec:wcm}, and for censored data from one of the random scale mixture models mentioned in Section \ref{subsec:flexiblemodels}. The neural network architecture used for parameter estimation is given in Table~\ref{tab:paresttable} of the Supplementary Material. Finally, the assessment of the NBEs is done on a test parameter set with $1000$ parameter vectors $\bm \theta,$ and corresponding test data realisations, each of size $n=1000.$

\begin{figure}[t!]
    \centering
    \includegraphics[width=\textwidth]{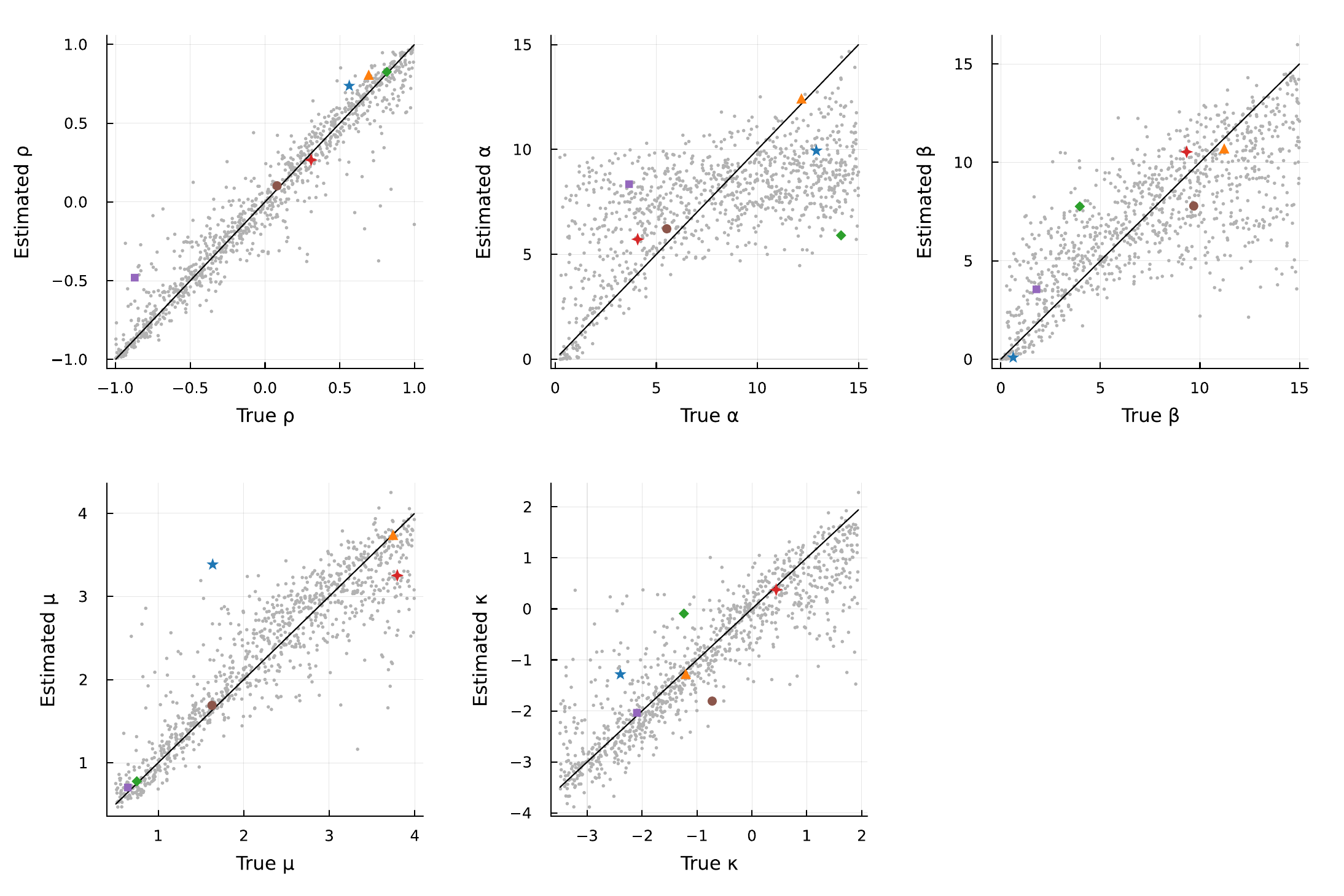}
    \caption{Assessment of the NBE when $c_b$ is the Gaussian copula with parameter $\rho,$ $c_t$ is Model E1 with parameters $\bm \lambda_t=(\alpha,\beta,\mu)',$ and with weighting function $\pi(x_1, x_2;\kappa)=(x_1x_2)^{\exp(\kappa)},$ $x_1, x_2 \in (0,1)$ for a sample size of $n=1000.$ The points highlighted in different shapes and colours refer to parameter configurations used for further diagnostics (see Figure~\ref{fig:chietaNBE}).}
    \label{fig:wcmassessment}
\end{figure}

In the first simulation study, we consider the weighed copula model from equation \eqref{eq:wcm}. In particular, we take $c_b$ to be the Gaussian copula density with correlation parameter $\lambda_b = \rho \in (-1,1),$ for which we take $\rho\sim \Unifd(-1,1),$ and $c_t$ to be the copula density of Model E1 with $\bm \lambda_t= (\alpha, \beta, \mu)'$. This is a configuration for which likelihood-based inference is simply infeasible, owing to the nesting of two complex models requiring numerical integrals and inversion of functions. Figure~\ref{fig:wcmassessment} shows the results; the true values for each parameter are compared with their estimated values given by the trained NBE. It can be seen that there is a bit of variability, in particular for the parameters $\alpha$ and $\beta;$ we note that these are the parameters with the largest prior range. We can see that the NBE generally estimates the weighting parameter $\kappa$ quite well, indicating that it is able to distinguish between the body and tail components of the copula. In addition, estimates of $\mu$ are quite accurate; this means that the two regimes of extremal dependence are reasonably well captured. In particular, $98.28\%$ of the data sets showing asymptotic dependence (i.e., $\mu > 1$) were estimated to be AD, whereas $86.15\%$ of the data sets showing asymptotic independence (i.e., $\mu\leq 1$) were estimated to be AI. Finally, estimates of $\rho$ exhibit the best results. Among negatively correlated data sets (i.e., $\rho < 0$), $96.06\%$ were estimated to be negatively correlated, whereas $92.28\%$ of the positively correlated data sets were correctly identified to be positively correlated. 

It is important to assess the uncertainty of the NBE; we consider two different approaches for this task. For the first, a non-parametric bootstrap procedure is adopted. Specifically, we generate $B=400$ bootstrap samples by resampling with replacement. Then, for each bootstrap sample, the model parameters are estimated using the trained NBE. From these estimates, $95\%$ confidence intervals are constructed by taking the $2.5\%$ and $97.5\%$ quantiles of the samples. This is done for each parameter configuration and data from the test set. While there are no guarantees on the quality of the confidence intervals obtained through this procedure, it is computationally inexpensive to implement owing to the amortised nature of the NBE. This makes the bootstrap procedure a practical choice for exploring the uncertainty of these estimators. For the second approach, we train an additional estimator designed to target credible intervals. To do so, a composite quantile loss function targeting jointly low and high posterior quantiles is used instead. Precisely, this loss function is defined as $L(\bm \theta, \hat{\bm \theta}(\cdot)) = \sum_{t=1}^{Q} \sum_{k=1}^p (\hat\theta_{k,t}(\cdot) -\theta_{k})(\mathbbm{1}_{(\hat\theta_{k,t}(\cdot)>\theta_{k})}-q_t)$ \linebreak for probability levels $q_t\in (0,1)$, $t=1,\ldots,Q,$ and parameter estimators $\hat\theta_{k,t}(\cdot),$ \linebreak $k=1,\ldots,p,$ $t=1,\ldots,Q.$ We train an NBE $\hat{\bm\theta}(\cdot;\bm\gamma)=(\hat\theta_{k,t}(\cdot;\bm\gamma);\,k=1,\ldots,p,$ \linebreak $t=1,\ldots,Q)'$ that outputs posterior quantiles for all parameters and probability levels, and note that the neural network is constrained in such a way that quantiles are properly ordered; see more details in \citet{SainsburyDaleetal2023}. With this estimator, herein referred to as the neural interval estimator, we are able to approximate marginal $95\%$ central credible intervals by taking $Q=2$ and setting $\{q_1, q_2\}=\{0.025, 0.975\}.$ We evaluate the performance of each approach by computing coverage probabilities, with the results presented in Table~\ref{tab:uncwcm}. The bootstrap procedure leads to lower than nominal coverage rates, with the extent of the miscalibration linked to the quality of estimates shown in Figure~\ref{fig:wcmassessment}: the best coverage is obtained for the parameters showing a better correspondence between true and estimated values, such as $\rho$ or $\mu.$ The parameters $\alpha$ and $\beta$ exhibit more bias in their estimation, which is reflected in worse coverage rates. On the other hand, the neural interval estimator is much better calibrated. In the context of NBEs, bootstrap-based confidence intervals have been used extensively to account for the estimation uncertainty; however, to the best of our knowledge, their coverage rates have not been systematically explored.


\begin{table}[!t]
    \centering
    \captionsetup{width=\textwidth}
    \caption{Coverage probability and average length of the $95\%$ uncertainty intervals obtained via a non-parametric bootstrap procedure and via the neural interval estimator averaged over $1000$ models fitted using a NBE (rounded to 2 decimal places).}
    \begin{tabular}{ccccc}
        \toprule
        \multirow{2}{*}{Parameter} & \multicolumn{2}{c}{Bootstrap procedure} & \multicolumn{2}{c}{Interval estimator} \\
        \cmidrule{2-5}
         & Coverage & Length & Coverage & Length \\
        \midrule
        $\rho$ & $0.80$ & $0.27$ & $0.97$ & $0.57$ \\
        $\alpha$ & $0.36$ & $3.12$ & $0.97$ & $11.84$ \\
        $\beta$ & $0.53$ & $3.42$ & $0.96$ & $10.32$ \\
        $\mu$ & $0.67$ & $0.63$ & $0.97$ & $1.67$ \\
        $\kappa$ & $0.70$ & $1.12$ & $0.97$ & $2.54$ \\
        \bottomrule
    \end{tabular}
    \label{tab:uncwcm}
\end{table}

\begin{figure}[t!]
    \centering
    \includegraphics[width=\textwidth]{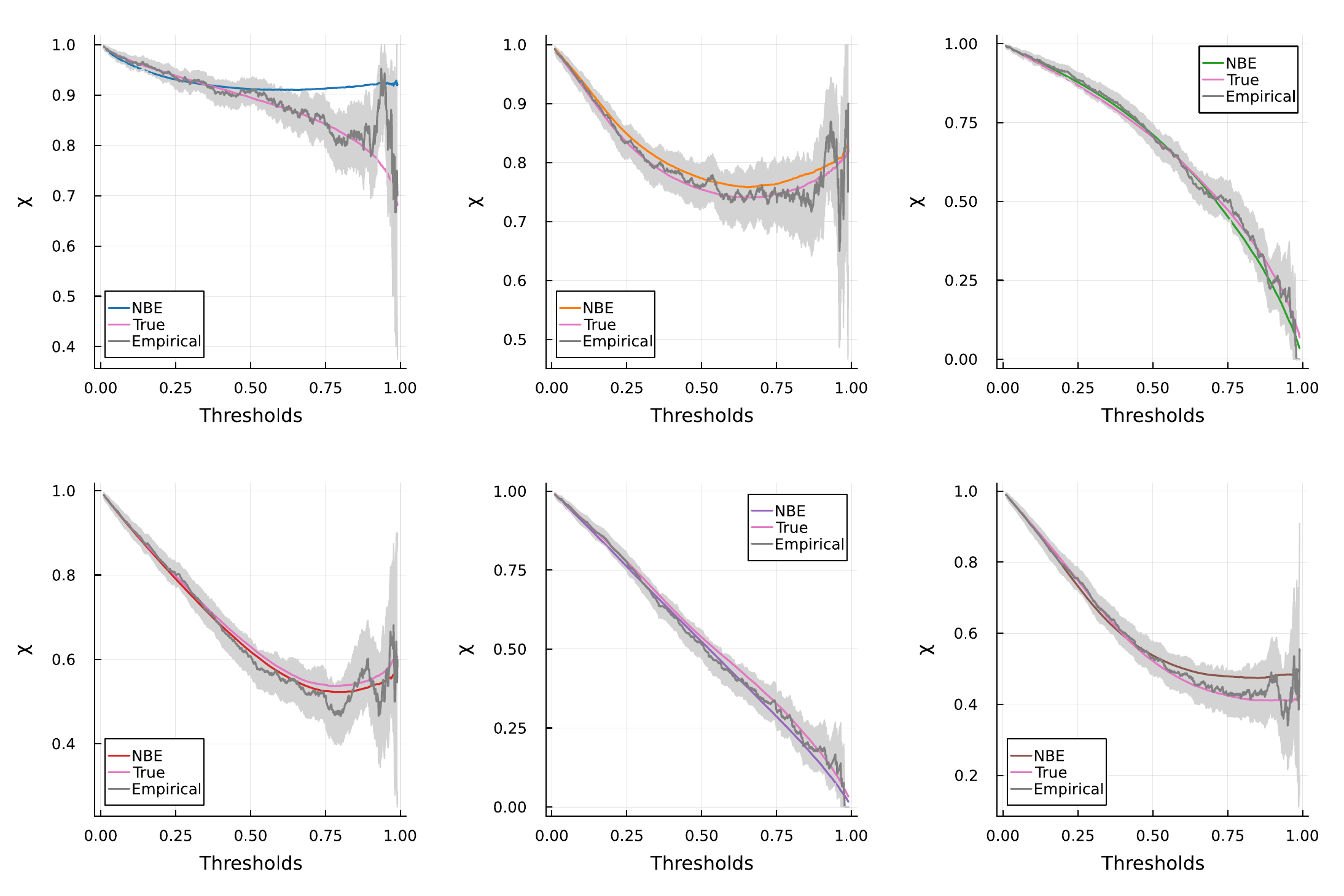}
    \caption{Empirical $\chi(u)$ (in grey), true $\chi(u)$ (in pink) and model-based $\chi(u)$ for the fitted weighted copula model with six different parameter configurations (in colour) for $u \in [0.01,0.99].$ The 95$\%$ confidence bands, representing uncertainty in the empirical estimates, were obtained by boostrapping.}
    \label{fig:chietaNBE}
\end{figure}

We also explore whether the apparent bias in parameter estimates leads to bias in dependence quantities of interest. To do so, we consider parameter and data sets for which the NBE severely under- or over-estimates at least one parameter and compare the empirical and model-based dependence measure $\chi(u)$ from equation~\eqref{eq:chi} at several thresholds $u\in[0.01,0.99].$ For this model configuration, model-based $\chi(u)$ values are estimated using a Monte Carlo approximation with $500\,000$ samples. The parameter configurations considered are highlighted with different shapes and colours in Figure~\ref{fig:wcmassessment}, and the results for $\chi(u)$ are shown in Figure~\ref{fig:chietaNBE}. Despite some parameters being greatly under/over-estimated, the dependence structure is still well captured overall, apart from the configuration shown in blue for which $\mu$ is massively over-estimated. Given that this is the parameter that controls the limiting extremal dependence structure of the data, directly determining the value of $\chi_{E1},$ this result is not very surprising. Finally, from the $B=400$ bootstrap samples obtained previously, we compute coverage probabilities of $95\%$ confidence intervals for $\chi(u)$ at levels $u = \{0.50, 0.80, 0.95\};$ these are shown in Table~\ref{tab:uncwcmchi}. As can be seen, the true value for $\chi(u)$ is within the confidence intervals in at least $79\%$ of the time, suggesting that this derived feature of the models is relatively well estimated and better calibrated than the individual parameters when their estimates obtained by the NBE exhibit bias and display poor coverage properties.

\begin{table}[!t]
    \centering
    \captionsetup{width=\textwidth}
    \caption{Coverage probability and average length of the $95\%$ confidence intervals for $\chi(u)$ at levels $u=\{0.50, 0.80, 0.95\}$ obtained via a non-parametric bootstrap procedure averaged over $1000$ models fitted using a NBE (rounded to 2 decimal places).}
    \begin{tabular}{ccc}
        \toprule
        $\chi(u)$ & Coverage & Length \\
        \midrule
        $\chi(0.50)$ & $0.86$ & $0.09$ \\
        $\chi(0.80)$ & $0.82$ & $0.11$ \\
        $\chi(0.95)$ & $0.79$ & $0.12$ \\
        \bottomrule
    \end{tabular}
    \label{tab:uncwcmchi}
\end{table}

\begin{figure}[t!]
    \centering
    \includegraphics[width=0.8\textwidth]{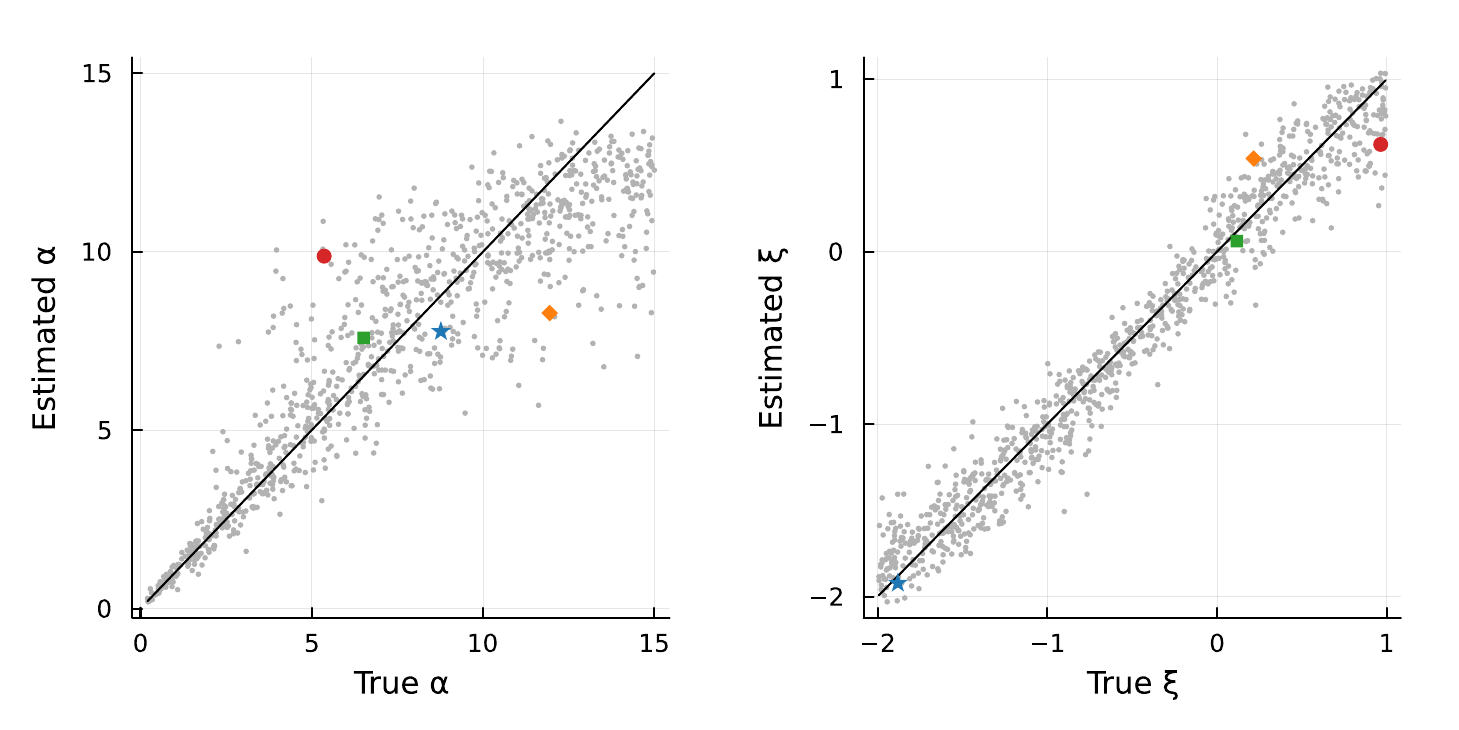}
    \caption{Assessment of the NBE for Model W with parameters $\bm \theta=(\alpha,\xi)'$ for a sample size of $n=1000.$ The points highlighted in different shapes and colours refer to parameter configurations used for further diagnostics (see Figure~\ref{fig:chietawads}).}
    \label{fig:wadsassessment}
\end{figure}

In the second simulation study, we consider Model W with the priors mentioned in Section \ref{subsec:generalsetting}. Since this model is suitable for censored data, we consider a variable censoring level $\tau$ drawn from the prior given in Section~\ref{subsec:generalsetting}. Figure~\ref{fig:wadsassessment} shows the performance of the trained NBE; estimates are quite accurate overall, but with some bias for large $\alpha$ close to the prior support upper bound. We observe from the right panel of Figure~\ref{fig:wadsassessment} that the two regimes of extremal dependence are well captured; in particular, $94.05\%$ of the samples exhibiting asymptotic dependence (i.e., $\xi> 0$) were estimated to be AD, while $98.19\%$ of samples exhibiting asymptotic independence (i.e., $\xi\leq 0$) were correctly estimated to be~AI.

\begin{figure}[t!]
    \centering
    \includegraphics[width=0.8\textwidth]{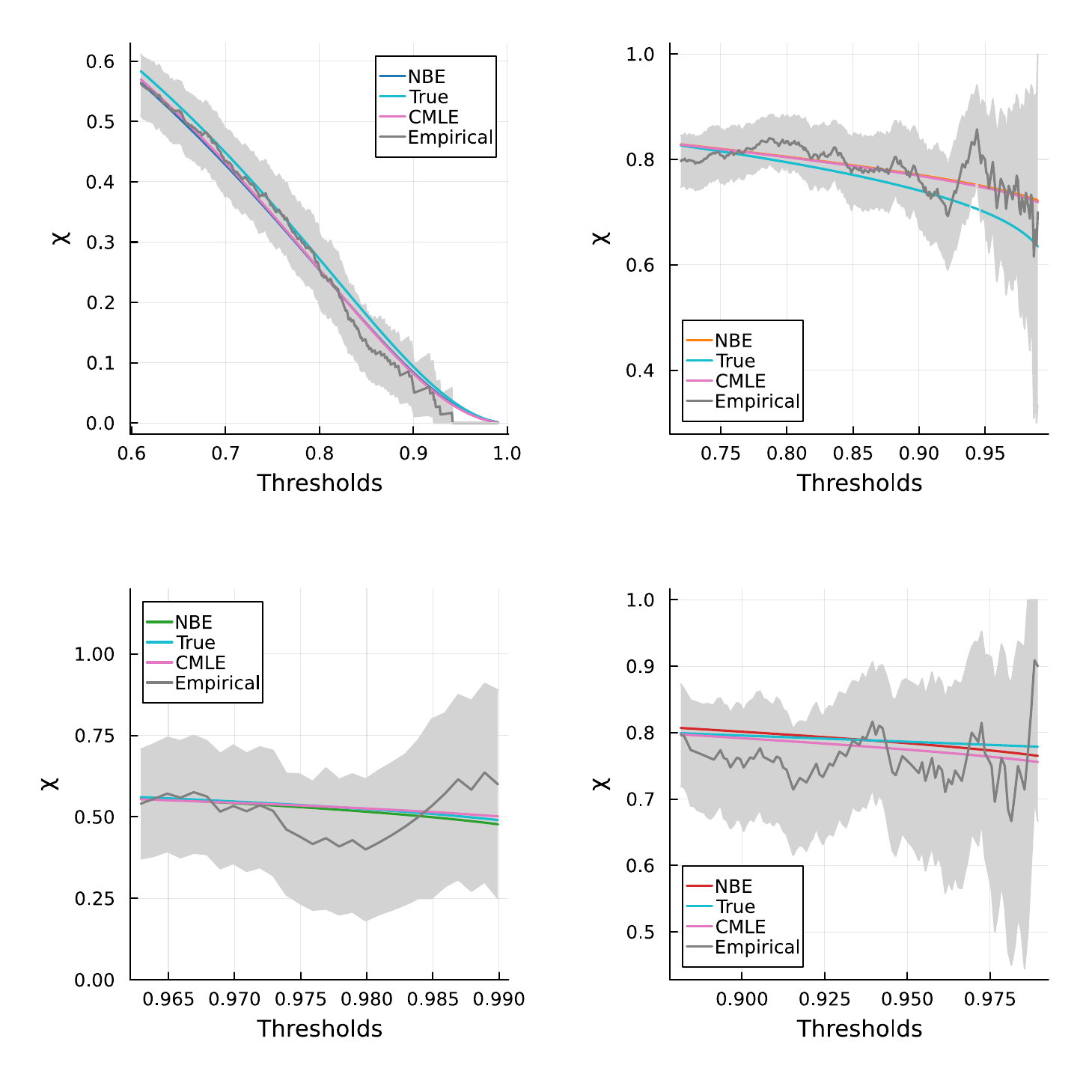}
    \caption{Empirical $\chi(u)$ (in grey), true $\chi(u)$ (in cyan) and model-based $\chi(u)$ estimated via the NBE  with four different parameter configurations (in colour) for $u \in [\tau,0.99],$ where $\tau$ is the corresponding censoring level. A comparison with model $\chi(u)$ estimated via censored maximum likelihood inference is given in pink. The 95$\%$ confidence bands, representing uncertainty in empirical estimates, were obtained by boostrapping.}
    \label{fig:chietawads}
\end{figure}

\begin{figure}[t!]
    \centering
    \includegraphics[width=\textwidth]{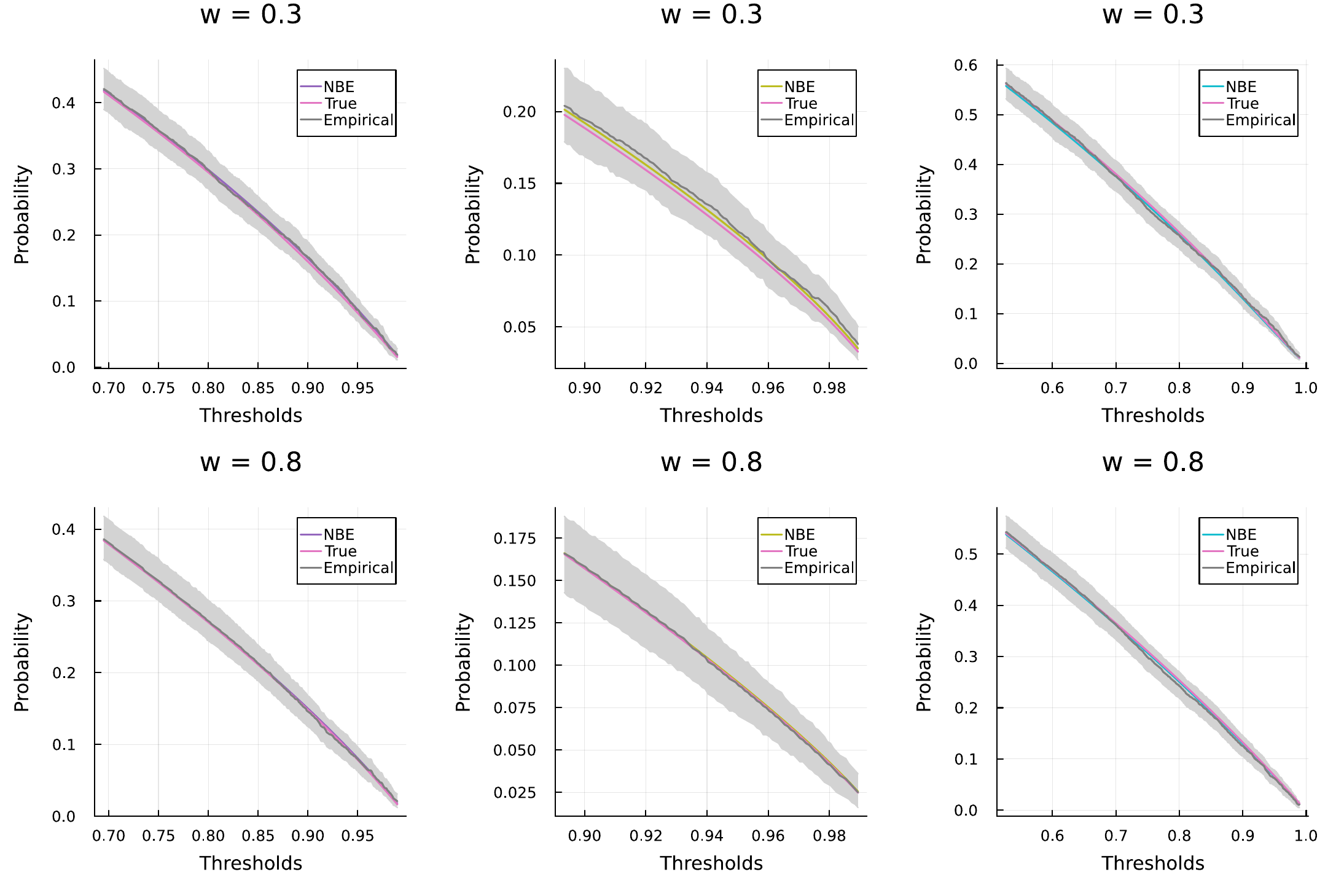}
    \caption{Empirical (in grey), true (in pink) and model-based $\Pr(X_1^E > -w\log(1-u), X_2^E > -(1-w)\log(1-u))$ estimated via the NBE  with three different parameter configurations (in colour) for $u \in [\tau,0.99],$ where $\tau$ is the corresponding censoring level. The 95$\%$ confidence bands, representing uncertainty in empirical estimates, were obtained by boostrapping.}
    \label{fig:adfwads}
\end{figure}

As with the first study, we assess the uncertainty of the NBE through non-parametric bootstrap and by training a neural interval estimator. Coverage probabilities of the $95\%$ uncertainty intervals and their average length are shown in Table~\ref{tab:uncwads}. The coverage probabilities are once again better using the trained neural interval estimator, with wider intervals on average. Figure~\ref{fig:chietawads} shows the extremal dependence measure $\chi(u)$ from equation~\eqref{eq:chi} computed as a further diagnostic for the four parameter configurations highlighted in Figure~\ref{fig:wadsassessment} at several thresholds $u\in[\tau,0.99],$ where $\tau$ is the censoring level. Despite the under/over-estimation by the NBE, e.g., shown by the vector of parameters in red, the extremal dependence behaviour is well captured. This is further supported by the fairly good coverage probabilities of $95\%$ confidence intervals for $\chi(u)$ at levels  $u = \{0.80, 0.95, 0.99\}.$ Coverage probabilities, reported in Table~\ref{tab:uncwadschi}, are computed from estimates obtained from new data sets for 1000 parameter configurations, each computed with a fixed censoring level $\tau = 0.8.$ As a final diagnostic, we compare the joint tail behaviour along different rays. Transforming the model variables, $U_1$ and $U_2,$ into standard exponentially distributed variables, $X_1^E$ and $X_2^E,$ respectively, we compute the joint probability $\Pr(X_1^E > -w\log(1-u), X_2^E > -(1-w)\log(1-u))$ and its empirical counterpart for two different rays $w = \{0.3, 0.8\}$ and $u\in[\tau,0.99].$ The results for three parameter configurations exhibiting asymptotic independence are given in Figure~\ref{fig:adfwads}. There is a very good agreement between the estimated and empirical joint probabilities for both rays considered, supporting the efficacy of the trained NBE. 

\begin{table}[!t]
    \centering
    \captionsetup{width=\textwidth}
    \caption{Coverage probability and average length of the $95\%$ uncertainty intervals obtained via a non-parametric bootstrap procedure and via the neural interval estimator averaged over $1000$ models fitted using a NBE (rounded to 2 decimal places).}
    \begin{tabular}{ccccc}
        \toprule
        \multirow{2}{*}{Parameter} & \multicolumn{2}{c}{Bootstrap procedure} & \multicolumn{2}{c}{Interval estimator} \\
        \cmidrule{2-5}
         & Coverage & Length & Coverage & Length \\
        \midrule
        $\alpha$ & $0.76$ & $3.59$ & $0.96$ & $6.68$ \\
        $\xi$ & $0.84$ & $0.49$ & $0.98$ & $0.81$ \\
        \bottomrule
    \end{tabular}
    \label{tab:uncwads}
\end{table}

\begin{table}[!t]
    \centering
    \captionsetup{width=\textwidth}
    \caption{Coverage probability and average length of the $95\%$ confidence intervals for $\chi(u)$ at levels $u=\{0.80, 0.95, 0.99\}$ obtained via a non-parametric bootstrap procedure averaged over $1000$ models fitted using a NBE (rounded to 2 decimal places).}
    \begin{tabular}{ccc}
        \toprule
        $\chi(u)$ & Coverage & Length \\
        \midrule
        $\chi(0.80)$ & $0.91$ & $0.06$ \\
        $\chi(0.95)$ & $0.89$ & $0.09$ \\
        $\chi(0.99)$ & $0.88$ & $0.09$ \\
        \bottomrule
    \end{tabular}
    \label{tab:uncwadschi}
\end{table}

\paragraph{Comparison with censored maximum likelihood estimation}

\begin{figure}[t!]
    \centering
    \begin{subfigure}[b]{\textwidth}
        \includegraphics[width=\textwidth]{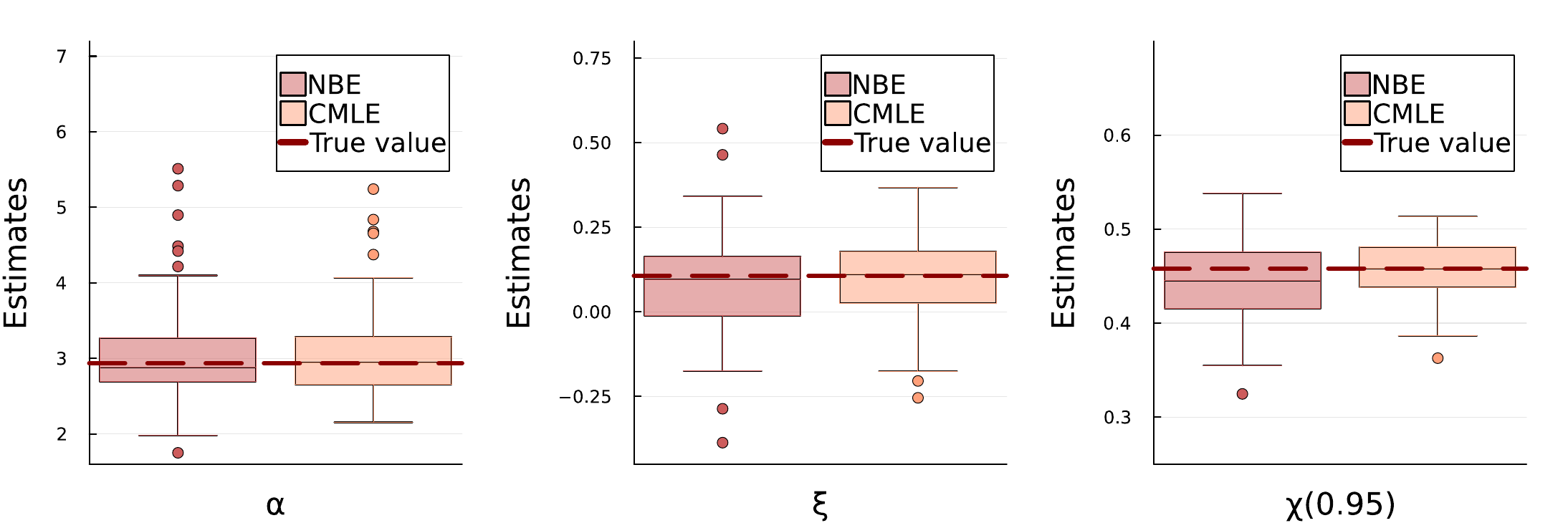}
        \caption{}
        \label{subfig:jointdat2}
    \end{subfigure}
    \hfill
    \begin{subfigure}[b]{\textwidth}
        \includegraphics[width=\textwidth]{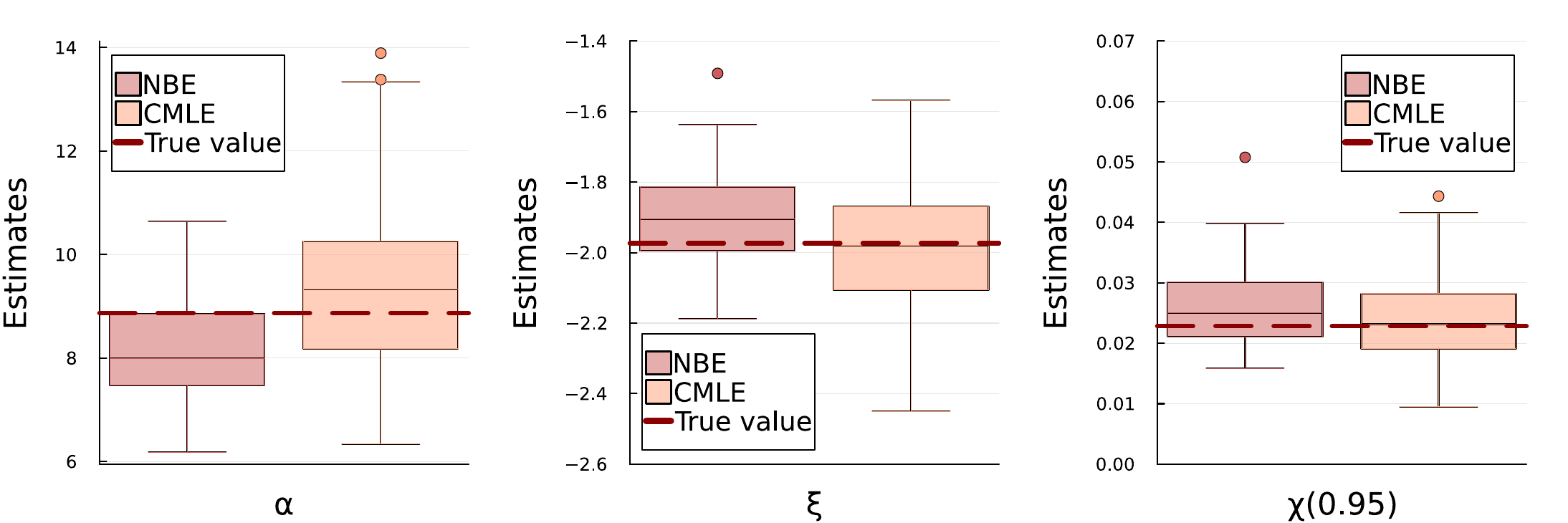}
        \caption{}
        \label{subfig:boxplotdat2}
    \end{subfigure}
    \caption{Comparison between parameter estimates $\hat{\bm\theta}=(\hat \alpha, \hat \xi)'$ given by CMLE (orange) and by NBE (red), and corresponding $\chi(0.95),$ for $100$ samples with $n=1000.$ The true parameters are given by the red line. (a) $\bm \theta_1 = (2.94, 0.11)'$ with censoring level $\tau_1 = 0.79$ and (b) $\bm \theta_2 = (8.87, -1.97)'$ with censoring level $\tau_2 = 0.60.$}
    \label{fig:cmlevsnbe}
\end{figure}

Finally, we compare estimates obtained by the NBE and those obtained by the censored maximum likelihood estimation (CMLE) procedure, using the same censoring scheme as outlined in \citet{Wadsworthetal2017}. We first compute the CMLE for the four parameter configurations highlighted in Figure~\ref{fig:wadsassessment}, and the corresponding estimates for $\chi(u)$ for $u\in [\tau, 0.99].$ Results based on the CMLE, shown by the pink lines in Figure~\ref{fig:chietawads}, are almost identical to those obtained using the NBE. Then, from the assigned prior distributions, we generate 5 different parameter vectors $\bm \theta = (\alpha, \xi)'$, censoring levels $\tau,$ with corresponding data sets from Model W, each with a sample size of $1000$. We then simulate data sets from each configuration $100$ times. Figure~\ref{fig:cmlevsnbe} shows boxplots comparing the sampling distribution of the estimators and the corresponding $\chi(0.95)$ for two parameter vectors: $\bm \theta_1 = (2.94, 0.11)'$ with censoring level $\tau_1 = 0.79$ and  $\bm \theta_2=(8.87, \allowbreak -1.97)'$ with $\tau_2 = 0.60$ (rounded to 2 decimal places). The remaining three cases are given in Section~\ref{supsesec:wadspaper} of the Supplementary Material. Estimates given by the NBE are slightly more biased, but often less variable than the CMLE. Despite this small bias, estimates provided by the NBE are still relatively accurate whilst being much faster to obtain. In particular, the NBE took on average $0.676$ seconds to evaluate, whereas the censored MLE took $183$ seconds on average---a 225-fold speed-up.

To assess the effect of treating the sample size $n$ and/or censoring level $\tau$ as a random variable, we perform a similar study considering fixed censoring level $(\tau = 0.8)$ with either fixed $(n=1000)$ or variable sample size. The results are presented in Sections~\ref{supsecsec:fixedwads} and \ref{supsecsec:fixedvarwads} of the Supplementary Material, respectively. Although the overall findings are similar, slightly higher coverage probabilities and shorter intervals, are obtained when $n$ and $\tau$ are fixed.

\subsection{Model selection assessment} \label{subsec:modelselsim}

Here we investigate the performance of the NBC for model selection, using the procedure outlined in Section~\ref{subsec:modelselection}. For each sampled model $m$ during the training step, a data set $\bm Z$ is generated with a random sample size $n,$ a random censoring level $\tau,$ and a random vector of model parameters $\bm \theta$ using the priors defined in Section \ref{subsec:generalsetting}. We demonstrate the performance of the trained NBC for the four models mentioned in Section \ref{subsec:flexiblemodels}, with censoring at level $\tau$. The neural network architecture used for the model selection procedure is given in Table~\ref{tab:modelseltable} of the Supplementary Material. We have compared all possible pairs among the four models, resulting in $K=2$, and all models simultaneously, resulting in $K=4$. Each model is selected with probability $1/K$, and the sample simulated from that model is of size $n=1000.$ The class which has the highest output probability is the assigned model for the data set. We also compare the NBC with model selection via the Bayesian information criterion (BIC). This requires optimisation of the CMLE, which is time-consuming. However, we wish to explore how well the NBCs perform in comparison to established likelihood-based methods.

\begin{figure}[t!]
    \centering
    \begin{subfigure}[b]{0.49\textwidth}
        \includegraphics[width=\textwidth]{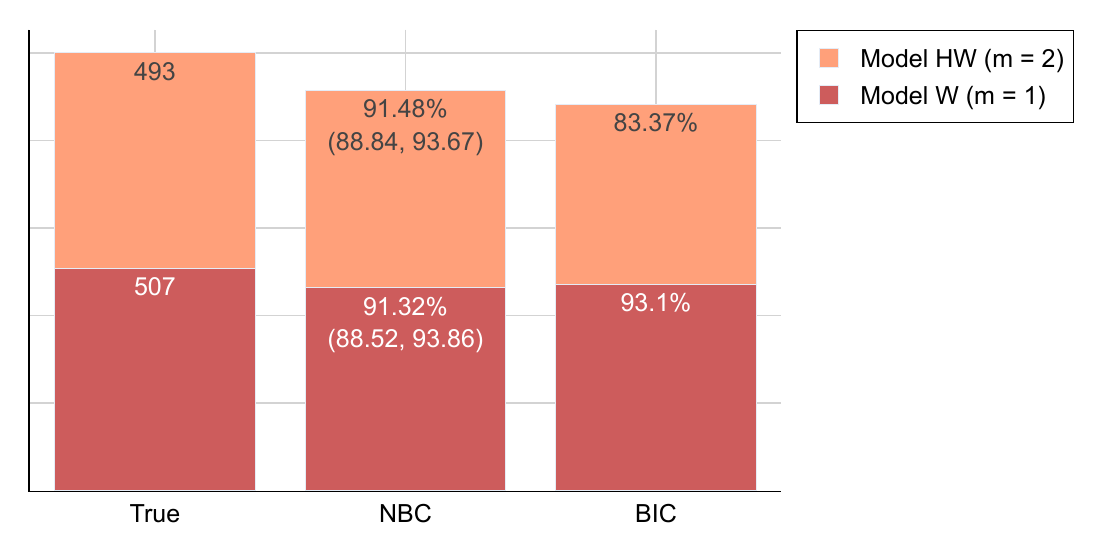}
        \label{subfig:wadshw}
    \end{subfigure}
    \hfill
    \begin{subfigure}[b]{0.49\textwidth}
        \includegraphics[width=\textwidth]{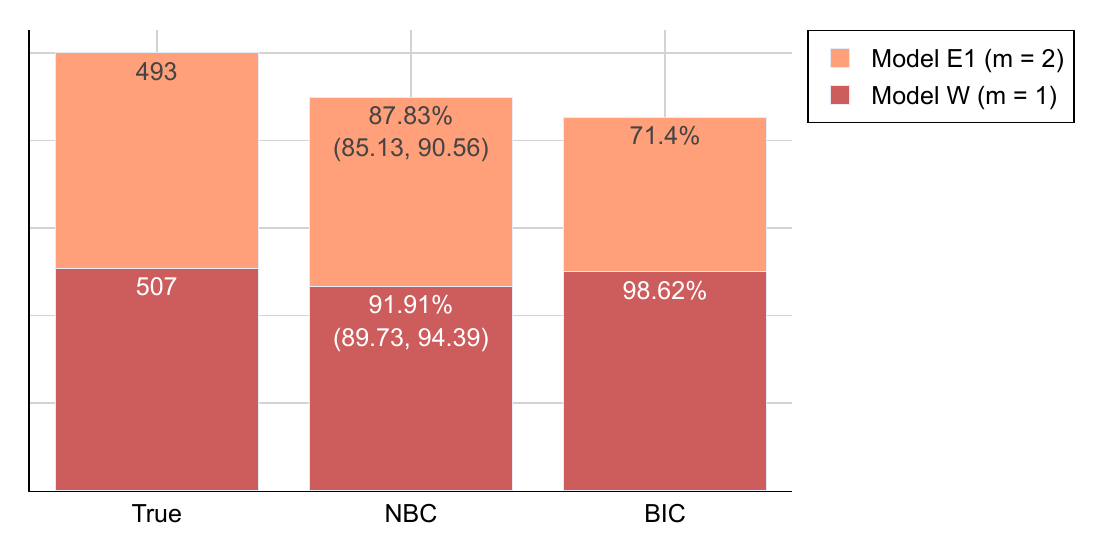}
        \label{subfig:wadsengelke1}
    \end{subfigure}
    \begin{subfigure}[b]{0.49\textwidth}
        \includegraphics[width=\textwidth]{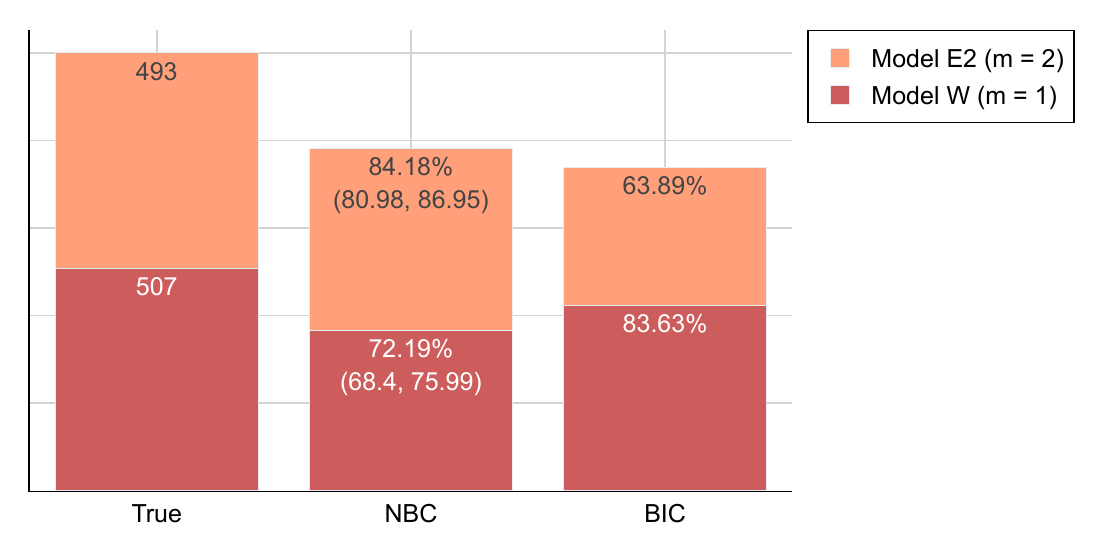}
        \label{subfig:wadsengelke2}
    \end{subfigure}
    \hfill
    \begin{subfigure}[b]{0.49\textwidth}
        \includegraphics[width=\textwidth]{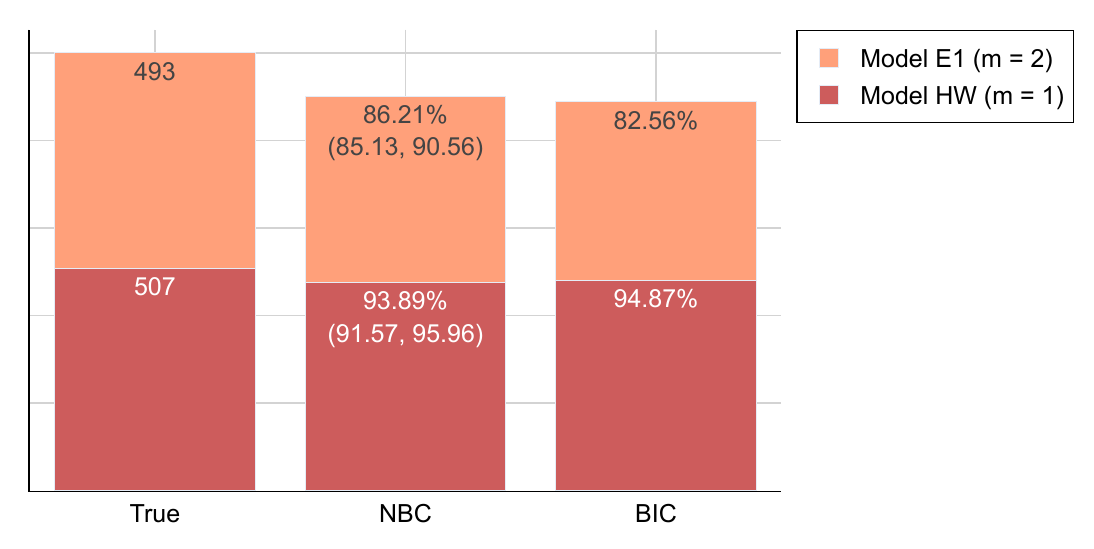}
        \label{subfig:hwengelke1}
    \end{subfigure}
    \begin{subfigure}[b]{0.49\textwidth}
        \includegraphics[width=\textwidth]{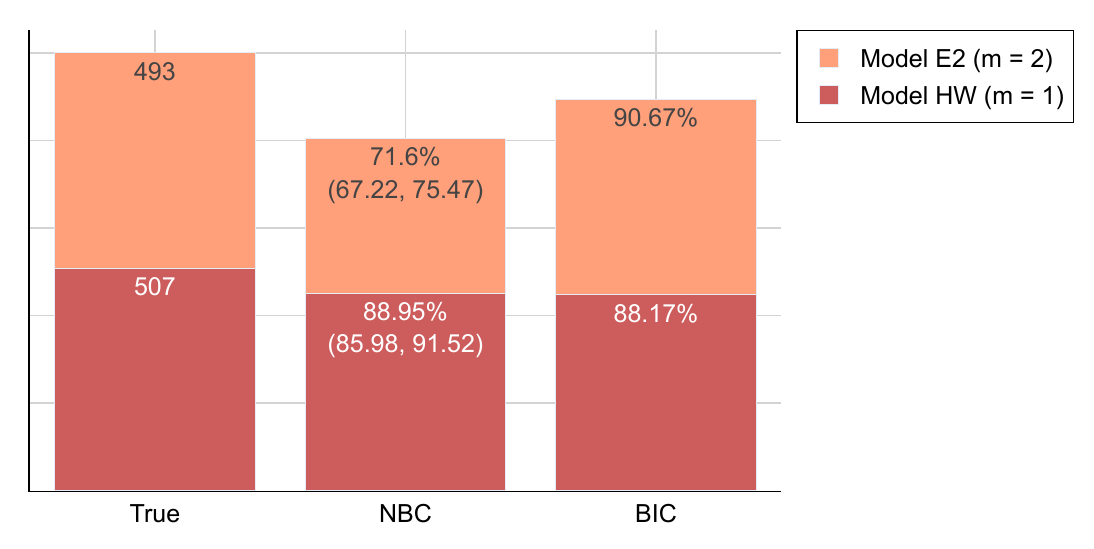}
        \label{subfig:hwengelke2}
    \end{subfigure}
    \hfill
    \begin{subfigure}[b]{0.49\textwidth}
        \includegraphics[width=\textwidth]{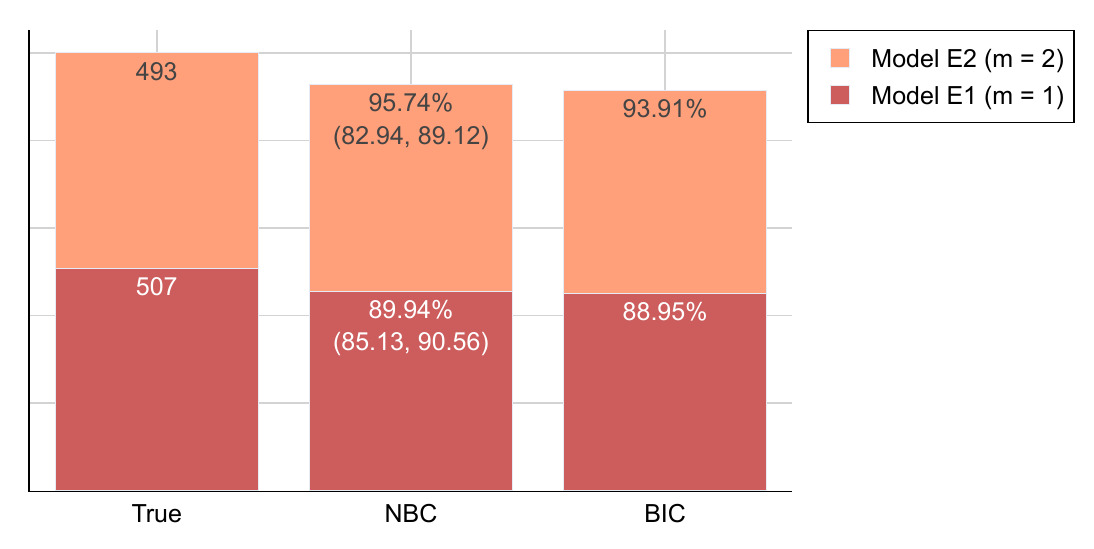}
        \label{subfig:engelke1engelke2}
    \end{subfigure}
    \caption{Proportion (in $\%$) of correctly identified data sets when $K=2$ through the NBC (middle) and through BIC (right) for the six pairs of models considered. The true counts of data sets generated from model $m = 1$ (red) and from model $m = 2$ (orange) are given in the left bar plot. The $95\%$ confidence intervals for the proportions of correctly identified data sets by the NBC are given in brackets.}
    \label{fig:binaryclassification}
\end{figure}

Figure~\ref{fig:binaryclassification} shows the assessment of the NBC when $K=2.$ For this study, the proportions of correctly identified models (i.e., the models with the highest posterior probability matching the one from which data were simulated) with the NBC are always above $71\%,$ with an average of $87\%,$ whilst through BIC these are above $63\%$ with an average of $86\%.$ Aside from the cases where the choice is between Model W and Model E2 or between Model HW and Model E2, the NBC and the likelihood-based BIC perform comparably, with the selection procedure through the NBC being quicker than using the BIC as no likelihood evaluation is needed. Finally, the uncertainty of the NBCs is assessed through a bootstrap procedure. Given $B=400$ samples, the $95\%$ confidence intervals for the proportions of correctly identified data sets are computed for each model and NBC; these results are reported in the middle bar plots of Figure~\ref{fig:binaryclassification}.

\begin{figure}[t!]
    \centering
    \includegraphics[width=\textwidth]{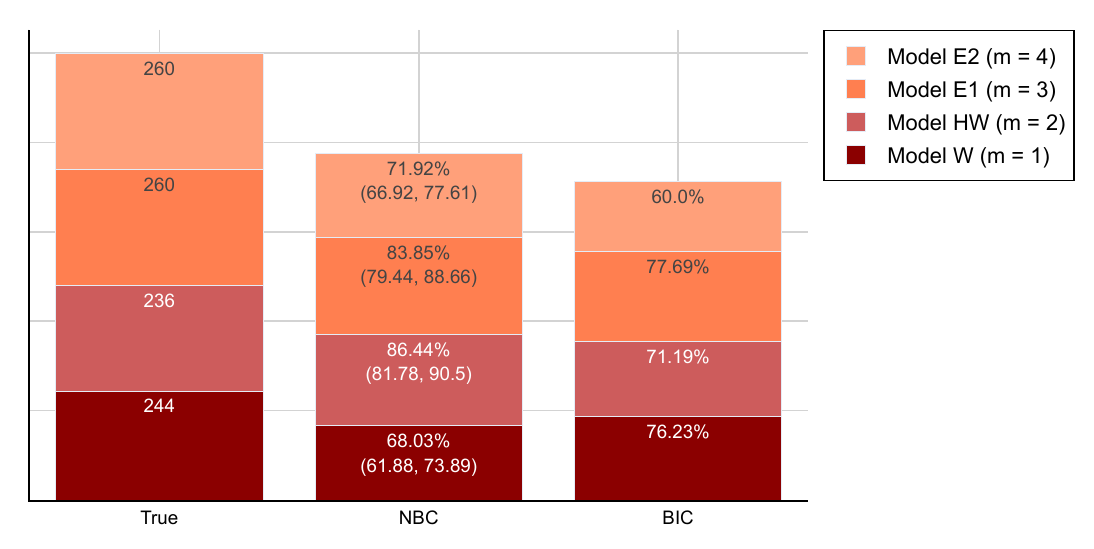}
    \caption{Proportion (in $\%$) of correctly identified data sets when $K=4$ through the NBCs (middle) and through BIC (right). The true counts of data sets generated from models $m = 1$ (red), $m = 2$ (light red), $m = 3$ (orange) and $m = 4$ (light orange) are given in the left bar plot. The $95\%$ confidence intervals for the proportions of correctly identified data sets by the NBC are given in brackets.}
    \label{fig:multiclassclassification}
\end{figure}

Results for the case when $K=4$ are displayed in Figure~\ref{fig:multiclassclassification}. Similarly to the previous case, the uncertainty of the NBC is assessed via a bootstrap procedure with $B=400$ bootstrap samples; the $95\%$ confidence intervals for the proportions obtained by the NBC are shown in the middle bar plot. The resulting proportions achieved through the NBC are all above $68\%$ with an average of $78\%,$ whereas via the BIC they are above $60\%$ with and average of $71\%.$ The proportions obtained via the BIC are generally lower than the ones obtained with the NBC apart from Model W, which indicates that the NBC correctly identifies the data sets more frequently than the BIC, and in a quicker way.

To assess the impact of the sample size on the performance of the NBC, and how this compares with the BIC, we have conducted a similar study, drawing $500$ model indices, and simulating with a range of sample sizes $n\in \{200, 500, 1400, 2000\}.$ The results are presented in Section~\ref{supsec:effectn} of the Supplementary Material. This shows that, across the training range, $n \in \{200,500,1400\}$, the NBC outperforms BIC for small $n$, while the BIC is slightly better for larger $n$. Outside of the training range, at $n=2000$, the NBC still exhibits good performance. Interestingly the BIC is worse for $n=2000$ than for $n=1400$. This non-intuitive result may reflect sampling variability, numerical problems, or convergence to local rather than global minima; however, further investigations did not reveal any convincing explanation for the reduced performance of the BIC for larger $n,$ and the computationally-intensive nature of the likelihood-based methods means that it is hard to conduct more repetitions to confirm such conjectures.

\subsection{Misspecified scenarios} \label{subsec:missscenarios}

We now examine the performance of the model selection and subsequent parameter estimation procedure in a misspecified scenario, where the underlying data do not come from one of the models considered. We consider two different situations and, in both cases, model selection is first done using the NBC, followed by estimation of the model parameters using the NBE trained for the selected models. A comparison with classical model selection and inference tools is also provided. 

\begin{figure}[t!]
    \centering
    \begin{subfigure}[b]{0.8\textwidth}
        \includegraphics[width=\textwidth]{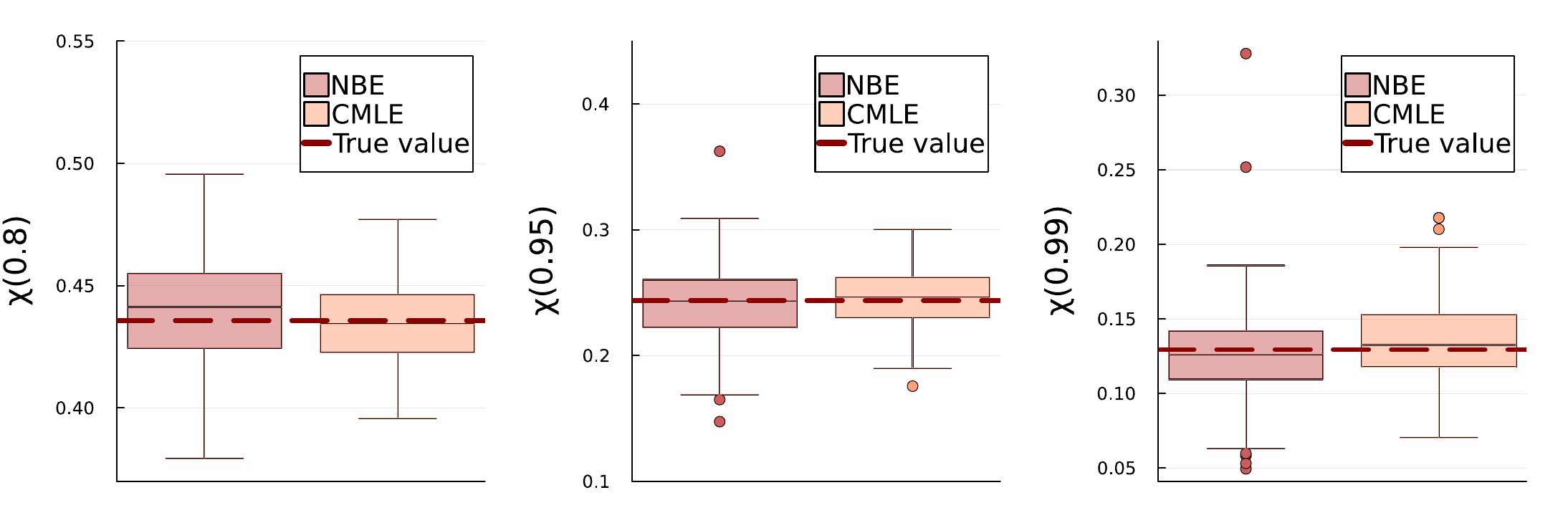}
        \caption{Gaussian.}
        \label{subfig:chietaNBEgaus}
    \end{subfigure}
    \hfill
    \begin{subfigure}[b]{0.8\textwidth}
        \includegraphics[width=\textwidth]{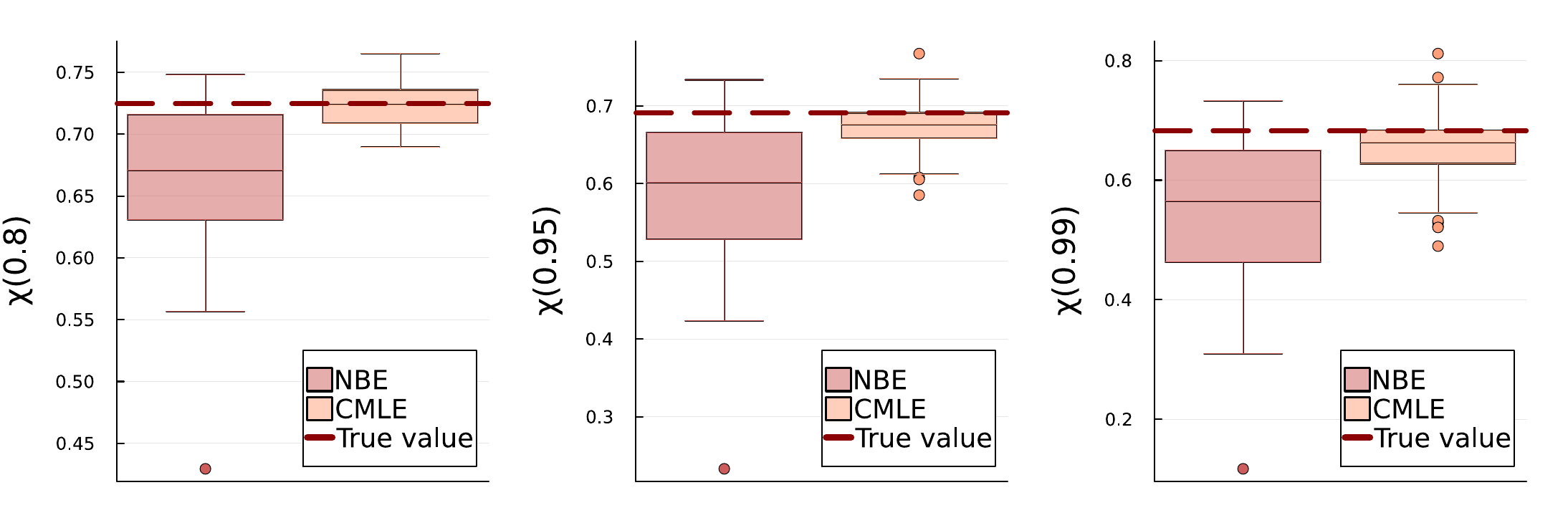}
        \caption{Logistic.}
        \label{subfig:chietaNBElog}
    \end{subfigure}
    \caption{Model-based estimates of $\chi(u)$ given by the NBE (red) and by the CMLE (orange) for levels $u\in\{0.80, 0.95, 0.99\}$ and for 100 samples of (a) a Gaussian copula with correlation parameter $\rho = 0.5$ and $\tau = 0.75,$ and of (b) a logistic distribution with dependence parameter $\alpha_L=0.4$ and $\tau = 0.8.$ For both cases, the true $\chi(u)$ value is given by the dashed red lines.}
    \label{fig:chimiss}
\end{figure}

For the first study, we generate $100$ samples, each with $n=1000,$ from a Gaussian copula with correlation parameter $\rho=0.5,$ and consider a censoring level of $\tau = 0.75.$ The proportion of times each model was selected through the NBC and BIC is given on the left of Table~\ref{tab:gausdata}, and the proportion of AD and AI samples identified by the NBE and CMLE is on the right. Model HW is the most suitable one according to either selection procedure, with the NBC selecting it $88\%$ of the time and the BIC $69\%$ of the time. This is to be expected given the nature of the underlying data and the model assumptions, i.e., we are assuming that $(V_1, V_2)'$ follows a Gaussian copula in Model HW, such that the Gaussian copula arises as a limiting special case of the model as $\delta\to 0$; this implies that Model HW is the least misspecified model in this case. Both inference procedures are able to capture the correct extremal dependence structure. In particular, according to the NBE and CMLE, $97\%$ and $96\%$ of the samples, respectively, exhibit AI, which is in agreement with the underlying data since Gaussian data are known to be AI. According to the NBE, the two samples identified as coming from Model E1 exhibit AD, with estimated values (and corresponding $95\%$ confidence intervals) $\mu = 1.127\, (0.925,\, 1.402)$ and $\mu = 1.247\, (0.828,\, 1.309).$ For the CMLE, three of the samples identified as coming from Model W are AD; for these, the estimates for $\xi$ are close to 0, specifically $\xi = 0.003\, (-0.269,\, 0.273),\, 0.021\, (-0.218,\, 0.378)$ and $0.043\, (-0.288,\, 0.266).$ As a further diagnostic, we compute $\chi(u)$ estimates at three levels $u\in\{0.80, 0.95, 0.99\}$ for the selected models, and compare with the true $\chi(u)$ at each level. The results are given in Figure~\ref{fig:chimiss}. The estimates obtained with either inference method are concentrated around the true value, indicating that both the NBE and CMLE are able to capture the true extremal dependence structure quite well. In Section~\ref{supsec:missscenarios} of the Supplementary Material we describe an individual example, where the proposed toolbox for model selection and inference is presented in detail. The results obtained through this individual example agree with the findings of this study.

\begin{table}[!t]
    \centering
    \captionsetup{width=\textwidth}
    \caption{Proportion of times each model was selected through the NBC and through BIC (left), and proportion of AD and AI samples identified by the NBE and CMLE (right). All the values are rounded up to 2 decimal places.} \label{tab:gausdata}
    \begin{tabular}{lcccclcc}
        \cmidrule[\heavyrulewidth]{1-3} \cmidrule[\heavyrulewidth]{6-8}
        Model & NBC & BIC & & & Method & AD &  AI \\
        \cmidrule{1-3} \cmidrule{6-8}
        Model W & $0.02$ & $0.30$ & & & NBE & $0.02$ & $0.98$ \\
        \addlinespace[1.5mm]
        Model HW & $0.88$ & $0.69$ & & & CMLE & $0.03$ & $0.97$  \\
        \cmidrule[\heavyrulewidth]{6-8}
        Model E1 & $0.02$ & $0.00$ & & & & & \\
        \addlinespace[1.5mm]
        Model E2 & $0.08$ & $0.01$ & & & & & \\
        \cmidrule[\heavyrulewidth]{1-3} 
    \end{tabular}
\end{table}

For the second case, we generate $100$ samples, each with $n=1000,$ from a logistic distribution \citep{Gumbel1960} with dependence parameter $\alpha_L = 0.4$ (the smaller $\alpha_L,$ the stronger the extremal dependence) and consider a censoring level of $\tau=0.8.$ The proportion of times each model was selected and the proportion of AD and AI samples identified by the NBE and CMLE are given in Table~\ref{tab:logdata}. For this case, the model selection procedures differ, with the NBC selecting Model HW $79\%$ of the time, and the BIC selecting Model W as the most suitable one in $42\%$ of cases. Additionally, the CMLE correctly identifies all of the samples to be AD, whereas the NBE misclassifies 16 of the samples fitted with Model HW as being AI. For these 16 samples, the estimated values for $\delta$ are near $0.5$ (precisely $\delta\in[0.45, 0.5]$), the boundary point for AI, in 12 samples. Moreover, in all 16 samples, the boundary point $\delta =0.5$ is within the $95\%$ uncertainty intervals. The estimates for $\chi(u)$ at three levels $u=\{0.80, 0.95, 0.99\}$ are shown in Figure~\ref{fig:chimiss}. Contrarily to the previous case, the NBE somewhat under-estimates the true $\chi(u)$ of the logistic distribution, while the CMLE under-estimates it slightly only at higher levels. Moreover, the estimates provided by the NBE exhibit higher variability than the CMLE. As with the first specification in the section, an individual example is presented in Section~\ref{supsec:missscenarios} of the Supplementary Material. Similarly to this study, Model HW is selected as the most suitable one to fit the data and seems to under-estimate the true $\chi(u)$ for $u\in [0.8, 0.99].$

\begin{table}[!t]
    \centering
    \captionsetup{width=\textwidth}
    \caption{Proportion of times each model was selected through the NBC and through BIC (left), and proportion of AD and AI samples identified by the NBE and CMLE (right). All the values are rounded up to 2 decimal places.} \label{tab:logdata}
    \begin{tabular}{lcccclcc}
        \cmidrule[\heavyrulewidth]{1-3} \cmidrule[\heavyrulewidth]{6-8}
        Model & NBC & BIC & & & Method & AD &  AI \\
        \cmidrule{1-3} \cmidrule{6-8}
        Model W & $0.18$ & $0.42$ & & & NBE & $0.84$ & $0.16$ \\
        \addlinespace[1.5mm]
        Model HW & $0.79$ & $0.21$ & & & CMLE & $1.00$ & $0.00$  \\
        \cmidrule[\heavyrulewidth]{6-8}
        Model E1 & $0.02$ & $0.08$ & & & & & \\
        \addlinespace[1.5mm]
        Model E2 & $0.01$ & $0.29$ & & & & &\\
        \cmidrule[\heavyrulewidth]{1-3} 
    \end{tabular}
\end{table}

\section{Case study: changes in geomagnetic field fluctuations} \label{section:application}

\subsection{Data and background}

The behaviour of the sun and the consequences of its interaction with the Earth's magnetic field and atmosphere are known as space weather events. Examples include phenomena such as the auroras, often known as Southern and Northern lights, or solar storms. These events can cause large fluctuations in the geomagnetic field, leading to geomagnetically induced currents (GICs), which are electrical currents generated at the surface of the planet by rapid changes in the magnetic field. Furthermore, GICs can cause disruptions on power grids, communication systems, railway systems, and other critical infrastructures. Thus, the modelling of extreme solar activity can help prevent the impacts of GICs. 

\begin{figure}[t!]
    \centering
    \includegraphics[width=\textwidth]{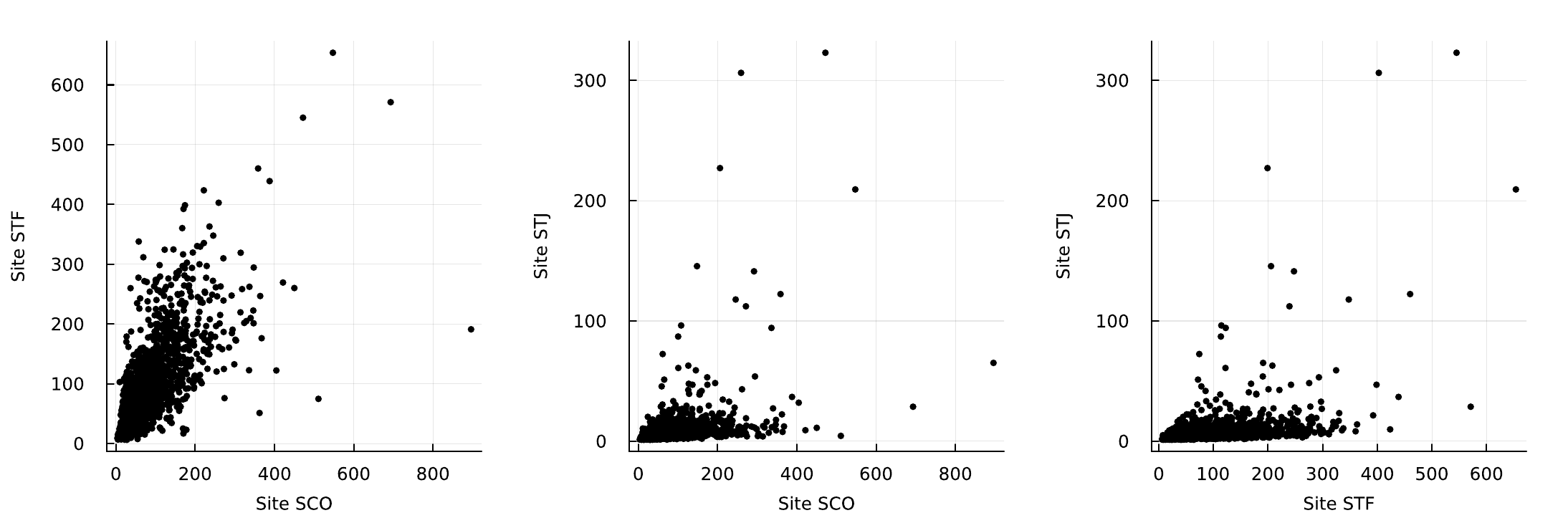}
    \caption{Daily maxima absolute one-minute changes in $\text{d}B_H\slash \text{d}t$ measurements between three pairs of locations: (SCO, STF) on the left, (SCO, STJ) in the middle, and (STF, STJ) on the right.}
    \label{fig:scatterapp}
\end{figure}

Following \citet{Rogersetal2020}, we use the rate of change of the horizontal component of geomagnetic field, $\text{d}B_H\slash \text{d}t,$ which is available through the SuperMAG interface \citep{Gjerloev2009}, as a measure of the magnitude of GICs. In particular, we apply the proposed toolbox to select and make inference on the pairwise extremal dependence structure between measurements at three pairs of locations in the northern hemisphere: two in Greenland and one in the east coast of Canada (see Table~\ref{tab:IAGA})---to assess whether a large magnitude of GICs occurring in one location has an effect on another location. We take daily maximum absolute one-minute changes in $\text{d}B_H\slash \text{d}t,$ which results in $7572$ complete observations. However, since the estimators are trained for sample sizes between $100$ and $1500,$ we take a subset of the data set with $n=1500,$ retaining every 5$\textsuperscript{th}$ observation in order to reduce the temporal dependence present in the data, and truncate the resulting data set to 1500 observations by removing the last few. Figure~\ref{fig:scatterapp} shows the scatterplots of pairwise daily maxima absolute one-minute changes in $\text{d}B_H\slash \text{d}t$ between the pairs of locations considered. We first transform the data to uniform margins using the semi-parametric approach of \citet{ColesTawn1991} with a generalised Pareto distribution (GPD) fit to the tail of the data. Thus, the cdf of each margin is estimated via
\begin{equation}
    \hat F(x)=\begin{cases}
        \tilde{F}(x), & \ift x \leq r, \\
        1-\hat\phi_r\left[1+\hat\xi\left(\frac{x-r}{\hat\sigma}\right)\right]^{-1\slash \hat\xi}_+, & \ift x > r,
    \end{cases}
\end{equation}
where $\tilde{F}(x)$ is the empirical distribution function, $\hat\phi_r$ is the estimated probability of exceeding a selecting high threshold $r,$ and $\hat\sigma$ and $\hat\xi$ are the estimated scale and shape parameters of the GPD, respectively. In particular, with $r$ as the $95\%$ marginal quantile of the data, we have $(r, \hat \sigma, \hat \xi) = (218.98, 59.67, 0.29)$ for SCO, $(r, \hat \sigma, \hat \xi) = (246.80, 44.57, 0.25)$ for STF, and $(r, \hat \sigma, \hat \xi) = (20.17, 13.67, 0.71)$ for STJ.

\begin{table}[!t]
    \centering
    \captionsetup{width=\textwidth}
    \caption{International Association of Geomagnetism and Aeronomy (IAGA) code, and location of the observatory for the three locations considered.}
    \begin{tabular}{clcc}
        \toprule
        IAGA code & Observatory (Country) & Latitude & Longitude \\
        \midrule
        SCO & Scoresby Sund 2 (Greenland) & $70.48$ & $-21.97$ \\
        STF & Sdr Stromfjord (Greenland) & $67.02$ & $-50.72$ \\
        STJ & St. John's (Canada) & $47.60$ & $-52.68$\\
        \bottomrule
    \end{tabular}
    \label{tab:IAGA}
\end{table}

\subsection{Statistical inference} 

\begin{figure}[t!]
    \centering
    \includegraphics[width=\textwidth]{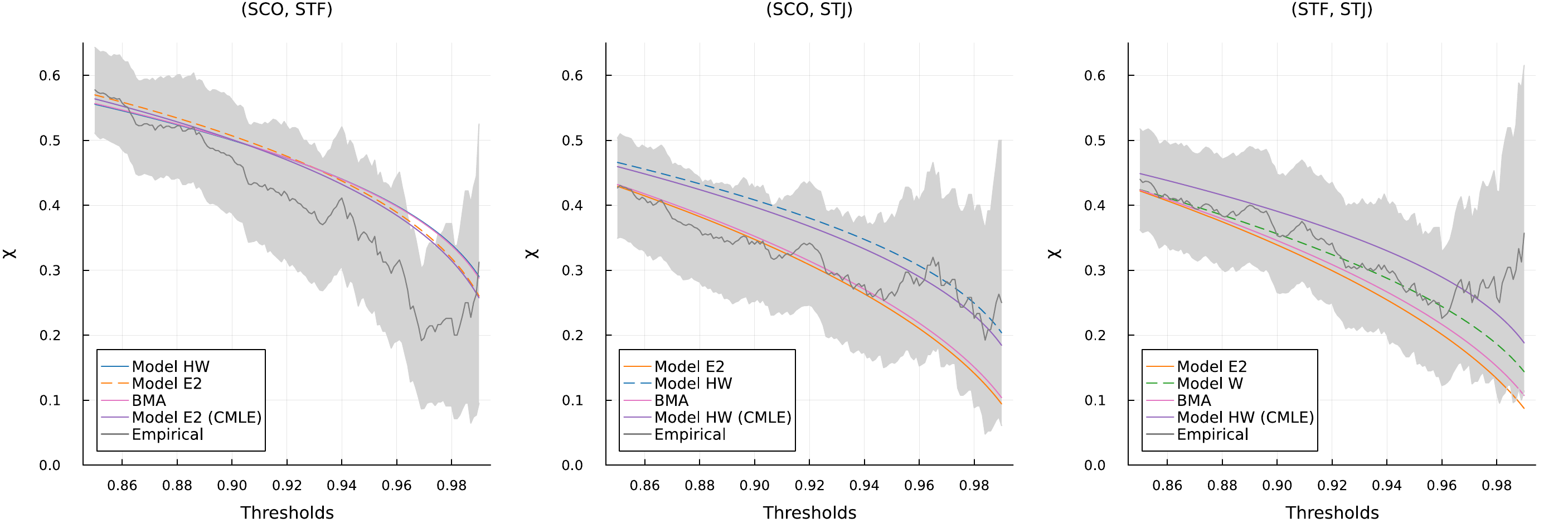}
    \caption{Empirical (in grey) and model $\chi(u)$ estimated via the NBE for $u \in [0.85,0.99]$ for the models with the two highest posterior probabilities. For each pair, the estimated $\chi(u)$ for the selected model is given by thick line, followed by its estimate for the model with the second highest probability given by the dashed line.  The BMA approximation of $\chi(u)$ is given by the pink line. Note that this may overlap with the estimated $\chi(u)$ for the model with the highest posterior probability. Estimated model $\chi(u)$ for the selected model through BIC is given by the purple line. The 95$\%$ confidence bands were obtained by block boostrapping.}
    \label{fig:application}
\end{figure}

We are interested in modelling the joint extremal behaviour of $\text{d}B_H\slash \text{d}t$ between each pair of locations, so we focus on the four models mentioned in Section~\ref{subsec:flexiblemodels}, and censor the non-extreme observations; we do so by taking $\tau = 0.85$ as the censoring threshold. We note, however, that we have considered a range of censoring levels and summarise these results in Section~\ref{supsec:application} of the Supplementary Material. We start by applying our NBC to select the best model and estimate its parameters through the corresponding trained NBE; the results are shown in Tables~\ref{tab:pair1}, \ref{tab:pair2} and \ref{tab:pair3} for pairs (SCO, STF), (SCO, STJ) and (STF, STJ), respectively. The fit of the preferred model is then assessed by comparing the estimated dependence measure $\chi(u)$ with its empirical counterparts for levels $u\in [0.85, 0.99].$ As extra comparisons, the estimated measures for the model with the second highest posterior probability are also obtained, and a Bayesian model averaged approximation of $\chi(u)$ is computed, by setting $\chi(u)^{{\rm BMA}}=\sum_{m=1}^K\hat p_m\chi_m(u),$ where $\hat p_m$ and $\chi_m(u)$ are the posterior probabilities and estimated $\chi(u)$ for $m=1,\ldots,4.$ Finally, a comparison with the results obtained through censored MLE is also provided; the results for $\chi(u)$ are shown in Figure~\ref{fig:application}, where the confidence bands for empirical estimates are obtained via block bootstrap with a block length of $10$ to reflect the remaining temporal dependence present in the data.

For pair (SCO, STF), Model HW is clearly the selected model with a probability of $0.922$ followed by Model E2 with probability $0.078.$ As can be seen by the estimates given by the parameters of interest, $\hat\delta$ and $\hat\xi,$ respectively, both models indicate the presence of asymptotic independence in the data. The same is not true for Model W. It can be seen on the left panel of Figure~\ref{fig:application} that models HW and E2 (in blue and orange, respectively) seem to capture the extremal dependence of data well. According to the BIC, however, Model E2 is the most appropriate to fit this pair. Like the NBE estimates, the CMLE estimates for this model indicate the presence of asymptotic independence, which is in agreement with the estimates obtained by the NBE and CMLE for models HW and E1. Furthermore, the estimated $\chi(u)$ given by the CMLE (in purple) almost overlaps with the obtained $\chi(u)$ with Model E2. When censoring below a range of thresholds, Model E2 or HW were consistently the preferred choices, both indicating asymptotic independence.

Model E2 is clearly selected as the best model to fit pair (SCO, STJ) with a posterior probability of $0.9.$ However, as shown by the middle panel of Figure~\ref{fig:application}, the estimated $\chi(u)$ measure (in orange) is somewhat below the empirical estimate for higher levels $u$. In turn, the estimated measure $\chi(u)$ given by Model HW (in blue) is closer to its empirical counterpart further in the tail, suggesting a better fit of the dependence structure in the limit. Although small, the probability for Model HW given by the NBC is the second highest, followed closely by Model W. For this pair, all four models indicate the presence of asymptotic independence through the estimates given by the NBE and CMLE. Contrarily to the NBC, BIC selects Model HW as the best model to fit the data, with a very small difference compared to model W. The estimated $\chi(u),$ highlighted in purple, is also able to capture the extremal dependence well, particularly further in the tail. Similarly to the first pair, Model HW or E2 were selected as the preferred fit for pair (SCO, STJ) for a range of censoring levels, both models indicating asymptotic independence.

For the final pair (STF, STJ), Model E2 is again selected as the preferred model with a probability of $0.672.$ Model W closely follows with a posterior probability of 0.308. However, Model E2 has the highest BIC value, indicating that, in the likelihood framework, any of the other models would provide a better fit for this pair. The right panel of Figure~\ref{fig:application} shows that Model E2 (in orange) leads to an estimate below the empirical $\chi(u)$ for all levels $u$ considered, whilst Model HW (obtained by CMLE and shown in purple) gives an estimate slightly above the empirical value of $\chi(u).$ On the other hand, Model W (in green) best captures the extremal dependence structure over most of the range. Despite this, all models indicate asymptotically independent variables; this is also in agreement with the estimates obtained by censored maximum likelihood inference. Interestingly enough, the estimated value of parameter $\mu$ of Model E1 given by either approach is very close to 1. When considering different censoring thresholds, the NBC consistently chose either Model HW or E2 as the best fit for the pair (STF, STJ), both models indicating asymptotic independence across all censoring levels.

\begin{table}[!t]
    \centering
    \captionsetup{width=\textwidth}
    \caption{Model selection procedure obtained through the NBC for censoring level $\tau = 0.85,$ and parameter estimates given by the trained NBE for pair (SCO, STF). The results through censoring MLE and BIC are given in the bottom table. All the values are rounded to 3 decimal places.}
    \begin{tabular}{cccc}
        \toprule
        Model & $\hat{\bm p}_{\text{NBC}}$ & $\hat{\bm \theta}_{\text{NBE}}$ & Extremal dependence \\
        \midrule
        W & $1.762\times 10^{-4}$ & $(\hat \alpha, \, \hat\xi) = (2.625,\, 0.121)$ & AD\\
        HW & $\bm{0.922}$ & $(\hat \delta, \, \hat\omega) = (0.258,\, 0.714)$ & AI\\
        E1 & $8.539\times 10^{-11}$ & $(\hat \alpha, \, \hat\beta, \, \hat \mu) = (3.771,\, 0.150, \, 0.939)$ & AI\\
        E2 & $0.078$ & $(\hat \alpha, \, \hat\xi) = (2.450,\, -0.179)$ & AI\\
        \bottomrule
        \toprule
        Model & BIC & $\hat{\bm \theta}_{\text{CMLE}}$ & Extremal dependence \\
        \midrule
        W & $226.785$ & $(\hat \alpha, \, \hat\xi) = (2.724,\, 0.093)$ & AD\\
        HW & $215.897$ & $(\hat \delta, \, \hat\omega) = (0.353,\, 0.676)$ & AI\\
        E1 & $214.556$ & $(\hat \alpha, \, \hat\beta, \, \hat \mu) = (0.487,\, 0.712, \, 0.338)$ & AI\\
        E2 & $\bm{213.428}$ & $(\hat \alpha, \, \hat\xi) = (2.321,\, -0.180)$ & AI\\
        \bottomrule
    \end{tabular}
    \label{tab:pair1}
\end{table}

\begin{table}[!t]
    \centering
    \captionsetup{width=\textwidth}
    \caption{Model selection procedure obtained through the NBC for censoring level $\tau = 0.85,$ and parameter estimates given by the trained NBE for pair (SCO, STJ). The results through censoring MLE and BIC are given in the bottom table. All the values are rounded to 3 decimal places.}
    \begin{tabular}{cccc}
        \toprule
        Model & $\hat{\bm p}_{\text{NBC}}$ & $\hat{\bm \theta}_{\text{NBE}}$ & Extremal dependence \\
        \midrule
        W & $0.029$ & $(\hat \alpha, \, \hat\xi) = (2.527,\, -0.271)$ & AI\\
        HW & $0.072$ & $(\hat \delta, \, \hat\omega) = (0.125,\, 0.621)$ & AI\\
        E1 & $2.352\times 10^{-7}$ & $(\hat \alpha, \, \hat\beta, \, \hat \mu) = (9.546,\, 2.171, \, 0.855)$ & AI\\
        E2 & $\bm{0.900}$ & $(\hat \alpha, \, \hat\xi) = (2.316\, -0.791)$ & AI\\
        \bottomrule
        \toprule
        Model & BIC & $\hat{\bm \theta}_{\text{CMLE}}$ & Extremal dependence \\
        \midrule
        W & $390.990$ & $(\hat \alpha, \, \hat\xi) = (2.364,\, -0.187)$ & AI\\
        HW & $\bm{390.742}$ & $(\hat \delta, \, \hat\omega) = (0.242,\, 0.584)$ & AI\\
        E1 & $396.119$ & $(\hat \alpha, \, \hat\beta, \, \hat \mu) = (2.634,\, 1.112, \, 0.735)$ & AI\\
        E2 & $393.674$ & $(\hat \alpha, \, \hat\xi) = (2.115,\, -0.645)$ & AI\\
        \bottomrule
    \end{tabular}
    \label{tab:pair2}
\end{table}

\begin{table}[!t]
    \centering
    \captionsetup{width=\textwidth}
    \caption{Model selection procedure obtained through the NBC for censoring level $\tau = 0.85,$ and parameter estimates given by the trained NBE for pair (STF, STJ). The results through censoring MLE and BIC are given in the bottom table. All the values are rounded to 3 decimal places.}
    \begin{tabular}{cccc}
        \toprule
        Model & $\hat{\bm p}_{\text{NBC}}$ & $\hat{\bm \theta}_{\text{NBE}}$ & Extremal dependence \\
        \midrule
        W & $0.308$ & $(\hat \alpha, \, \hat\xi) = (2.670,\, -0.317)$ & AI\\
        HW & $0.019$ & $(\hat \delta, \, \hat\omega) = (0.111,\, 0.632)$ & AI\\
        E1 & $5.876\times 10^{-5}$ & $(\hat \alpha, \, \hat\beta, \, \hat \mu) = (10.393,\, 3.228, \, 0.996)$ & AI\\
        E2 & $\bm{0.672}$ & $(\hat \alpha, \, \hat\xi) = (2.420,\, -0.849)$ & AI\\
        \bottomrule\toprule
        Model & BIC & $\hat{\bm \theta}_{\text{CMLE}}$ & Extremal dependence \\
        \midrule
        W & $398.908$ & $(\hat \alpha, \, \hat\xi) = (2.222,\, -0.153)$ & AI\\
        HW & $\bm{397.237}$ & $(\hat \delta, \, \hat\omega) = (0.025,\, 0.601)$ & AI\\
        E1 & $401.390$ & $(\hat \alpha, \, \hat\beta, \, \hat \mu) = (13.559,\, 3.023, \, 1.000)$ & AI\\
        E2 & $409.988$ & $(\hat \alpha, \, \hat\xi) = (2.072,\, -0.680)$ & AI\\
        \bottomrule
    \end{tabular}
    \label{tab:pair3}
\end{table}

A few conclusions can be drawn from these results. For instance, although the selected model may not always align most closely with the empirical $\chi(u)$, a good representation can still be achieved if the model with the second highest probability is considered instead; furthermore the majority of $\chi(u)$ estimates remained in the pointwise 95\% confidence intervals for the empirical estimates. When considering a BMA approximation of $\chi(u),$ the influence of the model selected through the NBC is clear. Additionally, with the proposed toolbox, we are not only able to infer about some model characteristics, but also able to assess the sensitivity to the censoring threshold by considering a range of different levels $\tau.$ In particular, from the results obtained for the range of $\tau$ considered (see Section~\ref{supsec:application} of the Supplementary Material), we can see that, for lower censoring levels, Model HW is clearly the preferred one to fit the three pairs, with posterior probabilities above 0.9 in all cases. This changes for censoring levels $\tau \geq 0.85,$ or $\tau \geq 0.9$ for pair (SCO, STF), for which Model E2 is the most suitable one to fit the data according to the NBC. In all cases, the selected models consistently indicate the presence of asymptotic independence in each pair. This sensitivity analysis to the censoring level is computationally expensive in a likelihood-based inference method, where the likelihood needs to be evaluated for all considered models across various censoring levels. Likelihood estimation can indeed be about 7300 times slower than using the NBE, especially for Models E1 or E2, where the code is not as efficient as for Models W or HW. This leads to the need for choosing a censoring level before the analysis, which might end up not being the most suitable one for the underlying application.
\section{Conclusion and discussion} \label{section:conclusion}

In some situations, the likelihood function of a model may be available but its evaluation may be computationally costly. In multivariate extremes, this is often due to the need for numerical inversion of functions, and numerical integration. This computational burden in the inference process poses limitations on the use of certain models in practice since modelling usually entails consideration of different candidate models and threshold levels. In this paper, we exploited the use of neural Bayes estimators, a modern likelihood-free approach which uses neural networks, to perform inference on the vector of model parameters. In particular, we focused on two types of models which are available in the multivariate extremes literature: a weighted copula model, which is able to represent both the body and tail regions of a data set, and models based on random scale constructions, which are flexible enough to capture the two regimes of extremal dependence, interpolating between the two in the interior of the parameter space. When likelihood evaluation is no longer required, model selection criteria such as the AIC or BIC are now unavailable for determining the best-fitting model. We overcame this by proposing a neural Bayes classifier which allows us to choose the most suitable model from candidates models in a fast and effective manner.

For the models where it is feasible---albeit relatively slow---to evaluate the likelihood function, we compared the performance of the neural Bayes estimators and NBC with (censored) maximum likelihood inference and BIC, respectively. Through simulation studies, we have shown that NBEs perform quite well overall, though they tend to be slightly more biased than the (censored) MLE in some cases; however, they also sometimes exhibit lower variability. Moreover, when estimating tail quantities of interest, such as the extremal measures $\chi(u)$ from equation~\eqref{eq:chi}, we did not find the bias to be problematic, as the overall dependence structure appeared to be well captured with the trained NBEs anyway. Nevertheless, the bias of the estimates is reflected by lower coverage probabilities when using bootstrap-based credible intervals. Whilst coverage probabilities significantly improved when we trained a neural interval estimator, obtaining a value close to nominal, it would be ideal to also obtain such values with bootstrap-based intervals as we would only need to train the estimator once using a single loss function. Despite this, since the neural Bayes estimates are considerably faster to obtain than the (censored) MLE, which may also sometimes converge to local rather than global minima, and the extremal dependence structure is generally well captured, the methodology may be considered preferable overall. 

Given the symmetry of all models considered,  an estimator that is invariant to permutations of the vector components $(Z_1, Z_2)$ would be desirable. However, this is not achieved with the DeepSets architectures mentioned in Section~\ref{subsec:nbe}, which results in different estimates when considering $(Z_1, Z_2)$ and $(Z_2, Z_1)$. We investigated this issue by creating an alternative estimator that is the average of the NBE obtained with inputs $(Z_1,Z_2)$ and $(Z_2,Z_1)$, but the results were almost indistinguishable from the basic approach presented.

When comparing the NBC with the BIC for model selection, both procedures performed similarly with $K=2$ candidate models. However, when $K=4,$ and $n=1000$, the NBC was able to correctly categorise more data sets than the BIC for all considered models, except Model W. As shown in Section~\ref{supsec:effectn} of the Supplementary Material, this conclusion also holds for smaller sample sizes. Interestingly in Section~\ref{section:application}, the BIC and NBC often displayed large disagreements in the selected model. While all selected models generally performed adequately, it could be of interest in further work to explore the effect of being on the boundary of our set of simulated sample sizes. We also demonstrated the possibility of using the NBC as part of  Bayesian model averaging-like approach to estimate measures of interest such as $\chi(u)$. However, full Bayesian model averaging requires complete posterior distributions rather than individual estimates.  Combining the NBC with neural posterior approximations such as normalising flows \citep{Radevetal2023} would be beneficial for amortised BMA, and presents an interesting area for future research.

We have shown that the proposed toolbox for model selection and inference performs well---though not always as accurately as classical likelihood inference---for misspecified scenarios where the data set does not originate from one of the models considered in this paper, with the extremal dependence structure being generally well captured. In particular, through repeated simulation, we showed that the NBC and NBE captured the true extremal dependence structure the majority of the times, especially for asymptotically independent data. Further investigation on the performance of NBEs in misspecified scenarios is an important future line of research. An alternative and interesting line for further research is to adapt the current framework to incorporate an unobserved model class $K+1$ into the model selection framework when there is sufficient evidence that the data is not generated by any of the $K$ candidate parametric models.

In all our examples, we assumed the parameters to be uniformly distributed a priori. However, we note this implies non-uniform priors in alternative parameterisations and we did not assess the effect of this choice on the performance of the trained NBE. Despite this, for studies involving the weighted copula model from Section~\ref{subsec:wcm}, we found that reparameterising some of the model parameters (see Section~\ref{supsec:wcm} of the Supplementary Material) helped the neural network learn about them in the training step of the NBE. More specifically, the assessment of the NBE showed less variability and higher coverage probabilities (with narrower range) of the $95\%$ uncertainty intervals for each model parameters obtained with bootstrap-based intervals. Additionally, the (censored) MLE only depends on the likelihood function as opposed to the NBE, which relies on the choice of a prior. While the (C)MLE is the most used likelihood-based inference method in the extremes literature, it would be of interest to incorporate the same prior belief and obtain a (censored) maximum a posteriori (MAP) estimate instead. Investigating and comparing the performance of the (C)MAP with the NBE in our bivariate copula context presents an interesting avenue for further research.

We restricted our analysis to the bivariate setting. Each of the random scale models described in Section~\ref{subsec:flexiblemodels} could be expanded to higher dimensions; however, this may not be as useful since the models considered are only suitable when all the variables are asymptotically dependent or asymptotically independent. On the other hand, the dynamic mixture model could usefully be extended to higher dimensions, and the benefits of NBEs may be even clearer in this setting as the complexity of the likelihood increases with dimension for this model.

The proposed toolbox has potential to be extended to other settings, such as online learning, change-point detection, and trend estimation problems. In particular, recent work has explored deep learning techniques for change-point detection (see, e.g., \citealp{Lietal2024}). Within our framework, a prior on the number and position of change-points could be used and a neural estimator could then be trained under a suitable loss function. This would allow estimating change-points simultaneously with the other model parameters. For capturing trends, we could incorporate covariates into the analysis. However, to amortise over covariate values, a suitable prior model for the covariates would be required. Depending on the application, this might not be straightforward to specify.

\bigskip

\noindent {\bf Declarations of Interest:} None.

\section*{Acknowledgments}
We are grateful to the two anonymous reviewers and editors for constructive comments and suggestions that have improved this article. This paper is based on work completed while L\'idia Andr\'e was part of the EPSRC funded STOR-i Centre for Doctoral Training (EP/S022252/1). The research of L\'idia Andr\'e was also partially supported financially by the Fonds de la Recherche Scientifique---FNRS, Belgium (grant number T.0203.21). We are thankful to Matthew Sainsbury-Dale and Andrew Zammit-Mangion for providing useful feedback to help improve this paper. We are also grateful to Matthew Sainsbury-Dale for all the help with the code and implementation of the bilinear layers used to handle censored multivariate data, and Jordan Richards for the support in the computational implementation of framework for the censored data. The ground magnetometer data were provided by SuperMAG (available from \url{https://supermag.jhuapl.edu/}) and we gratefully acknowledge Jim Wild, Emma Eastoe and Neil Rogers for sourcing them, as well as contributions from the SuperMAG collaborators: EMMA; The MACCS program; CARISMA and AALPIP. We are grateful to Neil Rogers for processing the data as described in \citet{Rogersetal2020}.

\bibliography{references}{}
\bibliographystyle{apalike}
\nocite{EMMA}
\nocite{MACCS}
\nocite{CARISMA}
\nocite{AALPIP}

\documentclass[12pt,a4paper]{article}
\usepackage[applemac]{inputenc}
\usepackage[T1]{fontenc}
\usepackage{amsmath,amsfonts,amssymb}
\usepackage[breaklinks=true]{hyperref}
\usepackage{mathtools}
\usepackage{graphicx,epsfig,float}
\usepackage{color,xcolor}
\usepackage{textcomp}
\usepackage{indentfirst}
\usepackage{multirow}
\usepackage{multicol}
\usepackage{enumerate}
\usepackage[font=small]{caption}
\usepackage{adjustbox}
\usepackage{amsthm}
\usepackage{graphics}
\usepackage{pstricks,pstricks-add,pst-math,pst-xkey}
\usepackage{bm}
\usepackage{nccmath}
\usepackage{url}
\usepackage{bbm}
\usepackage{setspace}
\usepackage[left=1in,right=1in,bottom=1.3in,top=1.3in]{geometry}
\usepackage{parskip}
\usepackage{breakcites}
\usepackage{algorithm}
\usepackage{algpseudocode}
\usepackage{tikz}
\usetikzlibrary{shapes,snakes}
\usetikzlibrary{automata, positioning}
\usetikzlibrary{matrix,arrows,decorations.pathmorphing,positioning,graphs,calc}
\usepackage{natbib}
\usepackage{subcaption}
\usepackage{subfiles}
\usepackage[title]{appendix}
\usepackage{titling}
\usepackage{mathtools}
\mathtoolsset{showonlyrefs}
\usepackage{placeins}
\usepackage{rotating}
\usepackage{booktabs}

\newcommand{\R}{{\mathbb{R}}}
\newcommand{\Lcal}{{\mathcal{L}}}
\newcommand{\ind}{\perp\!\!\!\perp} 
\theoremstyle{definition}
\newtheorem{definition}{Definition}[section]
\theoremstyle{plain}
\newtheorem{theorem}{Theorem}[section]
\newtheorem{corollary}{Corollary}[theorem]
\newtheorem{lemma}[theorem]{Lemma}
\theoremstyle{definition}
\newtheorem{example}{Example}

\DeclareMathOperator*{\argmax}{arg\,max}
\DeclareMathOperator*{\argmin}{arg\,min}
\DeclareMathOperator*{\U}{\text{Unif}}
\DeclareMathOperator*{\Betad}{\text{Beta}}
\DeclareMathOperator*{\Unifd}{\text{Unif}}
\DeclareMathOperator*{\Weibulld}{\text{Weibull}}
\DeclareMathOperator*{\Multinomiald}{\text{Multinomial}}
\DeclareMathOperator*{\Bernoullid}{\text{Bernoulli}}
\DeclareMathOperator*{\GPDd}{\text{GPD}}
\DeclareMathOperator*{\E}{\text{E}}
\DeclareMathOperator*{\ift}{\text{if }}

\definecolor{antiquefuchsia}{rgb}{0.57, 0.36, 0.51}

\providecommand{\keywords}[1]{{\small{\it Keywords:} #1}}
\def\LA#1{{\textcolor{magenta}{[LA: #1]}}}
\def\corr#1{{\textcolor{blue}{#1}}}
\def\todo#1{{\textcolor{red}{[{\bf To do:} #1]}}}
\def\writing#1{{\textcolor{purple}{#1}}}
\def\dint#1{\text{d}#1}

\def\JW#1{{\textcolor{blue}{[JW: #1]}}}
\def\RH#1{{\textcolor{orange}{[RH: #1]}}}

\onehalfspacing
\allowdisplaybreaks

\begin{document}

\title{Supplementary Material for \emph{Neural Bayes estimation and selection of complex bivariate extremal dependence models}}
\author{L. M. Andr\'e$^{1*}$, J. L. Wadsworth$^{2}$, R. Huser$^{3}$\\
\small $^{1}$ Namur Institute for Complex Systems, University of Namur, Rue Graf\'e 2, Namur 5000, Belgium\\
\small $^{2}$ School of Mathematical Sciences, Lancaster University, LA1 4YF, United Kingdom \\
\small $^{3}$ Statistics Program, Computer, Electrical and Mathematical Sciences and Engineering Division, \\
\small King Abdullah University of Science and Technology (KAUST), Saudi Arabia \\
\small $^*$ Correspondence to: \href{mailto:lidiamandre@gmail.com}{lidiamandre@gmail.com} }
\date{\today}

\maketitle
\pagenumbering{arabic}

\setcounter{section}{0}
\setcounter{figure}{0}
\setcounter{table}{0}

\renewcommand{\thefigure}{S\arabic{figure}}
\renewcommand{\thetable}{S\arabic{table}}
\renewcommand{\theequation}{S\arabic{equation}}
\renewcommand{\thesection}{S\arabic{section}}

\section{DeepSets architecture} \label{supsec:deepsets}

\begin{figure}[htb]
\begin{center}
\begin{tikzpicture}[scale=1]
\node (1) at (0,0) [ellipse, draw]{$\;\bm Z_1$};
\node (2) at (0,-2) [ellipse, draw]{$\bm Z_n$};
\node (3) at (2.5, 0) [draw]{\;$\bm\psi(\cdot)$\;};
\node (4) at (2.5, -2) [draw]{\;$\bm\psi(\cdot)$\;};
\node (5) at (5, -1) [draw, ellipse]{Average};
\node (6) at (7.5, -1) [circle, draw]{\;$\bm S$\;};
\node (7) at (10, -1) [draw]{\;$\bm \phi(\cdot)$\;};
\node (8) at (12.5, -1) [circle, draw]{\;$\bm{\hat \theta}(\cdot)$\;};
\node (9) at (0, -1) {$\vdots$};
\node (10) at (2.5, -1) {$\vdots$};
\draw[->] (1) to (3);
\draw[->] (2) to (4);
\draw[->] (3) to (5);
\draw[->] (4) to (5);
\draw[->] (5) to (6);
\draw[->] (6) to (7);
\draw[->] (7) to (8);
\end{tikzpicture}
\end{center}
\caption{In the first step, the data inputs $\bm Z_1, \ldots, \bm Z_n$ are transformed independently through neural network $\bm \psi(\cdot),$ They are then aggregated through the elementwise average, obtaining the summary statistic $\bm S.$ In the last step, neural network $\bm \phi(\cdot)$ maps the summary statistic $\bm S$ to an estimate of the vector of model parameters $\bm{\hat \theta}(\cdot).$}
\label{fig:deepsets}
\end{figure}
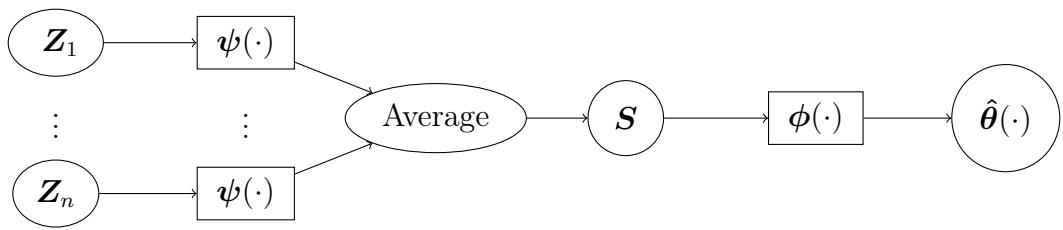

A schematic of the DeepSets architecture (recall Section~\ref{subsec:nbe} of the main paper) used is shown in Figure~\ref{fig:deepsets}. This is based on \citet{SainsburyDaleetal2024}.

\newpage

\section{Parameter estimation assessment}\label{sup:inference}

In this section, we present the simulation studies done for the remaining models considered in this work. In Section~\ref{supsec:wcm}, we show the performance of the NBEs for uncensored data in five configurations of the weighted copula model (WCM) from Section~\ref{subsec:wcm} of the main paper. Where feasible, a comparison with maximum likelihood inference is presented. In Section~\ref{supsec:wads}, we show the performance of the NBEs when the sample size and censoring level are kept fixed, and when the sample size is assumed variable but the censoring level is still fixed for Model W. Finally, in Sections~\ref{supsec:hw}, \ref{supsec:eng1} and \ref{supsec:eng2}, we present the results for the remaining three models from Section~\ref{subsec:flexiblemodels} of the main paper. In all of these cases, a comparison with censored maximum likelihood estimation is given, wherein the censoring scheme used is that of the main paper (i.e., outlined in \citet{Wadsworthetal2017}).

The neural network architecture used for parameter estimation (recall Section~\ref{subsec:parestimationsim} of the main paper) is given in Table~\ref{tab:paresttable}.

\begin{table}[!t]
    \centering
    \caption{Summary of the neural network architecture used to train the NBE. The input array to the first layer represents the dimension $d=2$ of data set $\bm Z;$ this differs for uncensored and censored data. For the censored case, a bilinear layer is used instead, and an extra dimension for the indicator vector $\bm I$ is needed. In addition, the input layer of $\bm \phi(\cdot)$ has an extra dimension in the case of censored data with random censoring level $\tau.$ The output array $[p]$ of the last layer represents the number of parameters in the model.}
    \begin{tabular}{ccc}
        \toprule
        Neural network & Input dimension & Output dimension \\
        \midrule
        \multirow{3}{*}{$\bm \psi(\cdot)$} & $[2]$ or $[2,2]$ & $[128]$ \\
        & $[128]$ & $[128]$ \\
        & $[128]$ & $[256]$ \\
        \midrule 
        \multirow{2}{*}{$\bm \phi(\cdot)$} & $[256]$ or $[257]$ & $[128]$ \\
        & $[128]$ & $[p]$ \\
        \bottomrule
    \end{tabular}
    \label{tab:paresttable}
\end{table}

\subsection{Weighted copula model} \label{supsec:wcm}

We consider now five additional configurations of the WCM. For the first two models, we assume $c_b$ and $c_t$ to be one-parameter copulas, while for the remaining three configurations (Sections~\ref{supsubsec:wcmwads}, \ref{supsubsec:wcmhw} and \ref{supsubsec:wcme2}) $c_b$ is assumed to be a Gaussian copula, and $c_t$ is one of the flexible copulas mentioned in Section~\ref{subsec:flexiblemodels} of the main paper. For these three models configurations, the likelihood is infeasible and hence no comparison with MLE is provided. In all the models, we take $\pi(x_1,x_2;\gamma) = (x_1x_2)^{\gamma}$ as the weighting function. Since preliminary analysis indicated that the neural network was struggling to learn $\gamma,$ we set $\kappa=\log\gamma$ and estimate $\kappa$ instead. Lastly, the model-based $\chi(u)$ estimates of the WCM are obtained using a Monte Carlo approximation with $500\,000$ samples.

\subsubsection{Model 1: $c_b$ is a Gaussian copula and $c_t$ is a logistic copula}

\begin{figure}[t!]
    \centering
    \includegraphics[width=\textwidth]{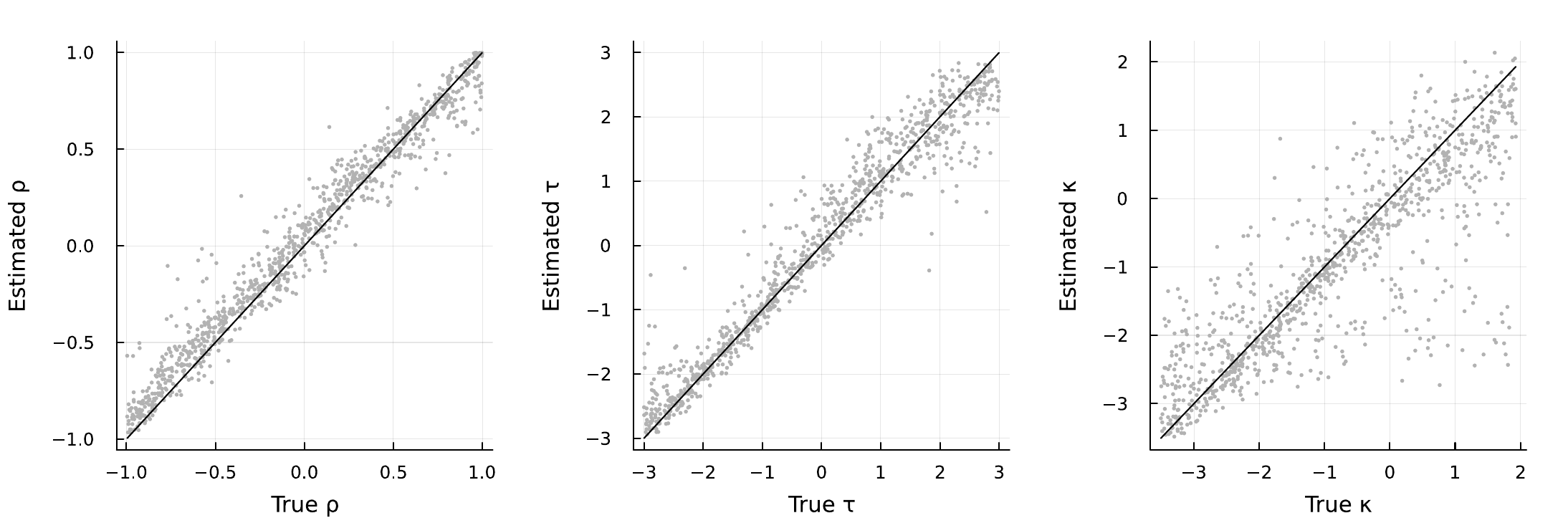}
    \caption{Assessment of the NBE when $c_b$ is a Gaussian copula with correlation parameter $\rho,$ $c_t$ is a logistic copula with parameter $\tau_L = \text{logit}(\alpha_L),$ and with weighting function $\pi(x_1, x_2;\kappa)=(x_1x_2)^{\exp(\kappa)},$ $x_1, x_2 \in (0,1)$ for a sample size of $n=1000.$}
    \label{fig:wcmassessment_mod1}
\end{figure}

For the first model, we consider the copula tailored to the body $c_b$ to be a Gaussian copula with correlation parameter $\rho\in (-1,1),$ and the copula tailored to the tail $c_t$ to be a logistic copula with $\alpha_L \in (0, 1].$ Similarly to the weighting function parameter $\gamma,$ we take an alternative parameterisation and set $\tau_L = \text{logit}(\alpha_L).$ Additionally, we set $\rho \sim \Unifd(-1, 1),$ \linebreak $\tau_L\sim \Unifd(-3,3),$ which results in $\alpha_L \in (0.05, 0.95),$ and $\kappa \sim \Unifd(-3.51, 1.95),$ which leads to $\gamma\in (-0.03, 7.03),$ as the priors for the parameters. The performance of the NBE is assessed in Figure~\ref{fig:wcmassessment_mod1} where the true values of the parameters are compared with their estimated values. It can be seen that parameter $\kappa$ exhibits a bit of variability, while parameters $\rho$ and $\tau_L$ are estimated quite well via the NBE. The coverage probabilities and average length of the $95\%$ uncertainty intervals obtained via a non-parametric bootstrap procedure (as described in Section~\ref{subsec:parestimationsim} of the main paper) are shown in Table~\ref{tab:uncwcm_mod1}. Similarly to the main paper, we compute the coverage probabilities of $95\%$ uncertainty intervals, and their average length, for $\chi(u)$ at levels $u=\{0.50, 0.80, 0.95\};$ the results are shown on the right of Table~\ref{tab:uncwcm_mod1}. According to these results, the true $\chi(u)$ is within the confidence intervals in more than $77\%$ of the time, which suggest that this measure is well derived from the NBE.

\begin{table}[!t]
    \centering
    \captionsetup{width=\textwidth}
    \caption{Coverage probability and average length of the $95\%$ uncertainty intervals for the parameters (left) and for $\chi(u)$ at levels $u=\{0.50, 0.80, 0.95\}$ (right)  obtained via a non-parametric bootstrap procedure averaged over $1000$ models fitted using a NBE (rounded to 2 decimal places).}
    \begin{tabular}{ccccccccc}
        \cmidrule[\heavyrulewidth]{1-3} \cmidrule[\heavyrulewidth]{7-9}
        Parameter & Coverage & Length & & & & $\chi(u)$ & Coverage &  Length \\
        \cmidrule{1-3} \cmidrule{7-9}
        $\rho$ & $0.71$ & $0.26$ & & & & $\chi(0.50)$ & $0.77$ & $0.05$ \\
        $\tau_L$ & $0.75$ & $0.76$ & & & & $\chi(0.80)$ & $0.79$ & $0.08$ \\
        $\kappa$ & $0.69$ & $1.33$ & & & & $\chi(0.95)$ & $0.78$ & $0.09$ \\
        \cmidrule[\heavyrulewidth]{1-3} \cmidrule[\heavyrulewidth]{7-9} 
    \end{tabular}
    \label{tab:uncwcm_mod1}
\end{table}

\subsubsection*{Comparison with maximum likelihood estimation}

\begin{sidewaysfigure}[!t]
    \centering
    \begin{subfigure}[b]{0.49\textwidth}
        \includegraphics[width=\textwidth]{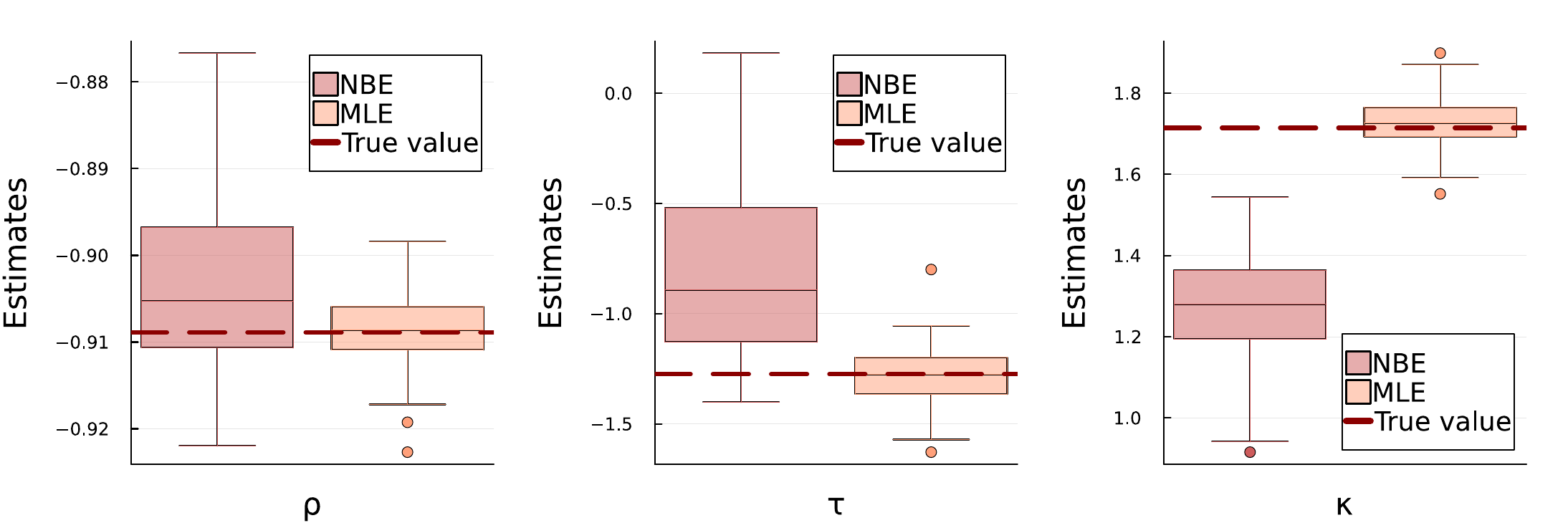}
        \caption{}
        \label{subfig:box1_mod1}
    \end{subfigure}
    \hfill
    \begin{subfigure}[b]{0.49\textwidth}
        \includegraphics[width=\textwidth]{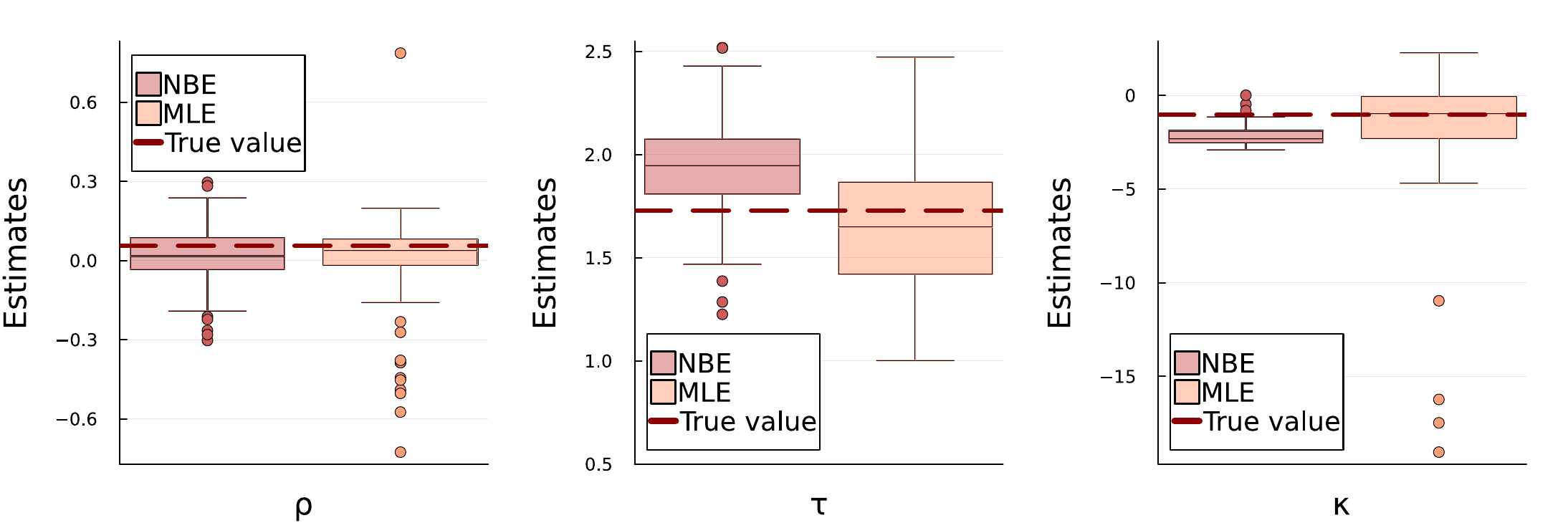}
        \caption{}
        \label{subfig:box2_mod1}
    \end{subfigure}
    \hfill
    \begin{subfigure}[b]{0.49\textwidth}
        \includegraphics[width=\textwidth]{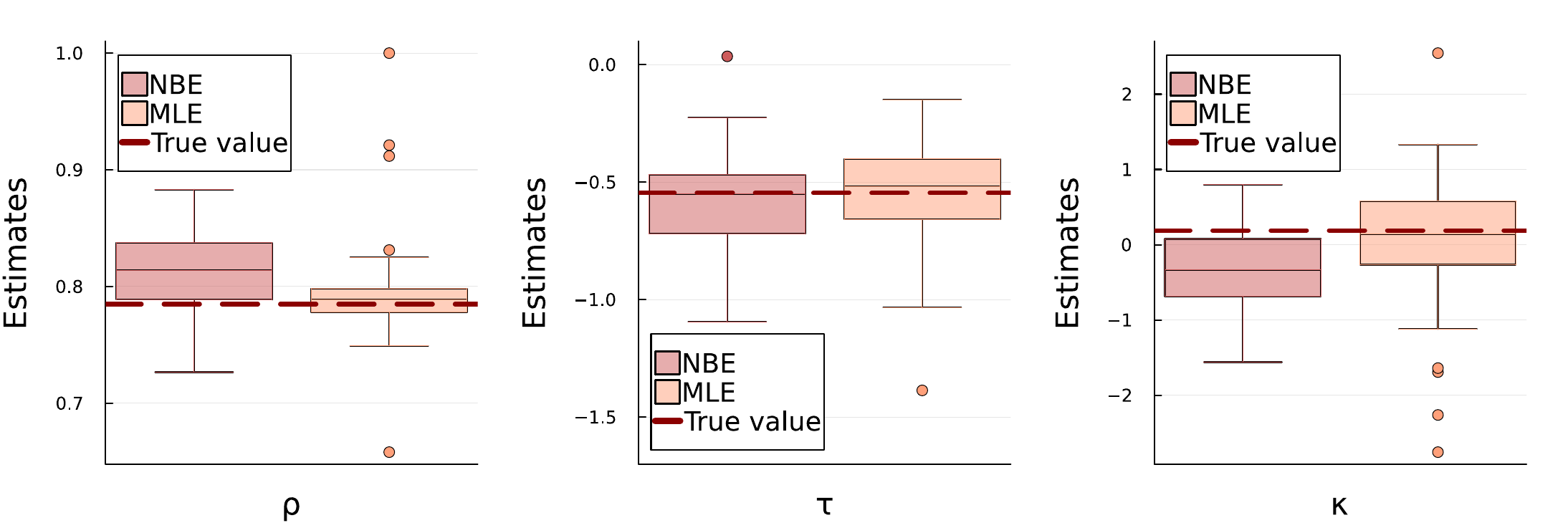}
        \caption{}
        \label{subfig:box3_mod1}
    \end{subfigure}
    \hfill
    \begin{subfigure}[b]{0.49\textwidth}
        \includegraphics[width=\textwidth]{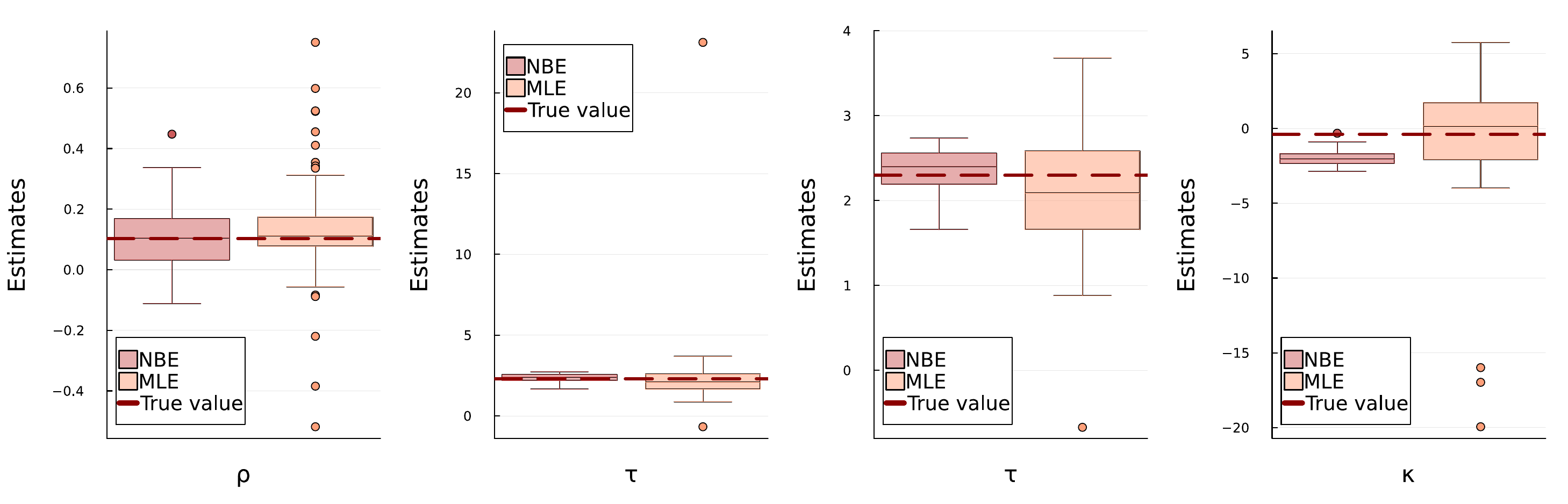}
        \caption{}
        \label{subfig:box4_mod1}
    \end{subfigure}
    \hfill
    \begin{subfigure}[b]{0.49\textwidth}
        \includegraphics[width=\textwidth]{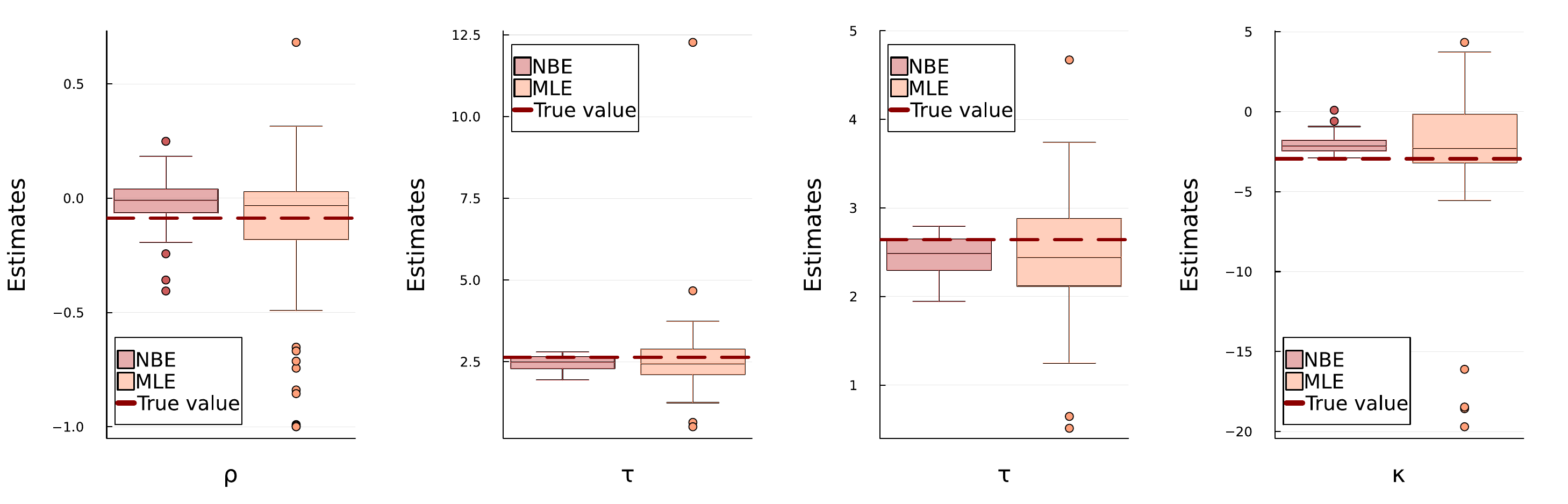}
        \caption{}
        \label{subfig:box5_mod1}
    \end{subfigure}
    \caption{Comparison between parameter estimates $\hat{\bm \theta}=(\hat\rho, \hat\tau_L, \hat\kappa)'$ given by MLE (orange) and by NBE (red) for $100$ samples with $n=1000.$ The true parameter values are given by the red line. (a) $\bm \theta = (0.91, -1.27, 1.71)',$ (b) $\bm \theta = (0.91, 1.73, -1.03)',$ (c) $\bm \theta = (0.91, -0.55, 0.19)',$ \linebreak (d) $\bm \theta = (0.91, 2.30, -0.38)'$ and (e) $\bm \theta = (0.91, 2.64, -2.95)'.$ For better visualisation, the larger outliers obtained through MLE were removed for $\hat\tau_L$ in (d) and (e).}
    \label{fig:mlevsnbe_mod1}
\end{sidewaysfigure}

Since the likelihood of this model is feasible, though computationally intensive, we compare the estimations obtained by the NBE to the MLEs. With the assigned priors, we generate five different parameter vectors $\bm \theta = (\rho, \tau_L, \kappa)'$ and corresponding data sets, each of which with $n = 1000.$ Additionally, each data set is simulated $100$ times. The results are shown in Figure~\ref{fig:mlevsnbe_mod1}; it can be seen that the NBE estimates are generally more biased, and sometimes more variable, than the MLEs. However, they are less likely to have big outliers as the neural network is trained in a bounded interval. Despite slightly more biased, the estimates obtained with the NBE are generally good. Furthermore, it is substantially faster to obtain an estimate through NBE than through maximum likelihood. In particular, on average, the MLE took 3 hours and 12 minutes to evaluate, while the NBE took $0.653$ seconds; this means that the NBE is about $17\,663$ times faster---a substantial improvement in computational time.

\subsubsection{Model 2: $c_b$ is a Frank copula and $c_t$ is a Joe copula}

\begin{figure}[t!]
    \centering
    \includegraphics[width=\textwidth]{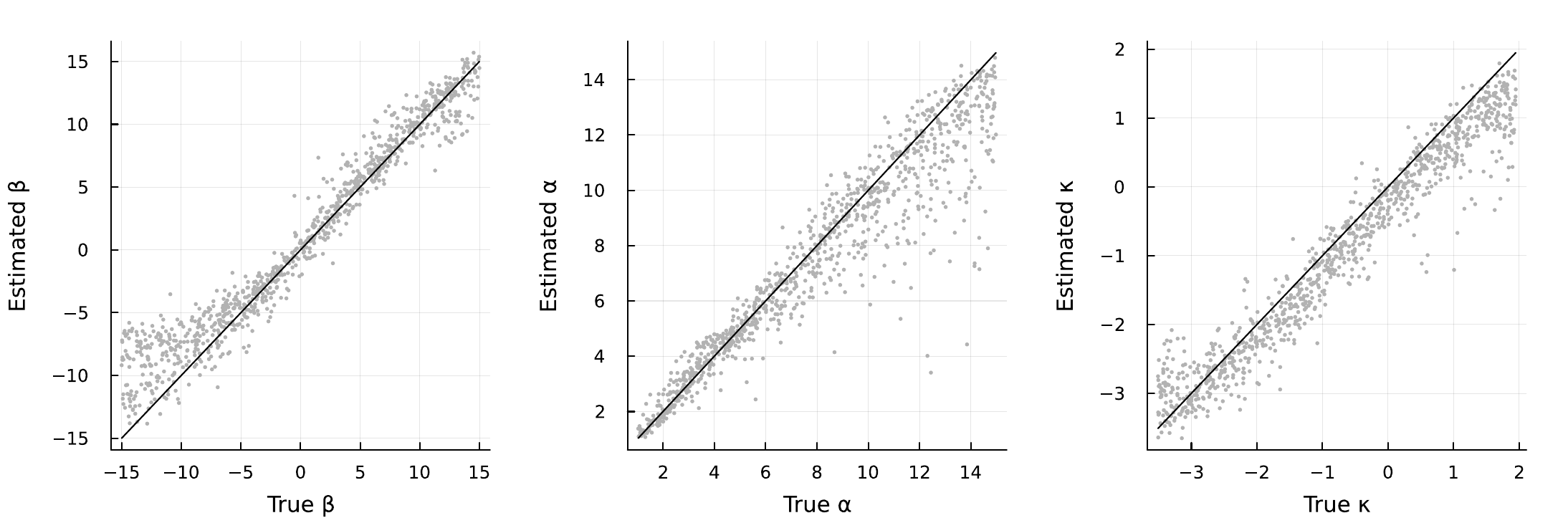}
    \caption{Assessment of the NBE when $c_b$ is a Frank copula with parameter $\beta,$ $c_t$ is a Joe copula with parameter $\alpha_J,$ and with weighting function $\pi(x_1, x_2;\kappa)=(x_1x_2)^{\exp(\kappa)},$ $x_1, x_2 \in (0,1)$ for a sample size of $n=1000.$}
    \label{fig:wcmassessment_mod2}
\end{figure}

For the second model, we consider $c_b$ to be a Frank copula \citep{Frank1979} with parameter $\beta_F\in \mathbb{R},$ and $c_t$ to be a Joe copula \citep{Joe1996} with $\alpha_J > 1.$ As priors for the model parameters, we take $\beta_F \sim \Unifd(-15, 15),$ $\alpha_J\sim \Unifd(1,15)$  and $\kappa \sim \Unifd(-3.51, 1.95).$ The performance of the NBE is assessed in Figure~\ref{fig:wcmassessment_mod2} where the true values of the parameters are compared with their estimated values. It can be seen that all the parameters are estimated quite well with the NBE, with $\beta_F$ and $\alpha_J$ showing a bit of variability for lower and higher values, respectively. The coverage probabilities and average length of the $95\%$ uncertainty intervals obtained via a non-parametric bootstrap procedure for the parameter estimates and for $\chi(u)$ at levels $u\in \{0.50, 0.80, 0.95\}$ are shown in Table~\ref{tab:uncwcm_mod2}. The lower coverage rates given on the left table reflect the bias shown by the parameter estimates. However, the results for $\chi(u)$ suggest that the NBE is able to capture the dependence structure of the data, especially for higher $u,$ with the true $\chi(u)$ being within the confidence intervals in more than $59\%$ of the time.

\FloatBarrier

\begin{table}[!t]
    \centering
    \captionsetup{width=\textwidth}
    \caption{Coverage probability and average length of the $95\%$ uncertainty intervals for the parameters (left) and for $\chi(u)$ at levels $u=\{0.50, 0.80, 0.95\}$ (right)  obtained via a non-parametric bootstrap procedure averaged over $1000$ models fitted using a NBE (rounded to 2 decimal places).}
    \begin{tabular}{ccccccccc}
        \cmidrule[\heavyrulewidth]{1-3} \cmidrule[\heavyrulewidth]{7-9}
        Parameter & Coverage & Length & & & & $\chi(u)$ & Coverage &  Length \\
        \cmidrule{1-3} \cmidrule{7-9}
        $\beta_F$ & $0.68$ & $3.03$ & & & & $\chi(0.50)$ & $0.61$ & $0.05$ \\
        $\alpha_J$ & $0.72$ & $1.95$ & & & & $\chi(0.80)$ & $0.59$ & $0.06$ \\
        $\kappa$ & $0.62$ & $0.92$ & & & & $\chi(0.95)$ & $0.64$ & $0.06$ \\
        \cmidrule[\heavyrulewidth]{1-3} \cmidrule[\heavyrulewidth]{7-9} 
    \end{tabular}
    \label{tab:uncwcm_mod2}
\end{table}

\subsubsection*{Comparison with maximum likelihood estimation}

\begin{sidewaysfigure}[!t]
    \centering
    \begin{subfigure}[b]{0.49\textwidth}
        \includegraphics[width=\textwidth]{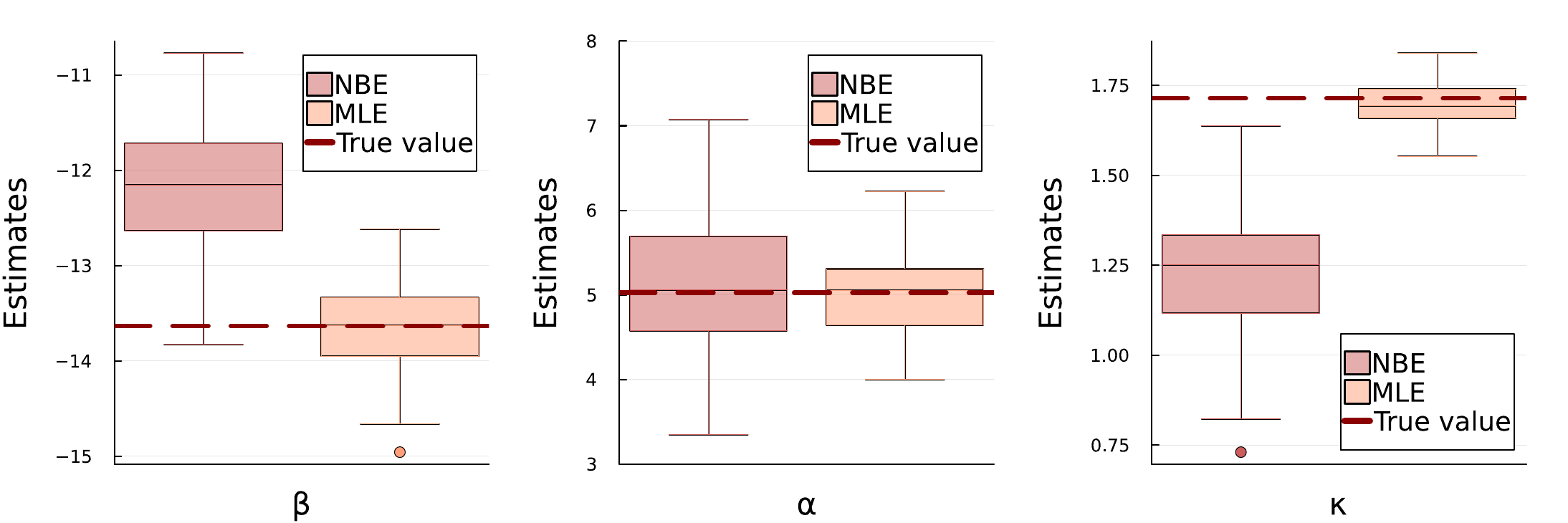}
        \caption{}
        \label{subfig:box1_mod2}
    \end{subfigure}
    \hfill
    \begin{subfigure}[b]{0.49\textwidth}
        \includegraphics[width=\textwidth]{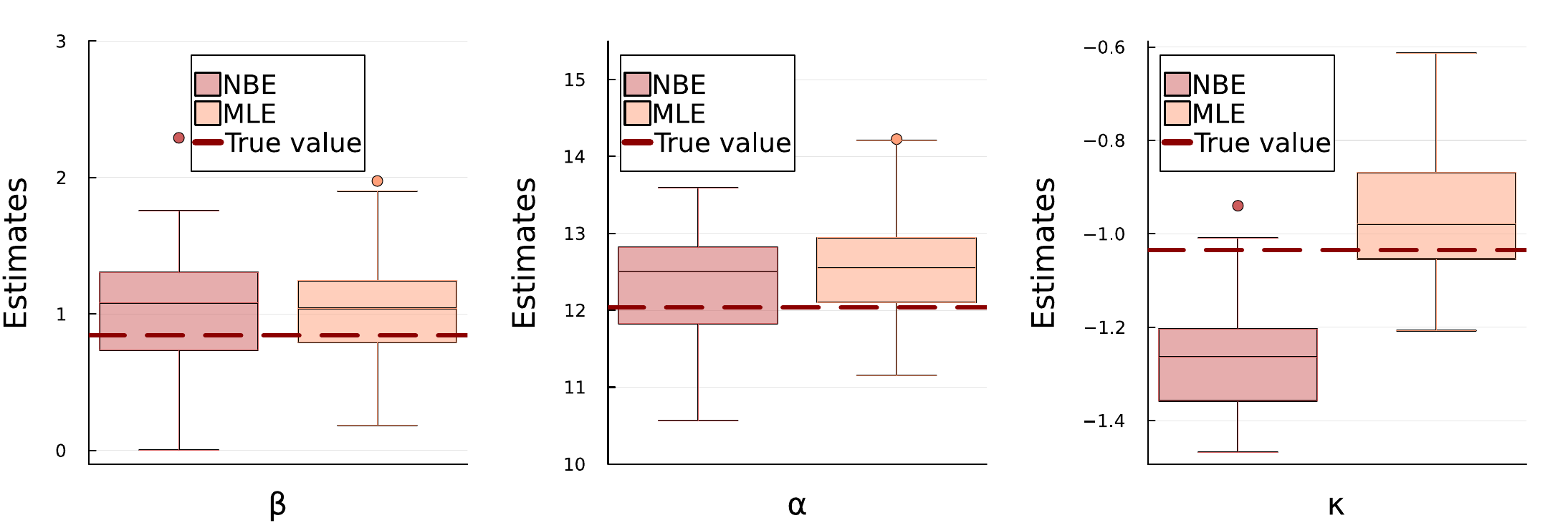}
        \caption{}
        \label{subfig:box2_mod2}
    \end{subfigure}
    \hfill
    \begin{subfigure}[b]{0.49\textwidth}
        \includegraphics[width=\textwidth]{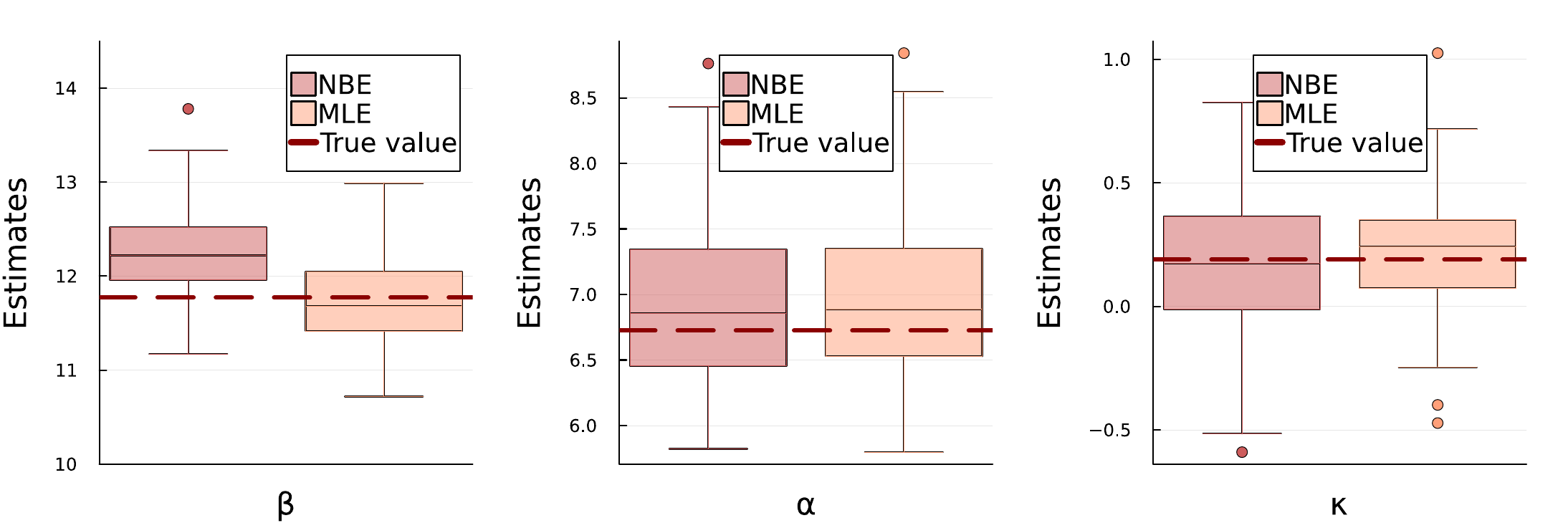}
        \caption{}
        \label{subfig:box3_mod2}
    \end{subfigure}
    \hfill
    \begin{subfigure}[b]{0.49\textwidth}
        \includegraphics[width=\textwidth]{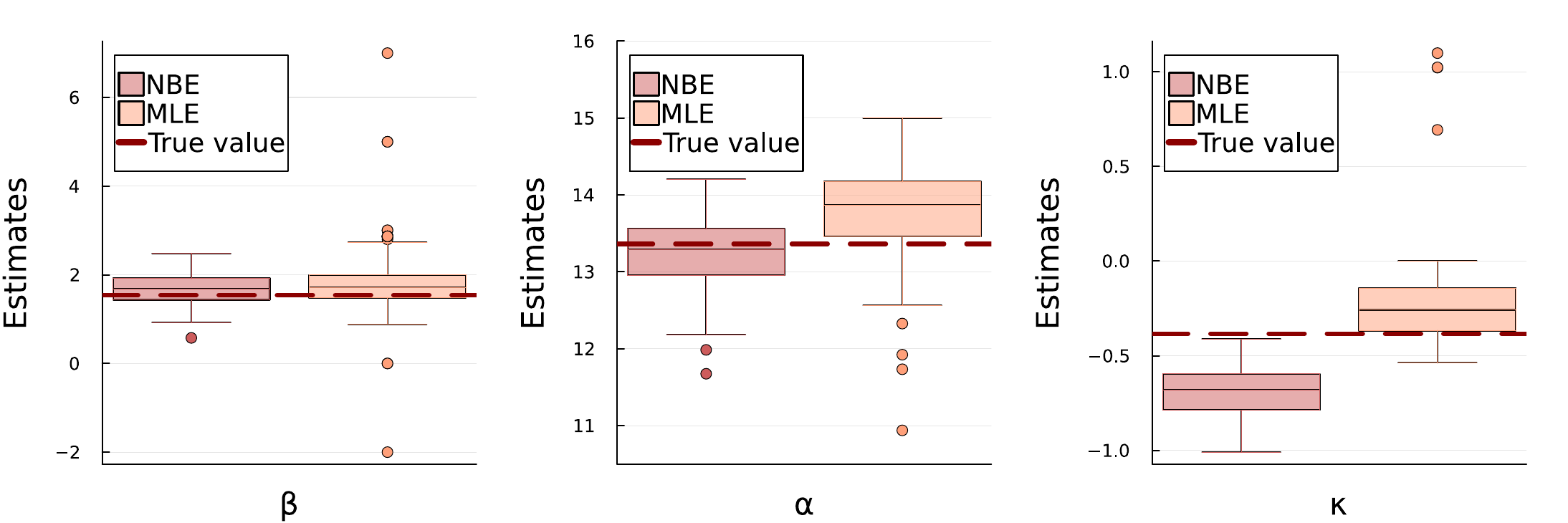}
        \caption{}
        \label{subfig:box4_mod2}
    \end{subfigure}
    \hfill
    \begin{subfigure}[b]{0.49\textwidth}
        \includegraphics[width=\textwidth]{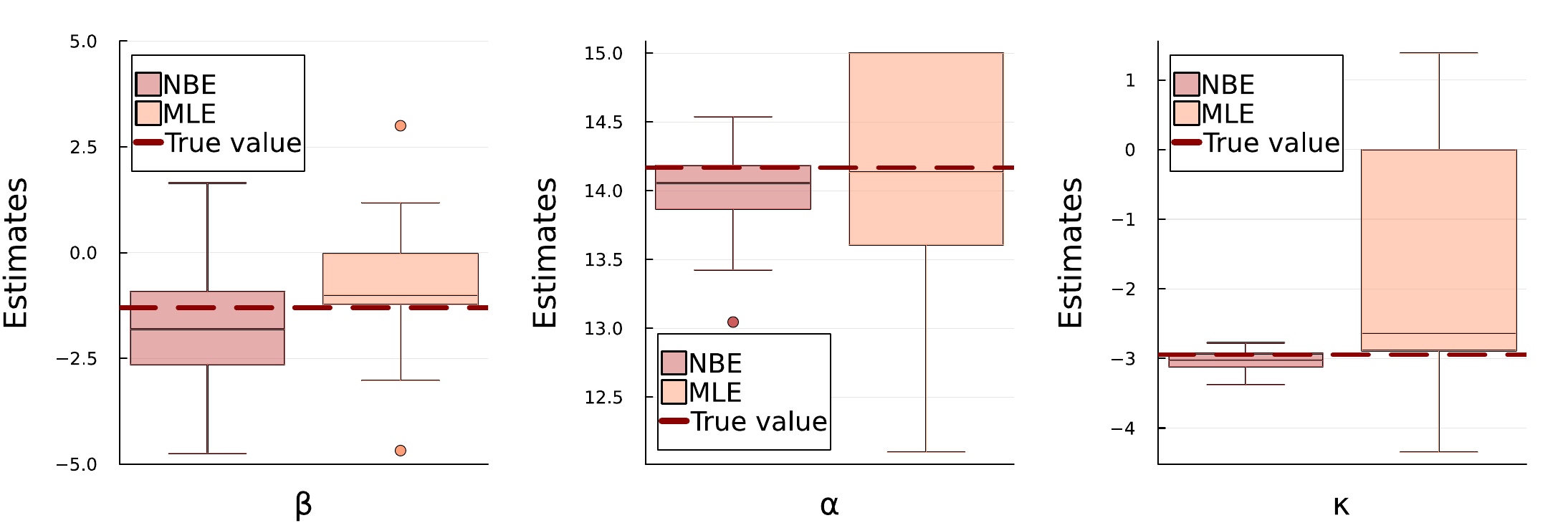}
        \caption{}
        \label{subfig:box5_mod2}
    \end{subfigure}
    \caption{Comparison between parameter estimates $\hat{\bm \theta}=(\hat\beta_F, \hat\alpha_J, \hat\kappa)'$ given by MLE (orange) and by NBE (red) for $100$ samples with $n=1000.$ The true parameter values are given by the red line. (a) $\bm \theta = (-13.63, 5.02,  1.71)',$ (b) $\bm \theta = (0.84, 12.04, -1.03)',$ \linebreak (c) $\bm \theta = (11.77, 6.73, 0.19)',$ (d) $\bm \theta = (1.54, 13.36, -0.38)'$ and (e) $\bm \theta = (-1.30, 14.17, -2.95)'.$}
    \label{fig:mlevsnbe_mod2}
\end{sidewaysfigure}

For this model the likelihood is also feasible (and computational expensive). Therefore, as before, we compare the estimations obtained by the NBE and by the MLE for five different parameter vectors $\bm \theta = (\beta_F, \alpha_J, \kappa)',$ generated with the pre-specified priors, and corresponding data sets (each with $n=1000)$. Again, each data set is simulated $100$ times; the results are shown in Figure~\ref{fig:mlevsnbe_mod2}. Similarly to the first model, the NBE estimates are generally more biased than the MLEs, are less prone to have large outliers, and are generally good. While, on average, the MLE took 52 minutes to evaluate, the NBE took $0.203$ seconds, which is about $15\,339$ times faster.

\FloatBarrier

\subsubsection{Model 3: $c_b$ is a Gaussian copula and $c_t$ is Model W} \label{supsubsec:wcmwads}

\begin{figure}[t!]
    \centering
    \includegraphics[width=0.9\textwidth]{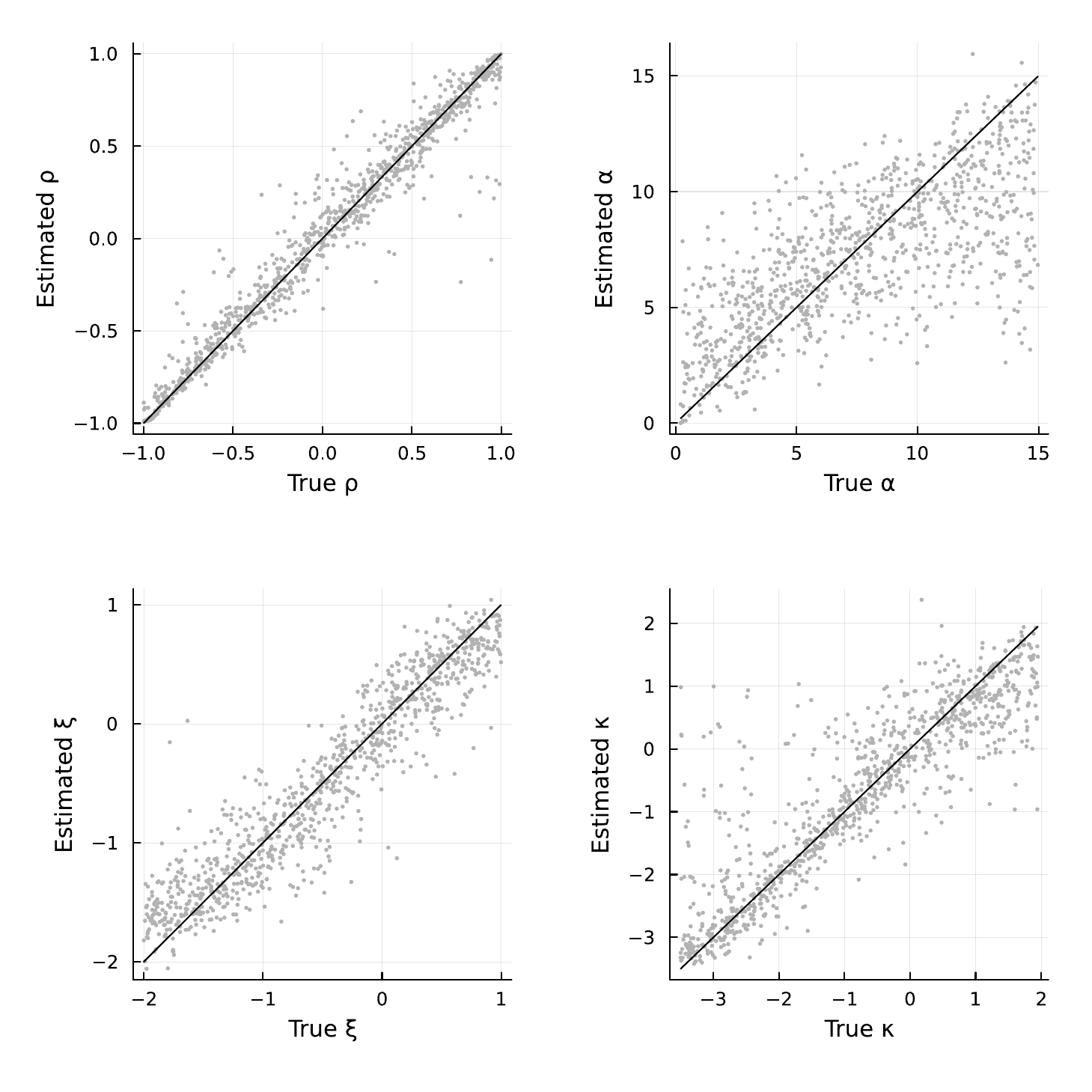}
    \caption{Assessment of the NBE when $c_b$ is a Gaussian copula with correlation parameter $\rho,$ $c_t$ is Model W with parameters $(\alpha, \xi)'$ and with weighting function $\pi(x_1, x_2;\kappa)=(x_1x_2)^{\exp(\kappa)},$ $x_1, x_2 \in (0,1)$ for a sample size of $n=1000.$}
    \label{fig:wcmassessment_wads}
\end{figure}

For the third model, we consider $c_t$ to be Model W, for which the priors for the parameters are those mentioned in Section~\ref{subsec:generalsetting} from the main paper. Figure~\ref{fig:wcmassessment_wads} displays the performance of the NBE. Despite the variability shown, especially by $\alpha$ and $\kappa,$ the NBE provides good estimates overall. The coverage probabilities and average length of the $95\%$ uncertainty intervals for the parameters and for $\chi(u)$ at levels $u=\{0.50, 0.80, 0.95\}$ obtained via a non-parametric bootstrap procedure are given in Table~\ref{tab:uncwcm_wads} on the left and right, respectively. The results for the parameter uncertainty are in agreement with Figure~\ref{fig:wcmassessment_wads}, where the coverage probability for $\alpha$ is the lowest and its average length the highest. However, as shown by the coverage probabilities for $\chi(u),$ this bias does not affect this dependence quantity. More specifically, the true value is within the confidence intervals in more than $85\%$ of the time.

\begin{table}[!t]
    \centering
    \captionsetup{width=\textwidth}
    \caption{Coverage probability and average length of the $95\%$ uncertainty intervals for the parameters (left) and for $\chi(u)$ at levels $u=\{0.50, 0.80, 0.95\}$ (right)  obtained via a non-parametric bootstrap procedure averaged over $1000$ models fitted using a NBE (rounded to 2 decimal places).}
    \begin{tabular}{ccccccccc}
        \cmidrule[\heavyrulewidth]{1-3} \cmidrule[\heavyrulewidth]{7-9}
        Parameter & Coverage & Length & & & & $\chi(u)$ & Coverage &  Length \\
        \cmidrule{1-3} \cmidrule{7-9}
        $\rho$ & $0.85$ & $0.24$ & & & & $\chi(0.50)$ & $0.91$ & $0.06$ \\
        \addlinespace[1.6mm]
        $\alpha$ & $0.60$ & $4.28$ & & & & $\chi(0.80)$ & $0.89$ & $0.09$ \\
        \addlinespace[1.6mm]
        $\xi$ & $0.71$ & $0.56$ & & & & $\chi(0.95)$ & $0.85$ & $0.11$ \\
        \cmidrule[\heavyrulewidth]{7-9} 
        $\kappa$ & $0.73$ & $1.16$ & & & & & & \\
        \cmidrule[\heavyrulewidth]{1-3} 
    \end{tabular}
    \label{tab:uncwcm_wads}
\end{table}

\subsubsection{Model 4: $c_b$ is a Gaussian copula and $c_t$ is Model HW} \label{supsubsec:wcmhw}

\begin{figure}[t!]
    \centering
    \includegraphics[width=0.9\textwidth]{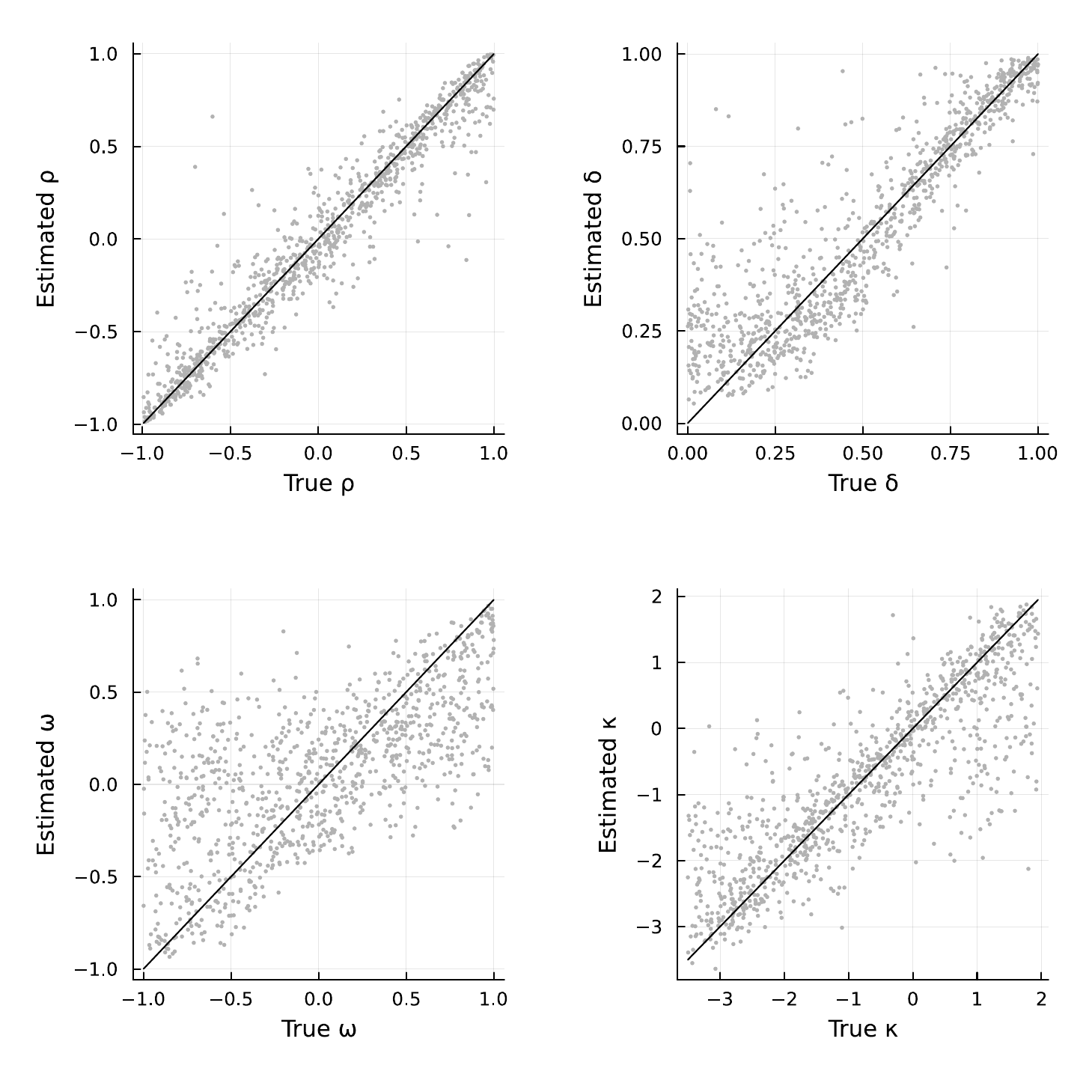}
    \caption{Assessment of the NBE when $c_b$ is a Gaussian copula with correlation parameter $\rho,$ $c_t$ is Model HW with parameters $(\delta, \omega)',$ and with weighting function $\pi(x_1, x_2;\kappa)=(x_1x_2)^{\exp(\kappa)},$ $x_1, x_2 \in (0,1)$ for a sample size of $n=1000.$}
    \label{fig:wcmassessment_hwGauss}
\end{figure}

For the forth model, we consider $c_t$ to be Model HW with the priors for the model parameters mentioned in Section~\ref{subsec:generalsetting} from the main paper. Figure~\ref{fig:wcmassessment_hwGauss} displays the performance of the NBE, showing that $\delta$ and $\omega$ seem to be over-estimated by the NBE for lower values. Table~\ref{tab:uncwcm_hwGauss} shows the coverage probabilities and average length of the $95\%$ uncertainty intervals obtained via a non-parametric bootstrap procedure for the parameters on the left, and for $\chi(u)$ at levels $u=\{0.50, 0.80, 0.95\}$ on the right. The results for the parameter estimates mirror the variability shown in Figure~\ref{fig:wcmassessment_hwGauss}, where the coverage probabilities for $\omega$ and $\delta$ are the lowest. The coverage probabilities of the $95\%$ uncertainty intervals for $\chi(u)$ show that the true value is within the confidence intervals in more than $86\%$ of the time, indicating that despite the bias shown by the estimation, this dependence measure is well calibrated.

\begin{table}[!t]
    \centering
    \captionsetup{width=\textwidth}
    \caption{Coverage probability and average length of the $95\%$ uncertainty intervals for the parameters (left) and for $\chi(u)$ at levels $u=\{0.50, 0.80, 0.95\}$ (right)  obtained via a non-parametric bootstrap procedure averaged over $1000$ models fitted using a NBE (rounded to 2 decimal places).}
    \begin{tabular}{ccccccccc}
        \cmidrule[\heavyrulewidth]{1-3} \cmidrule[\heavyrulewidth]{7-9}
        Parameter & Coverage & Length & & & & $\chi(u)$ & Coverage &  Length \\
        \cmidrule{1-3} \cmidrule{7-9}
        $\rho$ & $0.80$ & $0.28$ & & & & $\chi(0.50)$ & $0.91$ & $0.07$ \\
        \addlinespace[1.6mm]
        $\delta$ & $0.58$ & $0.15$ & & & & $\chi(0.80)$ & $0.88$ & $0.10$ \\
        \addlinespace[1.6mm]
        $\omega$ & $0.44$ & $0.47$ & & & & $\chi(0.95)$ & $0.86$ & $0.12$ \\
        \cmidrule[\heavyrulewidth]{7-9} 
        $\kappa$ & $0.71$ & $1.33$ & & & & & & \\
        \cmidrule[\heavyrulewidth]{1-3} 
    \end{tabular}
    \label{tab:uncwcm_hwGauss}
\end{table}

\subsubsection{Model 5: $c_b$ is a Gaussian copula and $c_t$ is Model E2} \label{supsubsec:wcme2}

\begin{figure}[t!]
    \centering
    \includegraphics[width=0.9\textwidth]{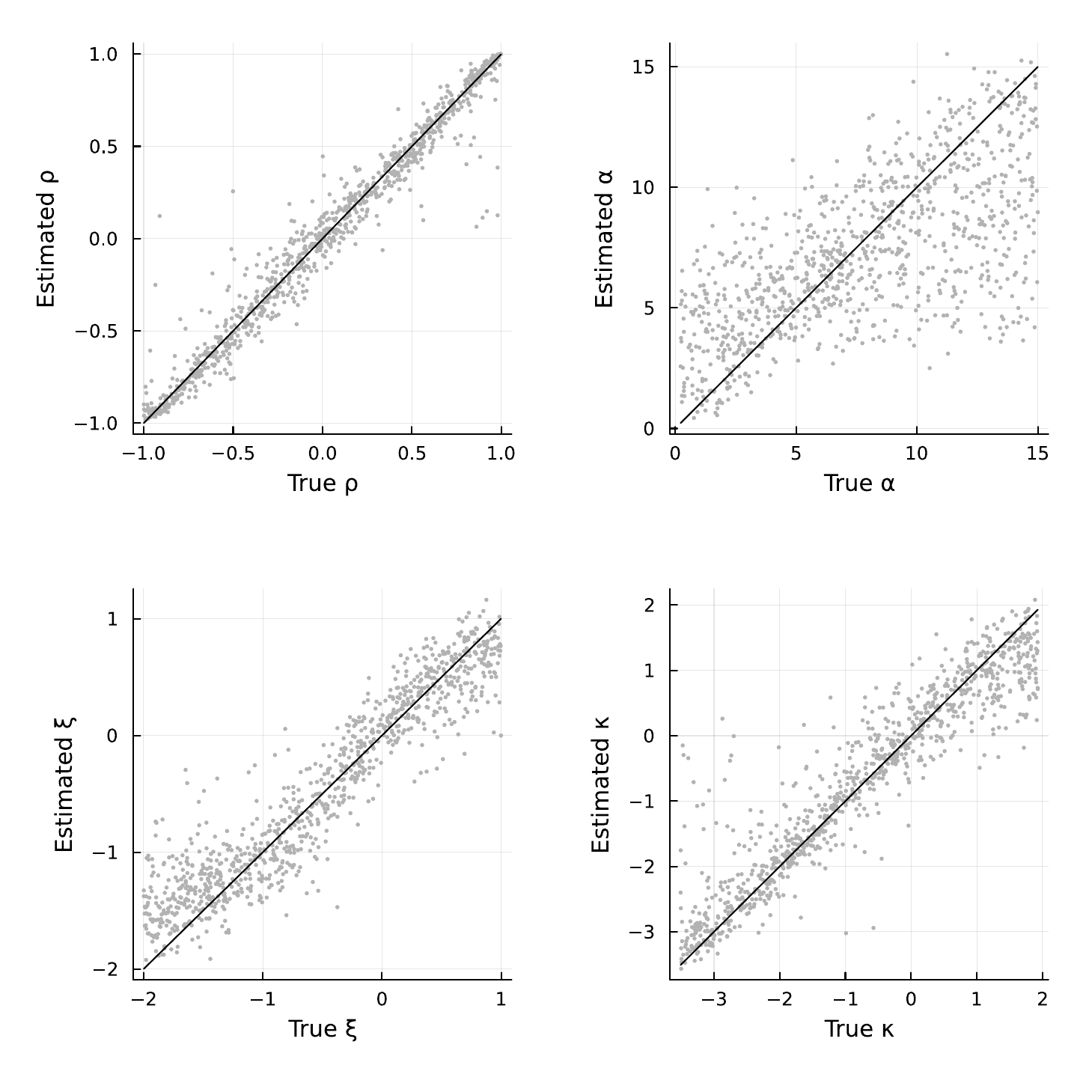}
    \caption{Assessment of the NBE when $c_b$ is a Gaussian copula with correlation parameter $\rho,$ $c_t$ is Model E2 with parameters $(\alpha, \xi)'$ and with weighting function $\pi(x_1, x_2;\kappa)=(x_1x_2)^{\exp(\kappa)},$ $x_1, x_2 \in (0,1)$ for a sample size of $n=1000.$}
    \label{fig:wcmassessment_E2}
\end{figure}

For the final model, we take $c_t$ to be Model E2 with the priors for the model parameters mentioned in Section~\ref{subsec:generalsetting} from the main paper. Figure~\ref{fig:wcmassessment_E2} shows the performance of the NBE. Similarly to Model 3, there is some variability in the NBEs, especially for $\alpha.$ This parameter is also the one with lowest coverage probability and wider interval for the parameters estimation procedure, as shown in left of Table~\ref{tab:uncwcm_E2}. Similarly to the previous models, the coverage probabilities for $\chi(u)$ at levels $u=\{0.50, 0.80, 0.95\},$ shown in the right of Table~\ref{tab:uncwcm_E2}, indicate that this measure is well captured by the NBE, with the true value lying within the confidence intervals in at least $83\%$ of the time.

\begin{table}[!t]
    \centering
    \captionsetup{width=\textwidth}
    \caption{Coverage probability and average length of the $95\%$ uncertainty intervals for the parameters (left) and for $\chi(u)$ at levels $u=\{0.50, 0.80, 0.95\}$ (right)  obtained via a non-parametric bootstrap procedure averaged over $1000$ models fitted using a NBE (rounded to 2 decimal places).}
    \begin{tabular}{ccccccccc}
        \cmidrule[\heavyrulewidth]{1-3} \cmidrule[\heavyrulewidth]{7-9}
        Parameter & Coverage & Length & & & & $\chi(u)$ & Coverage &  Length \\
        \cmidrule{1-3} \cmidrule{7-9}
        $\rho$ & $0.83$ & $0.22$ & & & & $\chi(0.50)$ & $0.88$ & $0.05$ \\
        \addlinespace[1.6mm]
        $\alpha$ & $0.57$ & $4.48$ & & & & $\chi(0.80)$ & $0.84$ & $0.07$ \\
        \addlinespace[1.6mm]
        $\xi$ & $0.72$ & $0.59$ & & & & $\chi(0.95)$ & $0.83$ & $0.10$ \\
        \cmidrule[\heavyrulewidth]{7-9} 
        $\kappa$ & $0.75$ & $0.96$ & & & & & & \\
        \cmidrule[\heavyrulewidth]{1-3} 
    \end{tabular}
    \label{tab:uncwcm_E2}
\end{table}

\clearpage

\subsection{Model W} \label{supsec:wads}

\subsubsection{Variable sample size and censoring level} \label{supsesec:wadspaper}

\subsubsection*{Comparison with censored maximum likelihood estimation}

\begin{figure}[h!]
    \centering
    \begin{subfigure}[b]{0.82\textwidth}
        \includegraphics[width=\textwidth]{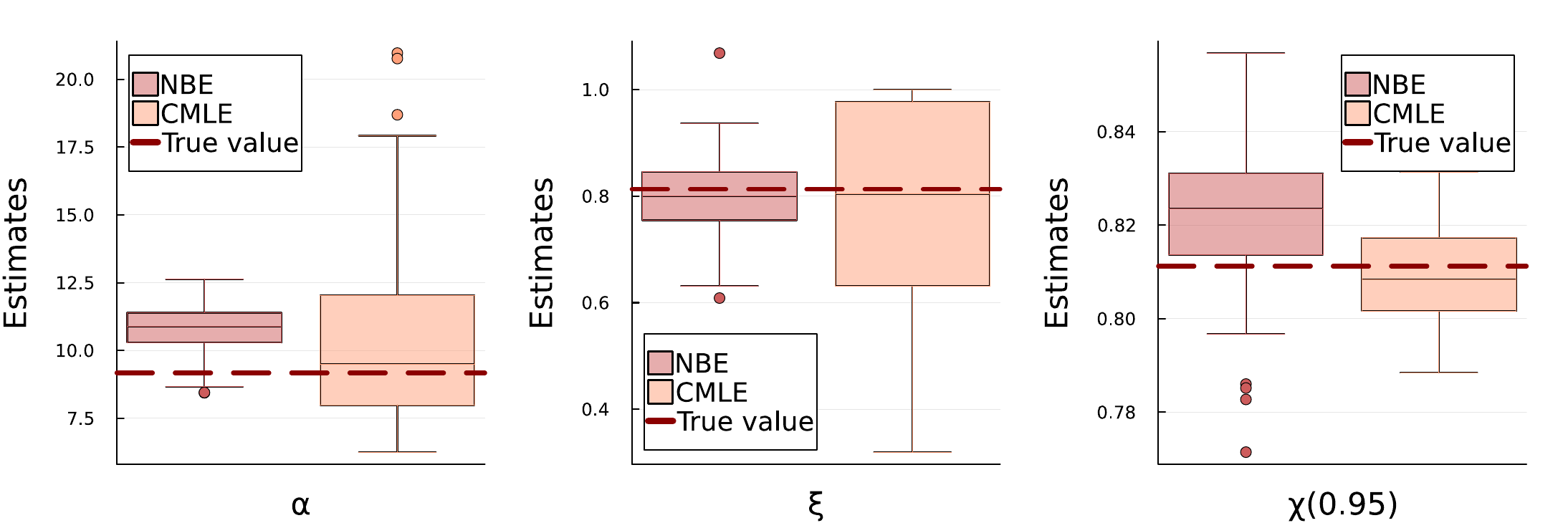}
        \caption{}
        \label{subfig:boxplotdat3}
    \end{subfigure}
    \hfill
    \begin{subfigure}[b]{0.82\textwidth}
        \includegraphics[width=\textwidth]{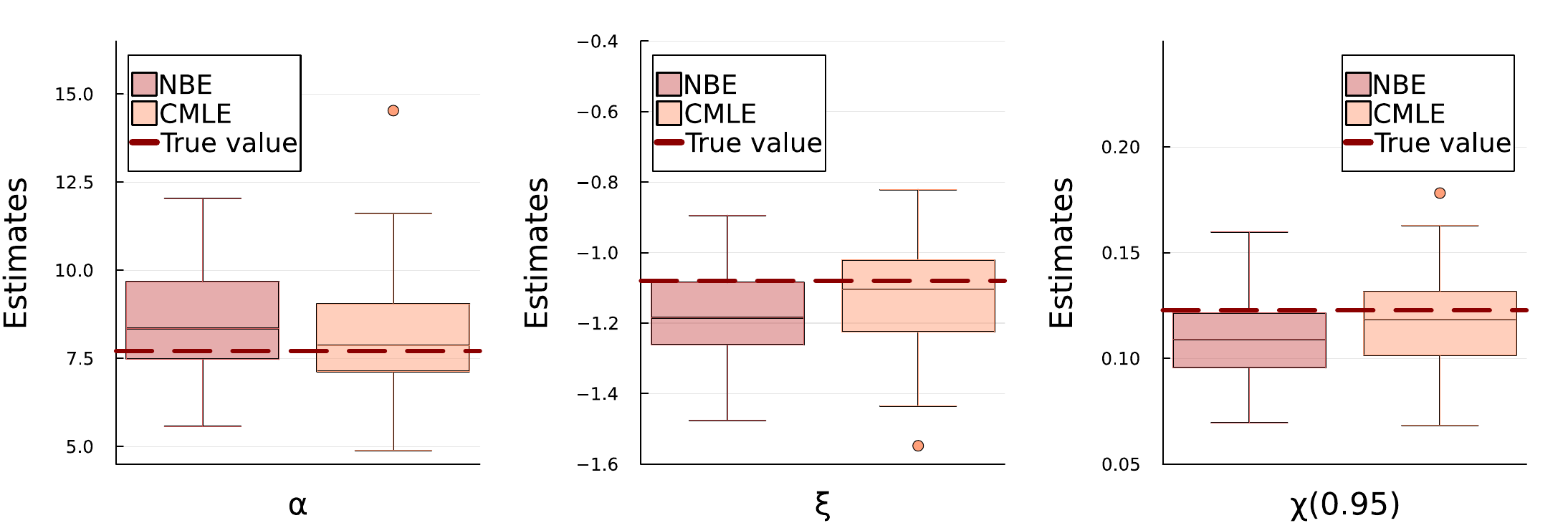}
        \caption{}
        \label{subfig:boxplotdat5}
    \end{subfigure}
    \hfill
    \begin{subfigure}[b]{0.82\textwidth}
        \includegraphics[width=\textwidth]{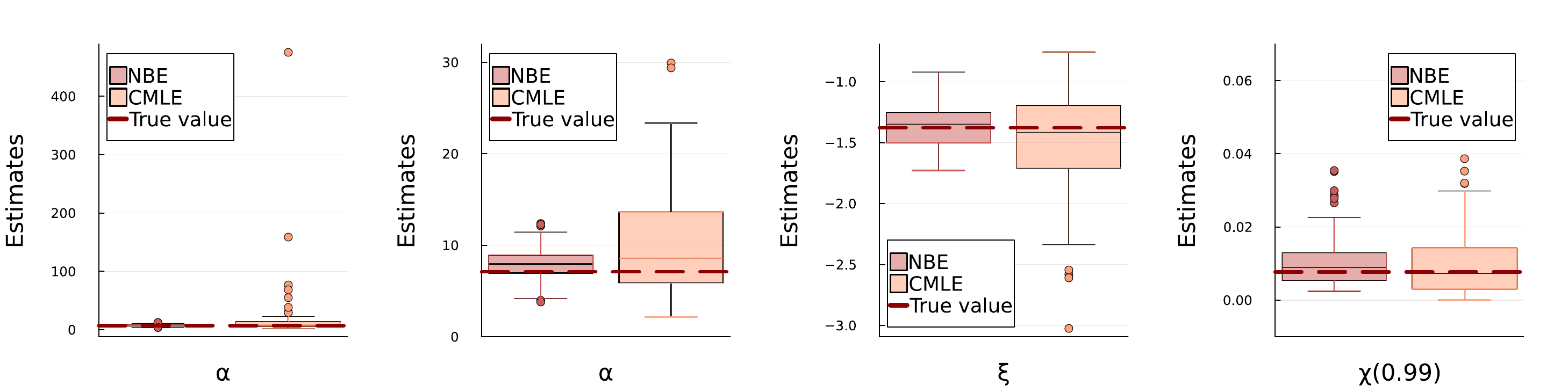}
        \caption{}
        \label{subfig:boxplotdat4}
    \end{subfigure}
    \caption{Comparison between parameter estimates $\hat{\bm\theta}=(\hat \alpha, \hat \xi)'$ given by CMLE (orange) and by NBE (red), and corresponding $\chi(u),$ for $100$ samples with $n=1000.$ The true parameters are given by the red line. (a) $\bm \theta_3 = (9.17, 0.81)'$ with $\tau_3 = 0.80$ and $u=0.95,$ (b) $\bm \theta_4 = (7.71, -1.08)'$ with $\tau_4 = 0.73$ and $u=0.95,$ and (c) $\bm \theta_5 = (7.10, -1.38)'$ with $\tau_5 = 0.98$ and $u=0.99.$ For better visualisation, the larger outliers obtained through MLE are removed for $\hat \alpha$ in (e).}
    \label{fig:cmlevsnbe_wads}
\end{figure}

The comparison between the NBE and CMLE for the remaining three parameter vectors considered in the simulation study of Section~\ref{subsec:parestimationsim} of the main paper is given in Figure~\ref{fig:cmlevsnbe_wads}.

\newpage 

For comparison with the simulation study of Model W given in Section~\ref{subsec:parestimationsim}, we now present the results for when both sample size $n$ and censoring level $\tau$ are kept fixed, and for when the sample size is assumed variable but the censoring level is kept fixed.

\subsubsection{Fixed sample size and censoring level} \label{supsecsec:fixedwads}

\begin{figure}[t!]
    \centering
    \includegraphics[width=\textwidth]{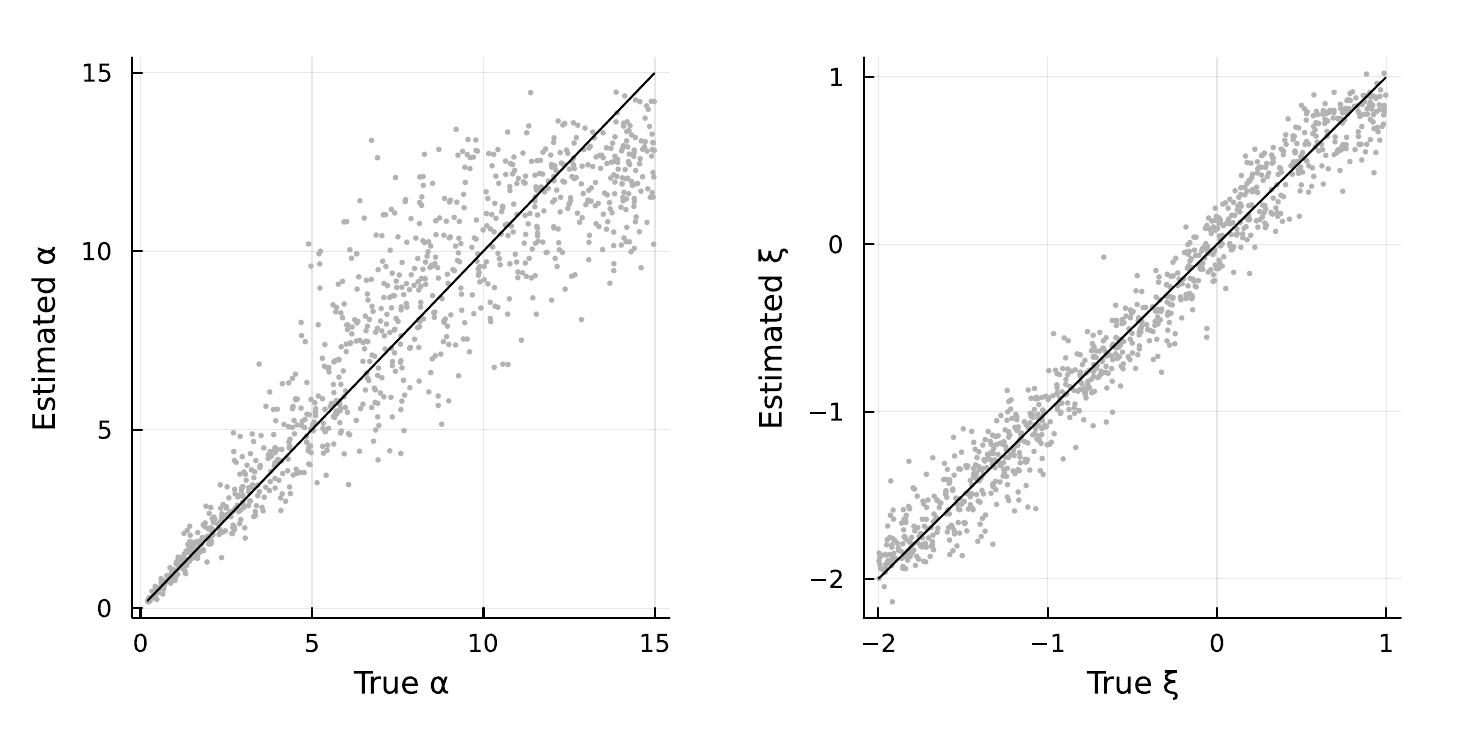}
    \caption{Assessment of the NBE for Model W with parameters $\bm \theta=(\alpha,\xi)'$ for a sample size of $n=1000$ and fixed censoring level $\tau = 0.8.$}
    \label{fig:assessment_wadsfixed}
\end{figure}

We first assume that the sample size and the censoring level are kept fixed at $n = 1000$ and $\tau = 0.8,$ respectively, and the performance of the NBE is shown in Figure~\ref{fig:assessment_wadsfixed}. Similarly to the case presented in Section~\ref{subsec:parestimationsim} of the main paper, the NBE exhibits some bias for larger values of $\alpha.$ This is also noticeable with the average length of the $95\%$ uncertainty intervals obtained via a non-parametric bootstrap procedure given in Table~\ref{tab:unc_wadsfixed}. It can also be seen that the coverage probabilities are slightly higher than the ones from Section~\ref{subsec:parestimationsim}. This might be due to the fact that there are less unknown variables in this configuration. Finally, the coverage probabilities of $95\%$ uncertainty intervals, and their average length, for $\chi(u)$ at levels $u=\{0.80, 0.95, 0.99\}$ are shown on the right of Table~\ref{tab:unc_wadsfixed}. The results are similar to the ones presented in the main paper, with a slightly higher coverage for larger $u.$

\begin{table}[!t]
    \centering
    \captionsetup{width=\textwidth}
    \caption{Coverage probability and average length of the $95\%$ uncertainty intervals for the parameters (left) and for $\chi(u)$ at levels $u=\{0.80, 0.95, 0.99\}$ (right)  obtained via a non-parametric bootstrap procedure averaged over $1000$ models fitted using a NBE (rounded to 2 decimal places).}
    \begin{tabular}{ccccccccc}
        \cmidrule[\heavyrulewidth]{1-3} \cmidrule[\heavyrulewidth]{7-9}
        Parameter & Coverage & Length & & & & $\chi(u)$ & Coverage &  Length \\
        \cmidrule{1-3} \cmidrule{7-9}
        $\alpha$ & $0.80$ & $3.98$ & & & & $\chi(0.80)$ & $0.91$ & $0.07$ \\
        \addlinespace[1.6mm]
        $\xi$ & $0.89$ & $0.52$ & & & & $\chi(0.95)$ & $0.92$ & $0.09$ \\
        \cmidrule[\heavyrulewidth]{1-3}
        & & & & & & $\chi(0.99)$ & $0.92$ & $0.10$ \\
        \cmidrule[\heavyrulewidth]{7-9} 
    \end{tabular}
    \label{tab:unc_wadsfixed}
\end{table}


\subsubsection*{Comparison with censored maximum likelihood estimation}

\begin{figure}[t!]
    \centering
    \begin{subfigure}[b]{0.49\textwidth}
        \includegraphics[width=\textwidth]{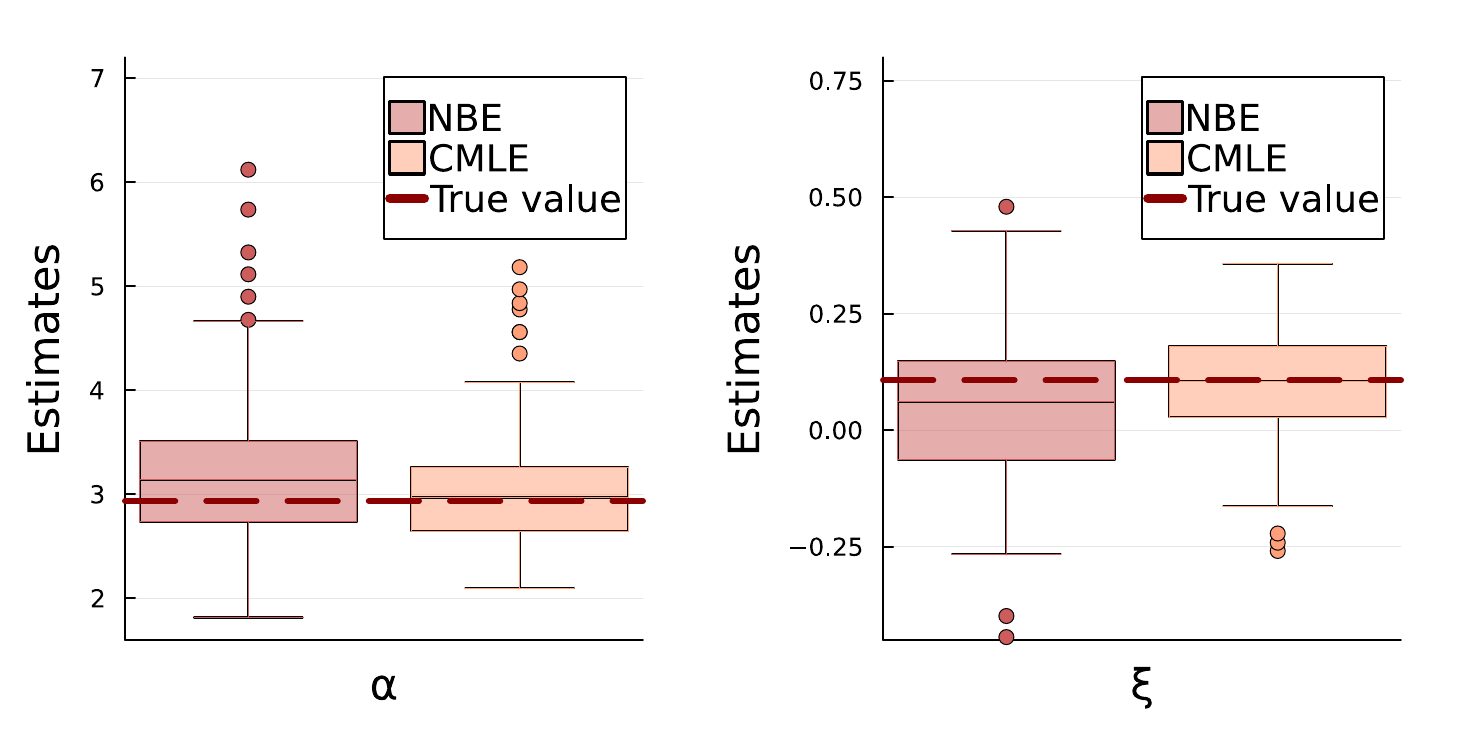}
        \caption{}
        \label{subfig:box1_wadsfixed}
    \end{subfigure}
    \hfill
    \begin{subfigure}[b]{0.49\textwidth}
        \includegraphics[width=\textwidth]{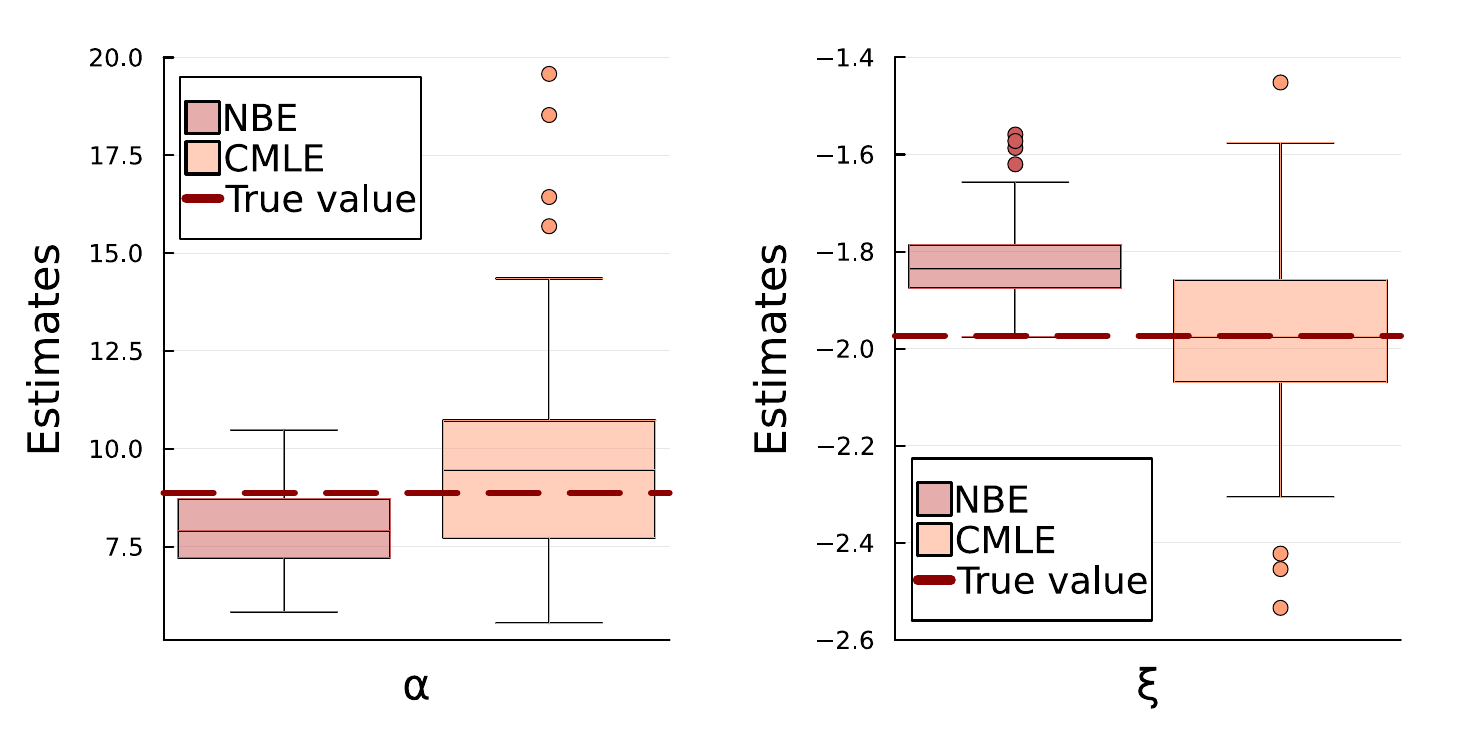}
        \caption{}
        \label{subfig:box2_wadsfixed}
    \end{subfigure}
    \hfill
    \begin{subfigure}[b]{0.49\textwidth}
        \includegraphics[width=\textwidth]{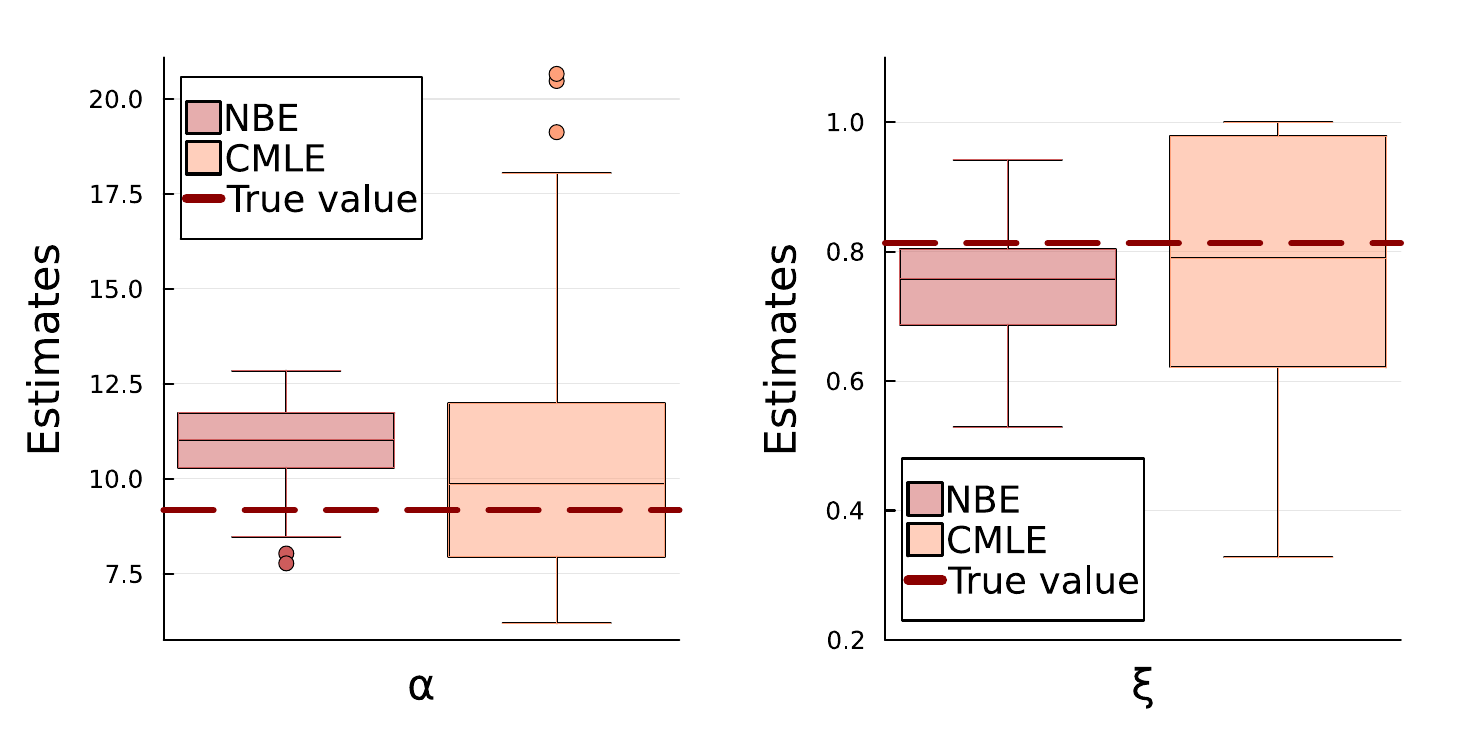}
        \caption{}
        \label{subfig:box3_wadsfixed}
    \end{subfigure}
    \hfill
    \begin{subfigure}[b]{0.49\textwidth}
        \includegraphics[width=\textwidth]{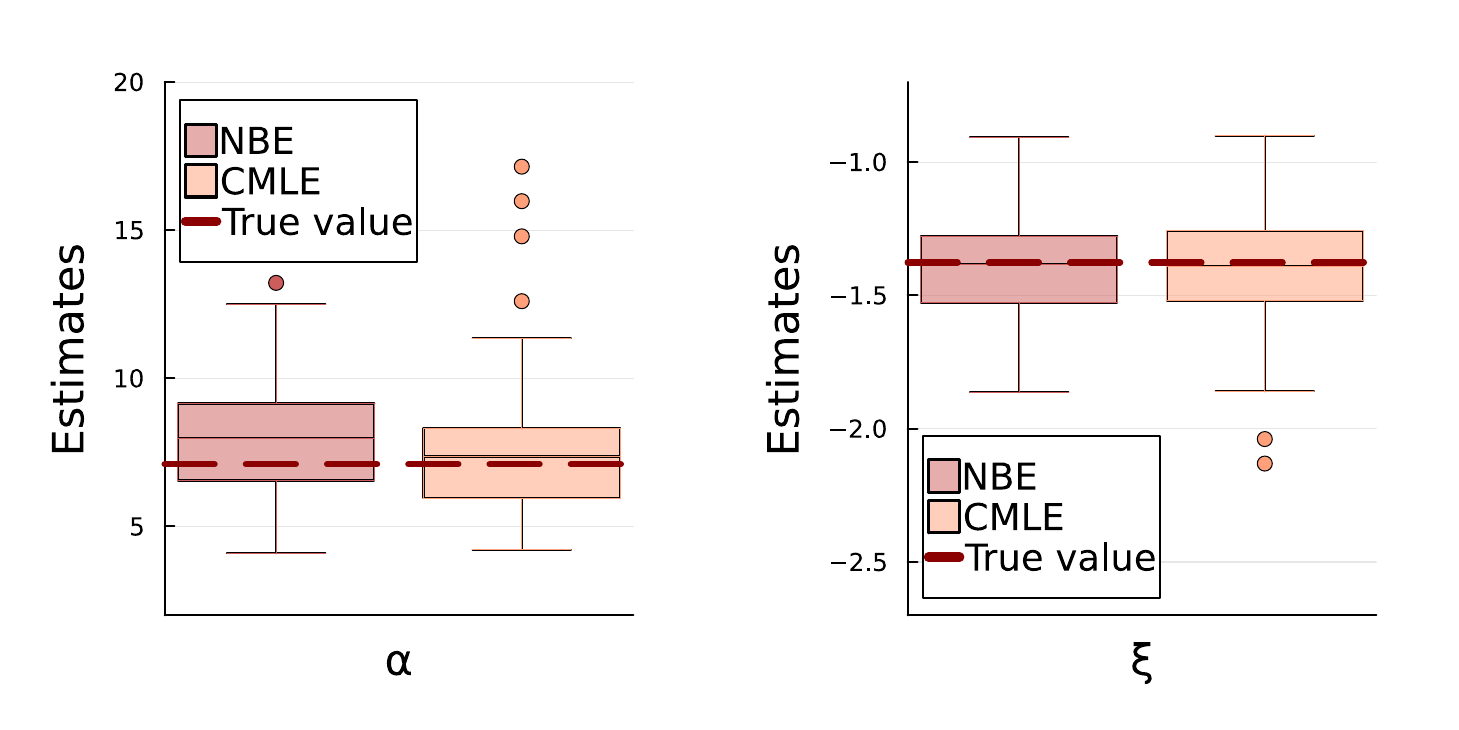}
        \caption{}
        \label{subfig:box4_wadsfixed}
    \end{subfigure}
    \hfill
    \begin{subfigure}[b]{0.49\textwidth}
        \includegraphics[width=\textwidth]{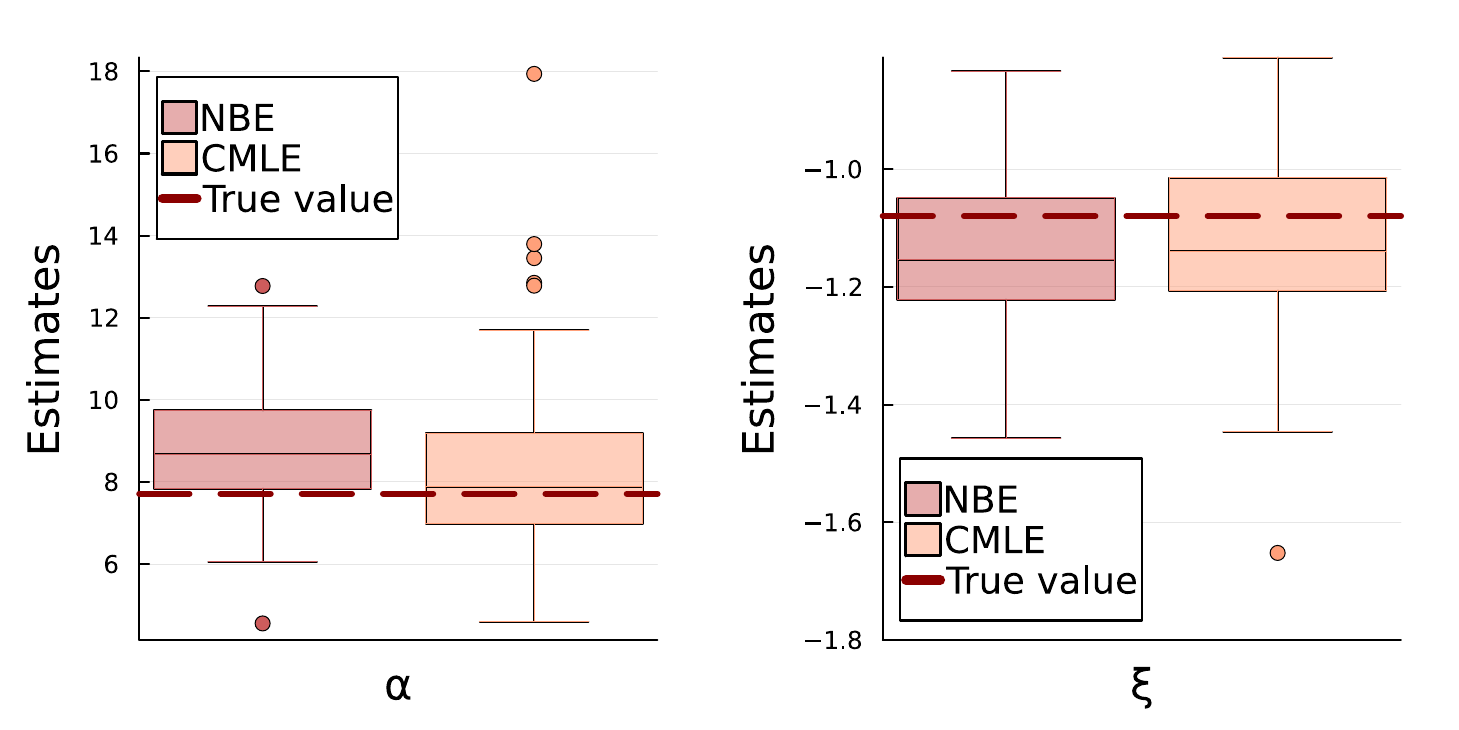}
        \caption{}
        \label{subfig:box5_wadsfixed}
    \end{subfigure}
    \caption{Comparison between parameter estimates $\hat{\bm\theta}=(\hat \alpha, \hat \xi)'$ given by CMLE (orange) and by NBE (red) for $100$ samples with $n=1000.$ The true parameters are given by the red line. (a) $\bm \theta = (2.94, 0.11)',$ (b) $\bm \theta = (8.87, -1.97)',$ (c) $\bm \theta = (9.17, 0.81)',$ (d) $\bm \theta = (7.10, -1.38)'$ and (e) $\bm \theta = (7.71, -1.08)'.$}
    \label{fig:mlevsnbe_wadsfixed}
\end{figure}

We compare the estimations obtained by the NBE and by the MLE for the five parameter vectors $\bm \theta = (\alpha, \xi)'$ considered in Section~\ref{subsec:parestimationsim} with now fixed $\tau = 0.8.$ Likewise before, each data set is simulated $100$ times and has a sample size of $n=1000.$ The results are shown in Figure~\ref{fig:mlevsnbe_wadsfixed}; these are fairly similar to those obtained when $n$ and $\tau$ are assumed unknown, and given in the main paper. This is also the configuration for which censored MLE is faster; for instance, on average, the CMLE took $159.298$ seconds, while the NBE was 731 times faster with an average time of 0.218 seconds.

\subsubsection{Variable sample size and fixed censoring level} \label{supsecsec:fixedvarwads}

\begin{figure}[t!]
    \centering
    \includegraphics[width=\textwidth]{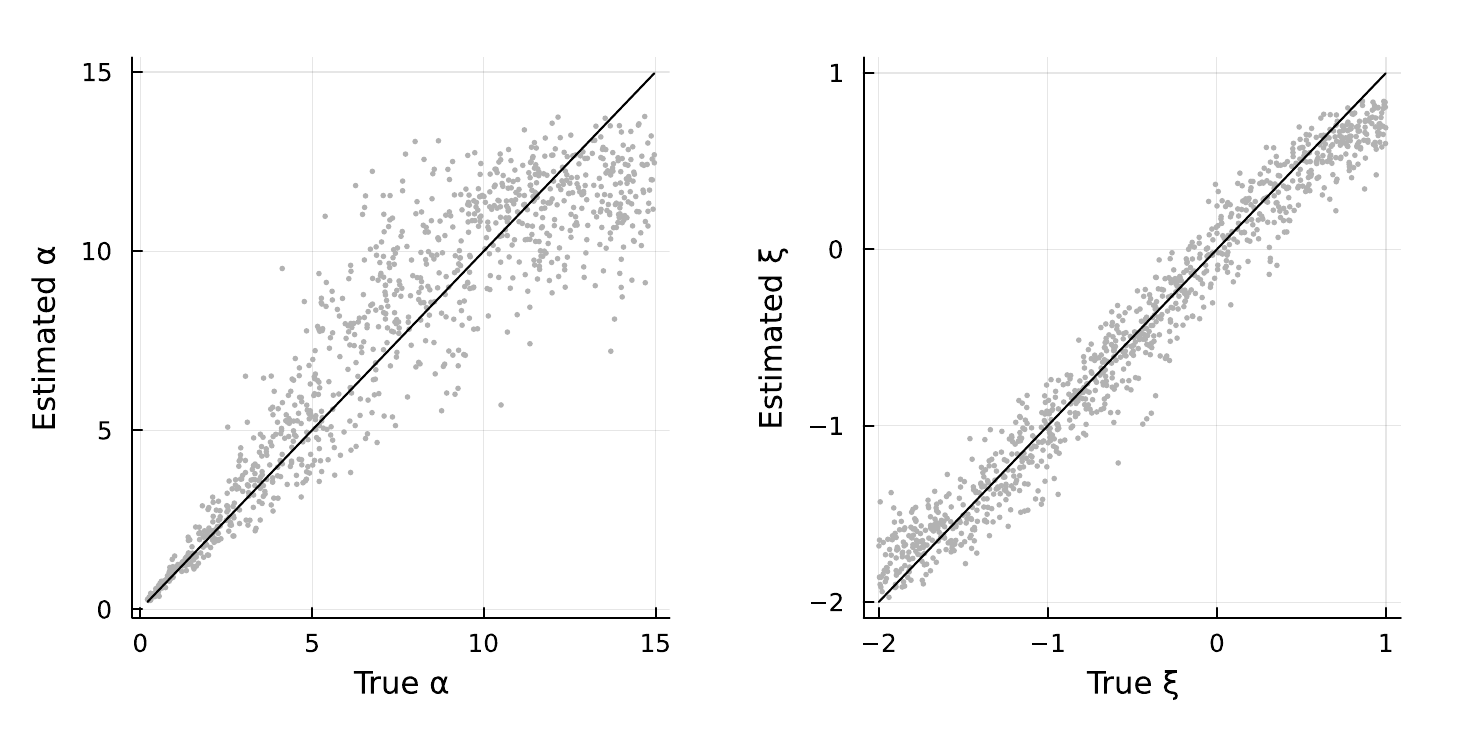}
    \caption{Assessment of the NBE for Model W with parameters $\bm \theta=(\alpha,\xi)'$ for a sample size of $n=1000$ and fixed censoring level $\tau = 0.8.$}
    \label{fig:assessment_wadsfixedvarn}
\end{figure}

We now assume the sample size is unknown but we keep the censoring level fixed at $\tau = 0.8.$ The performance of the NBE is given in Figure~\ref{fig:assessment_wadsfixedvarn}, where a similar behaviour to the results obtained either when $n$ is assumed fixed or when $\tau$ is also assumed unknown. The coverage probabilities of the $95\%$ uncertainty intervals obtained via (non-parametric) bootstrap, shown in Table~\ref{tab:unc_wadsfixedvarn}, are now slightly lower than the ones from the case when $n$ is assumed fixed at $1000.$ The coverage probabilities of $95\%$ uncertainty intervals, and their average length, for $\chi(u)$ at levels $u=\{0.80, 0.95, 0.99\}$ are shown on the right of Table~\ref{tab:unc_wadsfixedvarn}, and are similar in magnitude to the corresponding results presented in the main paper.

\begin{table}[!t]
    \centering
    \captionsetup{width=\textwidth}
    \caption{Coverage probability and average length of the $95\%$ uncertainty intervals for the parameters (left) and for $\chi(u)$ at levels $u=\{0.80, 0.95, 0.99\}$ (right)  obtained via a non-parametric bootstrap procedure averaged over $1000$ models fitted using a NBE (rounded to 2 decimal places).}
    \begin{tabular}{ccccccccc}
        \cmidrule[\heavyrulewidth]{1-3} \cmidrule[\heavyrulewidth]{7-9}
        Parameter & Coverage & Length & & & & $\chi(u)$ & Coverage &  Length \\
        \cmidrule{1-3} \cmidrule{7-9}
        $\alpha$ & $0.73$ & $3.53$ & & & & $\chi(0.80)$ & $0.90$ & $0.06$ \\
        \addlinespace[1.6mm]
        $\xi$ & $0.80$ & $0.46$ & & & & $\chi(0.95)$ & $0.88$ & $0.08$ \\
        \cmidrule[\heavyrulewidth]{1-3}
        & & & & & & $\chi(0.99)$ & $0.85$ & $0.09$ \\
        \cmidrule[\heavyrulewidth]{7-9} 
    \end{tabular}
    \label{tab:unc_wadsfixedvarn}
\end{table}


\subsubsection*{Comparison with censored maximum likelihood estimation}

\begin{figure}[t!]
    \centering
    \begin{subfigure}[b]{0.49\textwidth}
        \includegraphics[width=\textwidth]{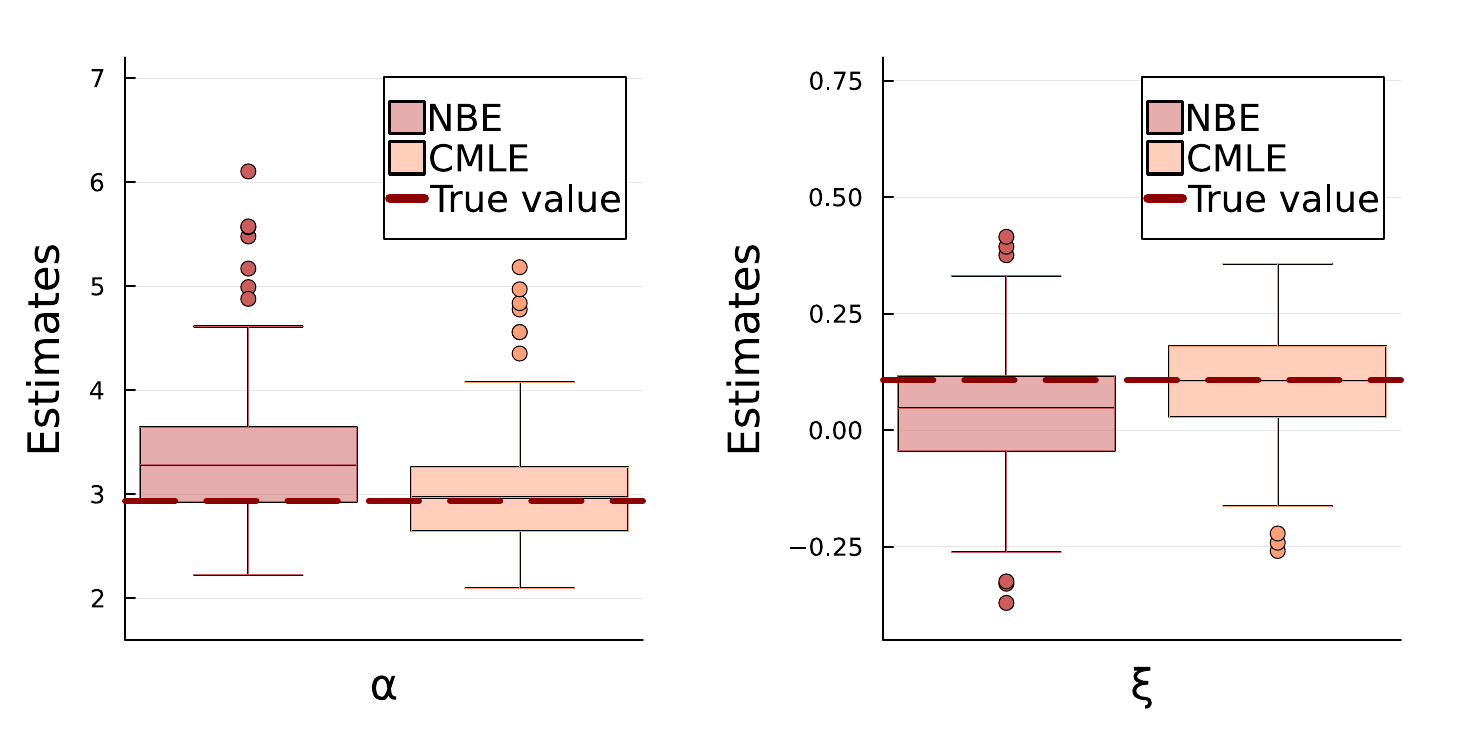}
        \caption{}
        \label{subfig:box1_wadsfixedvarn}
    \end{subfigure}
    \hfill
    \begin{subfigure}[b]{0.49\textwidth}
        \includegraphics[width=\textwidth]{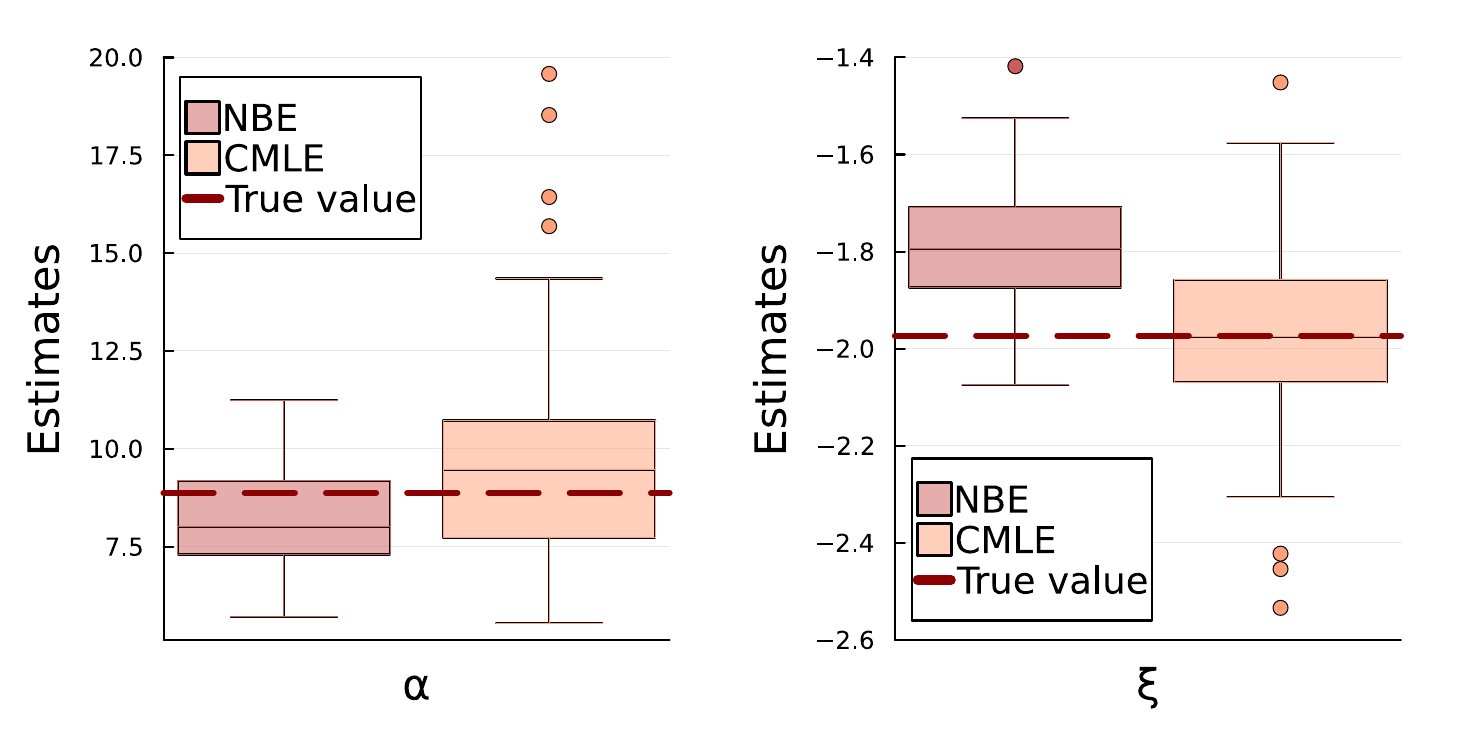}
        \caption{}
        \label{subfig:box2_wadsfixedvarn}
    \end{subfigure}
    \hfill
    \begin{subfigure}[b]{0.49\textwidth}
        \includegraphics[width=\textwidth]{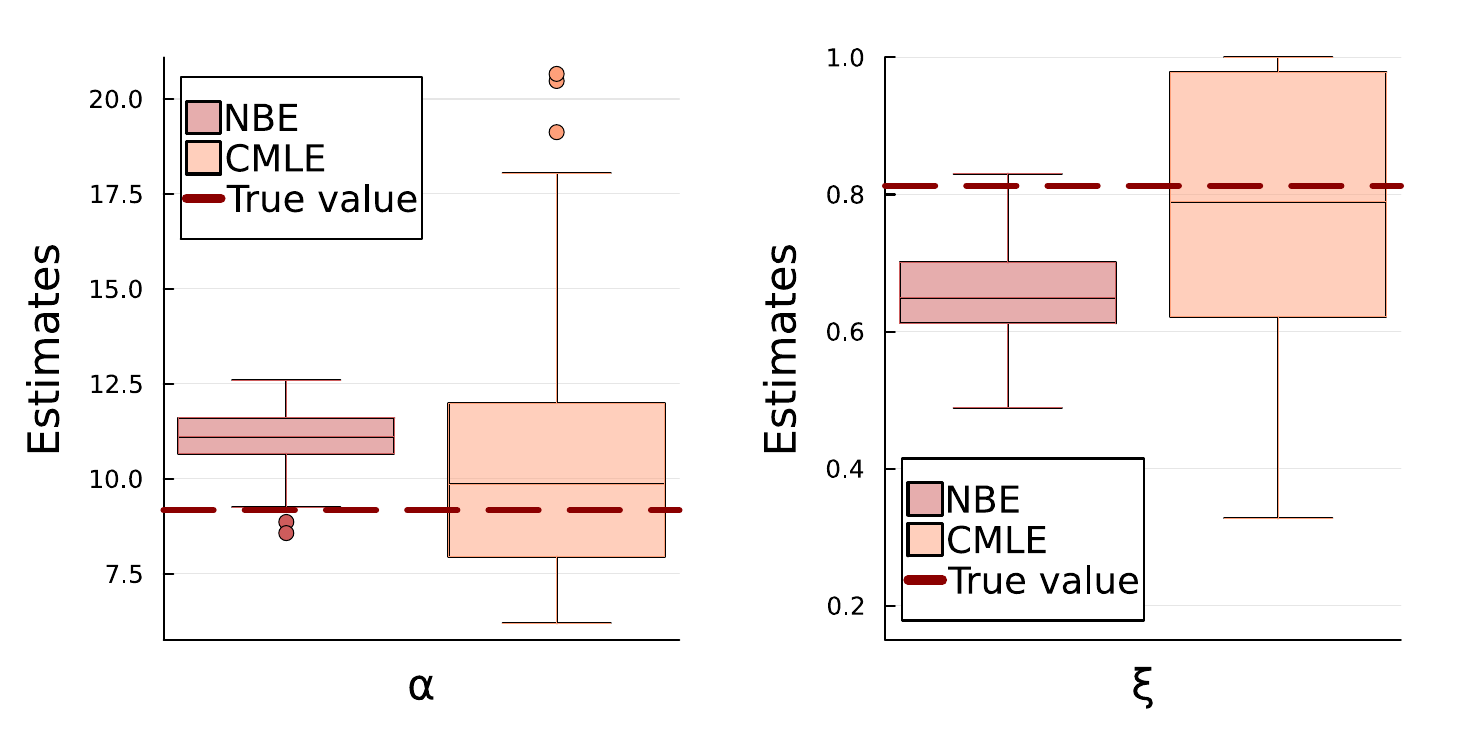}
        \caption{}
        \label{subfig:box3_wadsfixedvarn}
    \end{subfigure}
    \hfill
    \begin{subfigure}[b]{0.49\textwidth}
        \includegraphics[width=\textwidth]{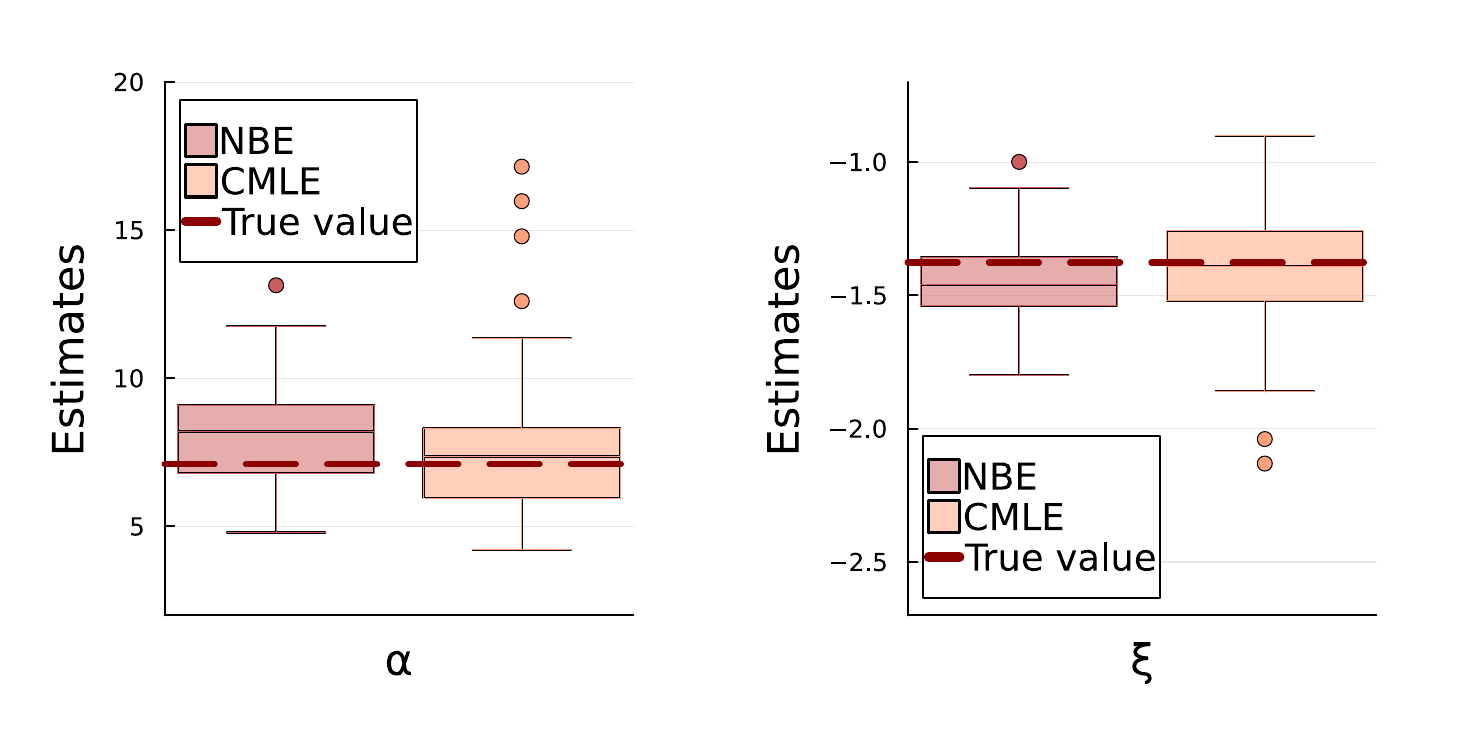}
        \caption{}
        \label{subfig:box4_wadsfixedvarn}
    \end{subfigure}
    \hfill
    \begin{subfigure}[b]{0.49\textwidth}
        \includegraphics[width=\textwidth]{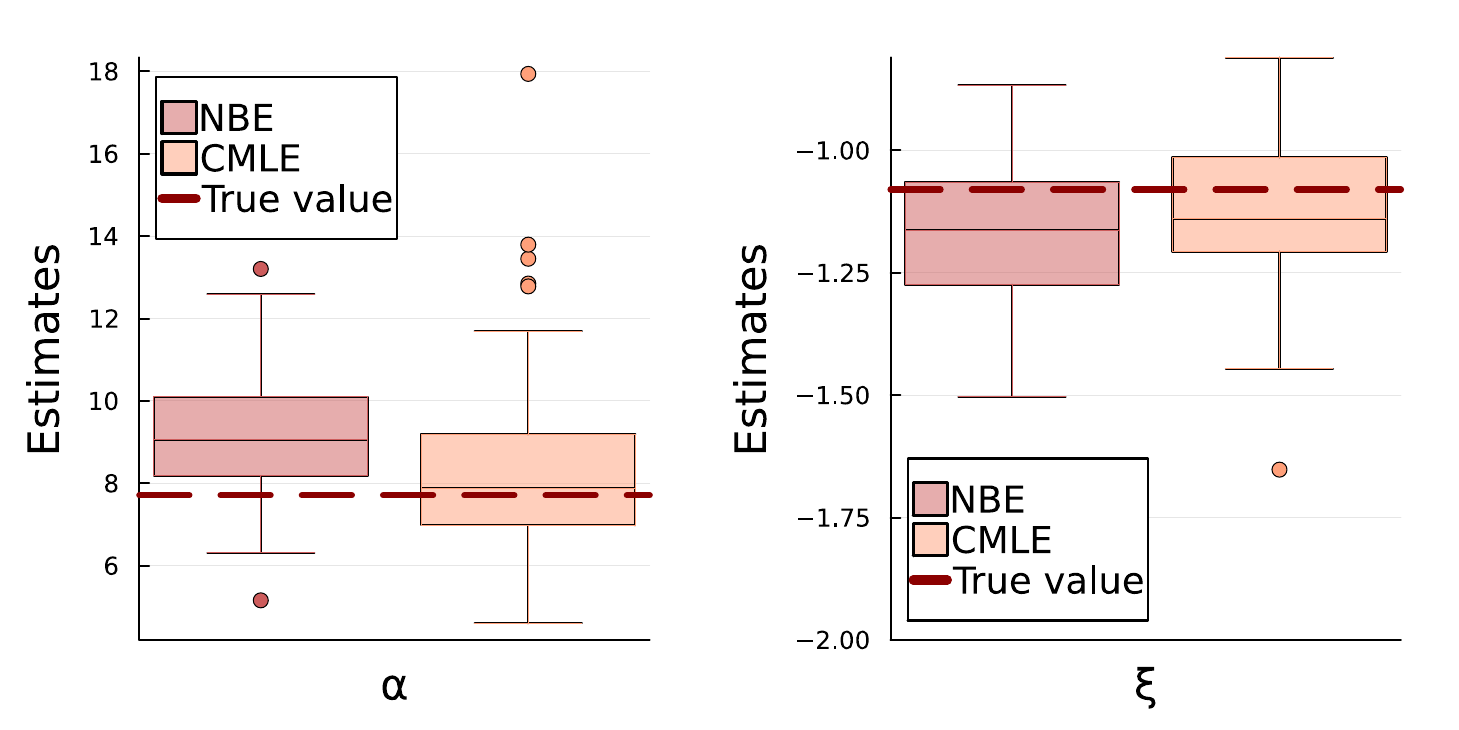}
        \caption{}
        \label{subfig:box5_wadsfixedvarn}
    \end{subfigure}
    \caption{Comparison between parameter estimates $\hat{\bm\theta}=(\hat \alpha, \hat \xi)'$ given by CMLE (orange) and by NBE (red) for $100$ samples with $n=1000.$ The true parameters are given by the red line. (a) $\bm \theta = (2.94, 0.11)',$ (b) $\bm \theta = (8.87, -1.97)',$ (c) $\bm \theta = (9.17, 0.81)',$ (d) $\bm \theta = (7.10, -1.38)'$ and (e) $\bm \theta = (7.71, -1.08)'.$}
    \label{fig:mlevsnbe_wadsfixedvarn}
\end{figure}

The same five parameter vectors $\bm \theta = (\alpha, \xi)'$ considered for the cases where $n$ is fixed at 1000 and the one presented in Section~\ref{subsec:parestimationsim} are used to compare the NBE and CMLE estimates. Similarly to the previous case, we fix $\tau = 0.8,$ and each data set with $n=1000$ is simulated 100 times. No evident differences to the estimates obtained when $n$ is assumed fixed and when $n$ and $\tau$ are assumed unknown are visible from the results in Figure~\ref{fig:mlevsnbe_wadsfixedvarn}. For this case, the average time to get a NBE is of 0.470 seconds, which is about 339 faster than CMLE on average.

\subsubsection{General conclusions}

The results with fixed censoring level $(\tau=0.8)$ with fixed $(n=1000)$ and variable sample size exhibit similar findings. In the case where both $\tau$ and $n$ are fixed, however, the obtained bootstrap-based intervals have better coverage. When comparing the estimates given by the NBEs with the ones obtained by classical inference techniques, fixing one or both $n$ and $\tau$ did not improve the performance of the estimators.

\clearpage

\subsection{Model HW} \label{supsec:hw}

\begin{figure}[t!]
    \centering
    \includegraphics[width=\textwidth]{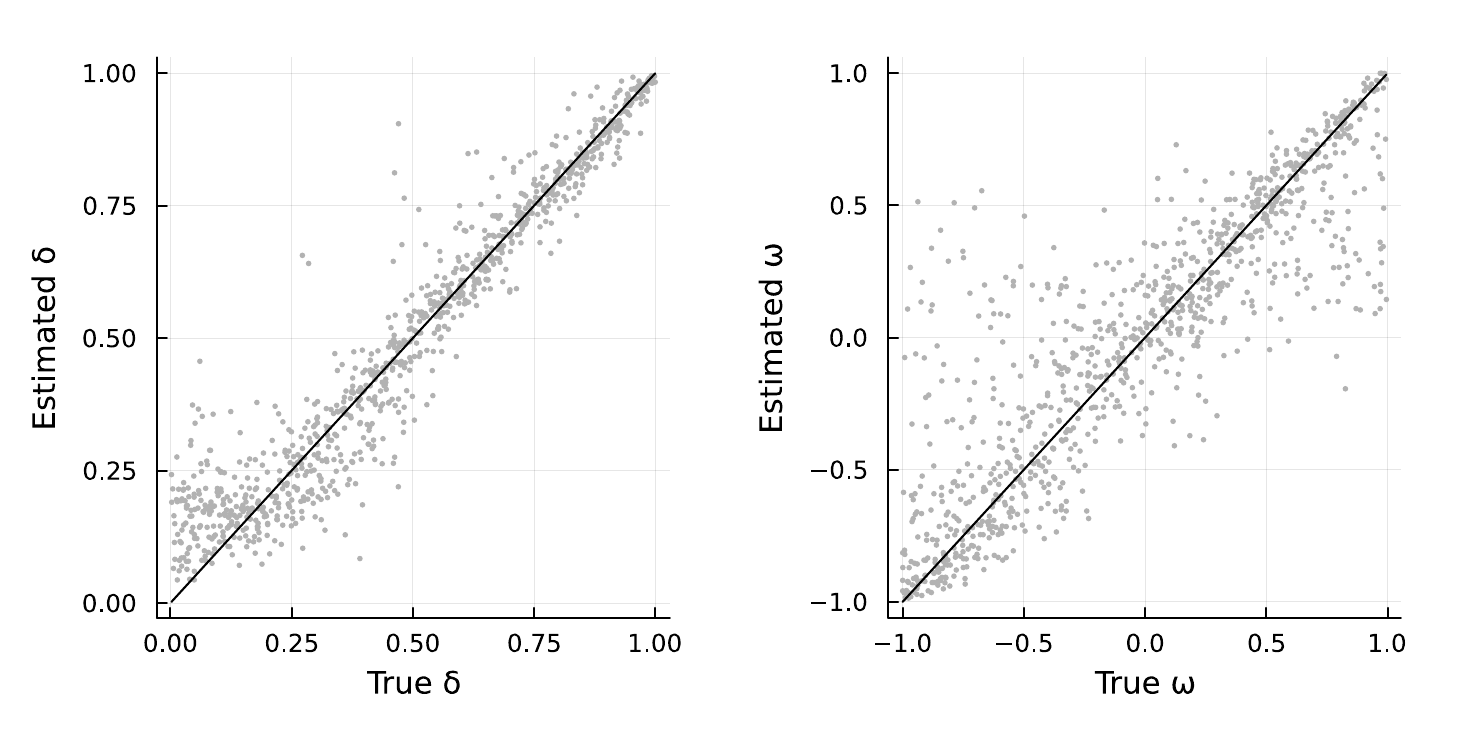}
    \caption{Assessment of the NBE for Model HW, where $\bm V$ follows a bivariate Gaussian copula, with parameters $\bm \theta=(\delta,\omega)'$ for a sample size of $n=1000.$}
    \label{fig:assessment_hwGauss}
\end{figure}

We assess the performance of the NBE for Model HW. The results are shown in Figure~\ref{fig:assessment_hwGauss} and Table~\ref{tab:unc_hwGauss}. Similarly to Model W, there is some variability in the estimates, in particular for lower values of $\delta$ and $\omega.$ The coverage probabilities of $95\%$ uncertainty intervals, and their average length, for $\chi(u)$ at levels $u=\{0.80, 0.95, 0.99\},$ shown on the right of Table~\ref{tab:unc_wadsfixedvarn}, indicate that even with biased results, the NBE is able to characterise the extremal dependence at high levels of $u.$ We note that, similarly to the study involving Model W given in the main paper, the coverage probabilities for $\chi(u)$ are achieved with new data sets for 1000 parameter configurations, each generated with a fixed censoring level $\tau = 0.8.$

\begin{table}[!t]
    \centering
    \captionsetup{width=\textwidth}
    \caption{Coverage probability and average length of the $95\%$ uncertainty intervals for the parameters (left) and for $\chi(u)$ at levels $u=\{0.80, 0.95, 0.99\}$ (right)  obtained via a non-parametric bootstrap procedure averaged over $1000$ models fitted using a NBE (rounded to 2 decimal places).}
    \begin{tabular}{ccccccccc}
        \cmidrule[\heavyrulewidth]{1-3} \cmidrule[\heavyrulewidth]{7-9}
        Parameter & Coverage & Length & & & & $\chi(u)$ & Coverage &  Length \\
        \cmidrule{1-3} \cmidrule{7-9}
        $\delta$ & $0.71$ & $0.14$ & & & & $\chi(0.80)$ & $0.90$ & $0.08$ \\
        \addlinespace[1.6mm]
        $\omega$ & $0.75$ & $0.48$ & & & & $\chi(0.95)$ & $0.90$ & $0.09$ \\
        \cmidrule[\heavyrulewidth]{1-3}
        & & & & & & $\chi(0.99)$ & $0.90$ & $0.10$ \\
        \cmidrule[\heavyrulewidth]{7-9} 
    \end{tabular}
    \label{tab:unc_hwGauss}
\end{table}

\subsubsection*{Comparison with censored maximum likelihood estimation}

\begin{figure}[t!]
    \centering
    \begin{subfigure}[b]{0.49\textwidth}
        \includegraphics[width=\textwidth]{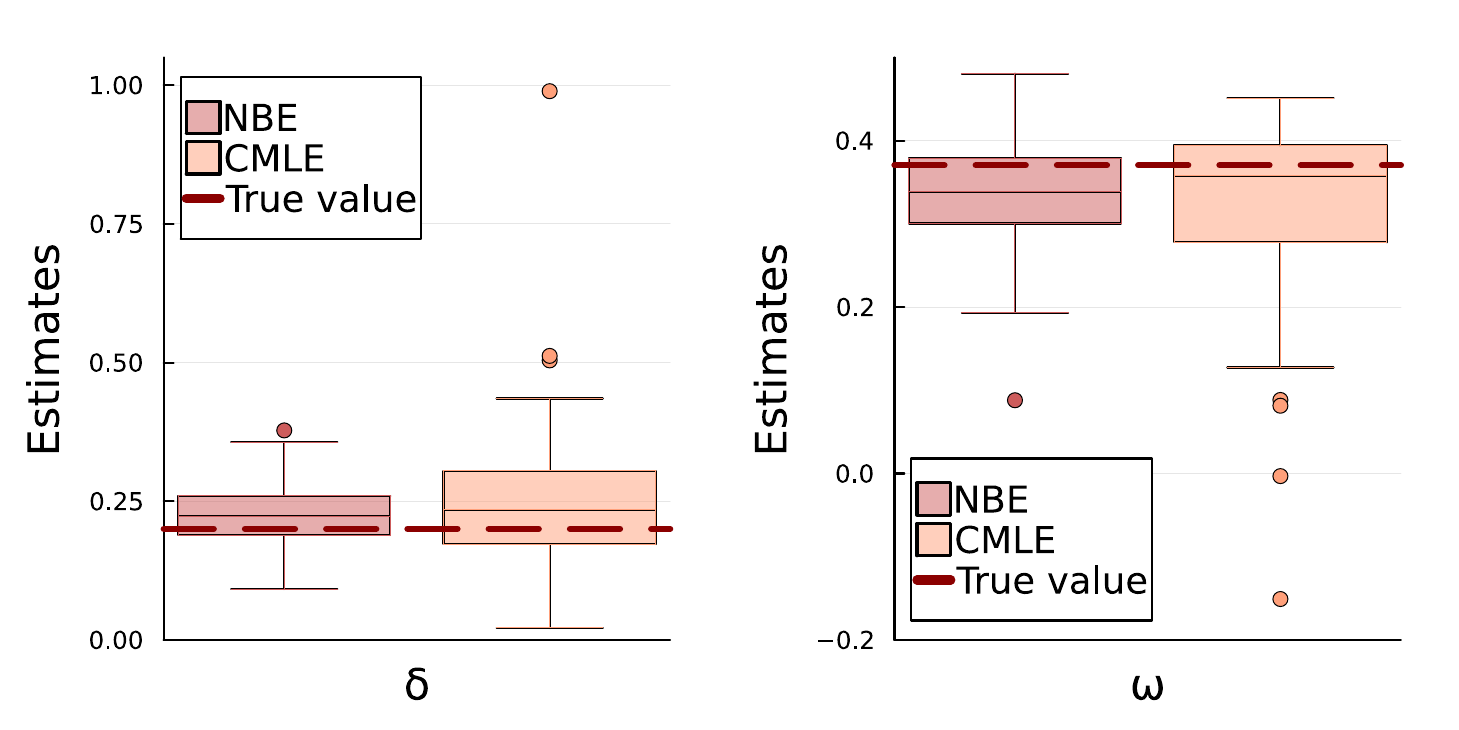}
        \caption{}
        \label{subfig:box1_hwGauss}
    \end{subfigure}
    \hfill
    \begin{subfigure}[b]{0.49\textwidth}
        \includegraphics[width=\textwidth]{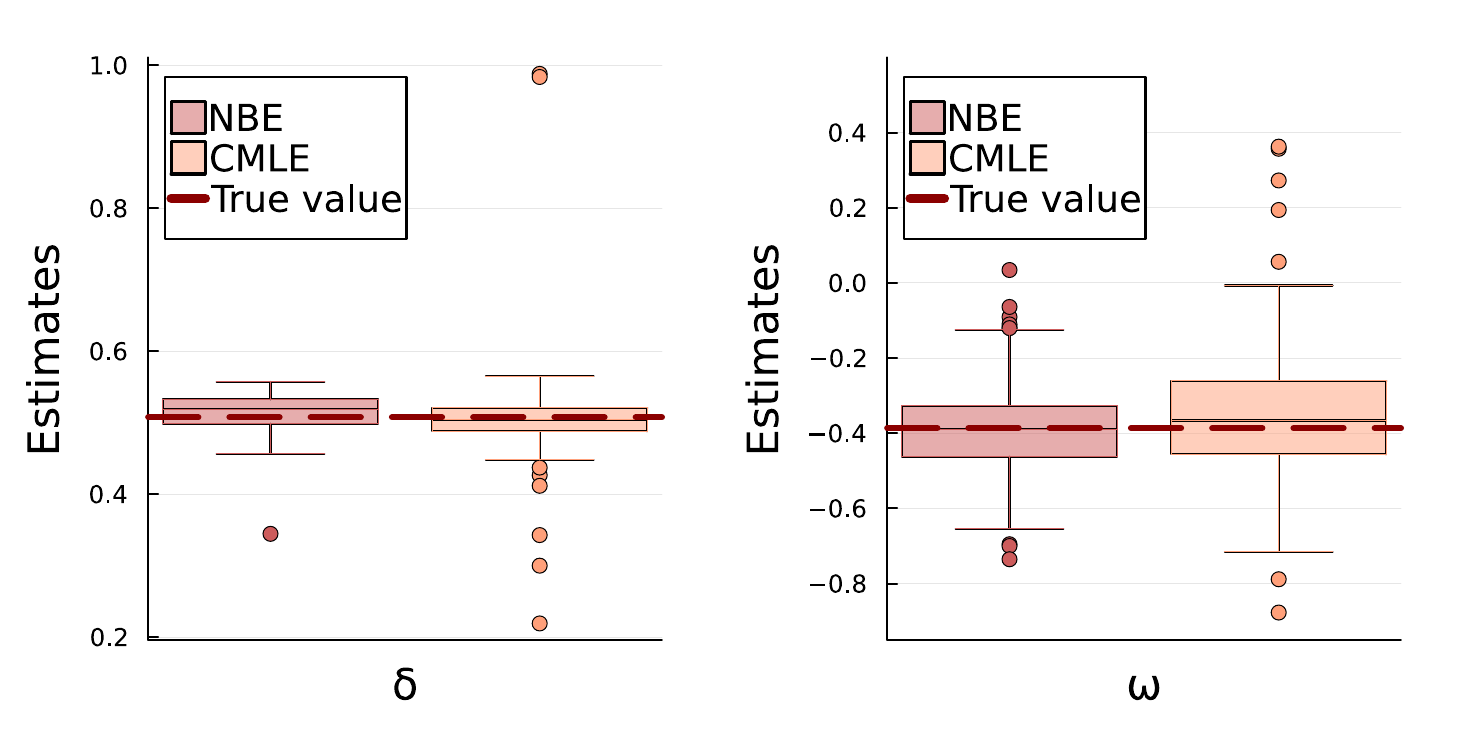}
        \caption{}
        \label{subfig:box2_hwGauss}
    \end{subfigure}
    \hfill
    \begin{subfigure}[b]{0.49\textwidth}
        \includegraphics[width=\textwidth]{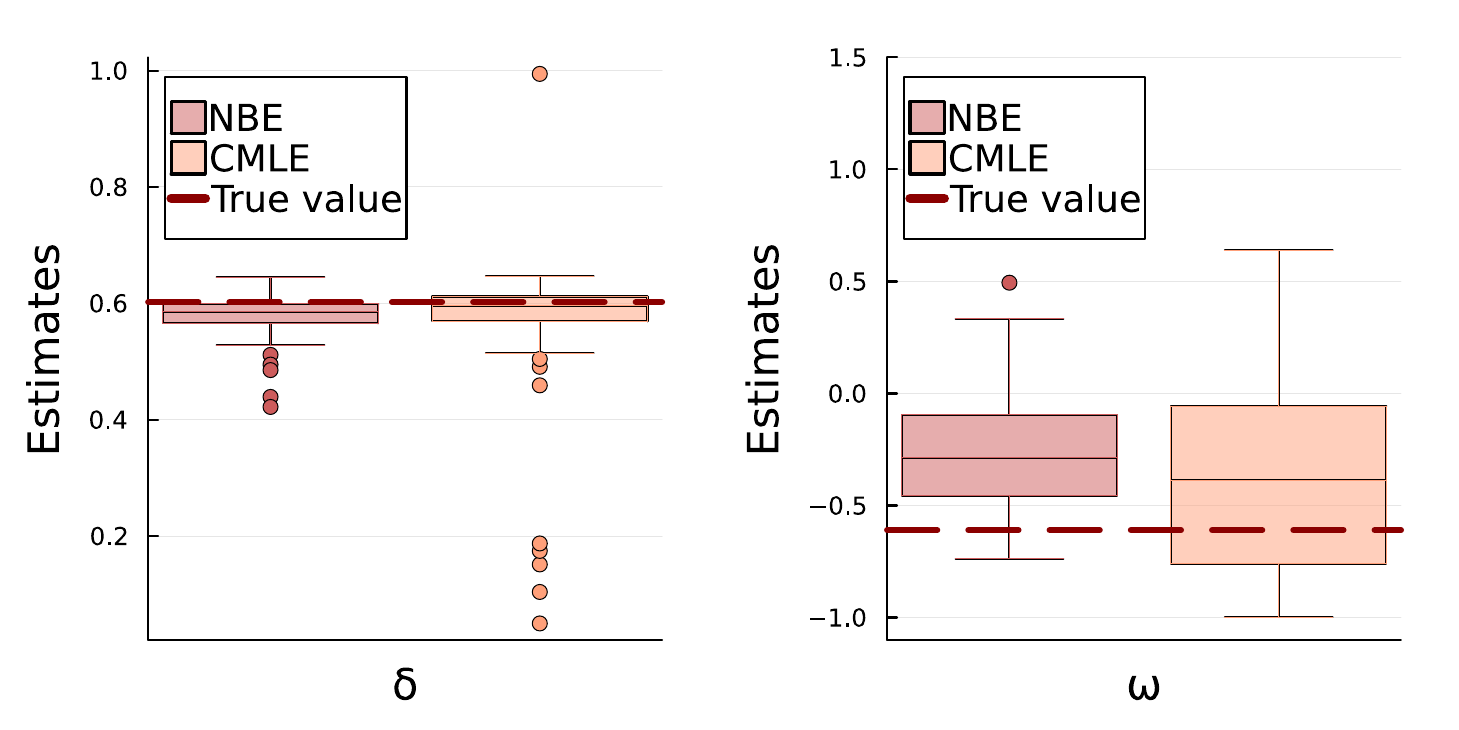}
        \caption{}
        \label{subfig:box3_hwGauss}
    \end{subfigure}
    \hfill
    \begin{subfigure}[b]{0.49\textwidth}
        \includegraphics[width=\textwidth]{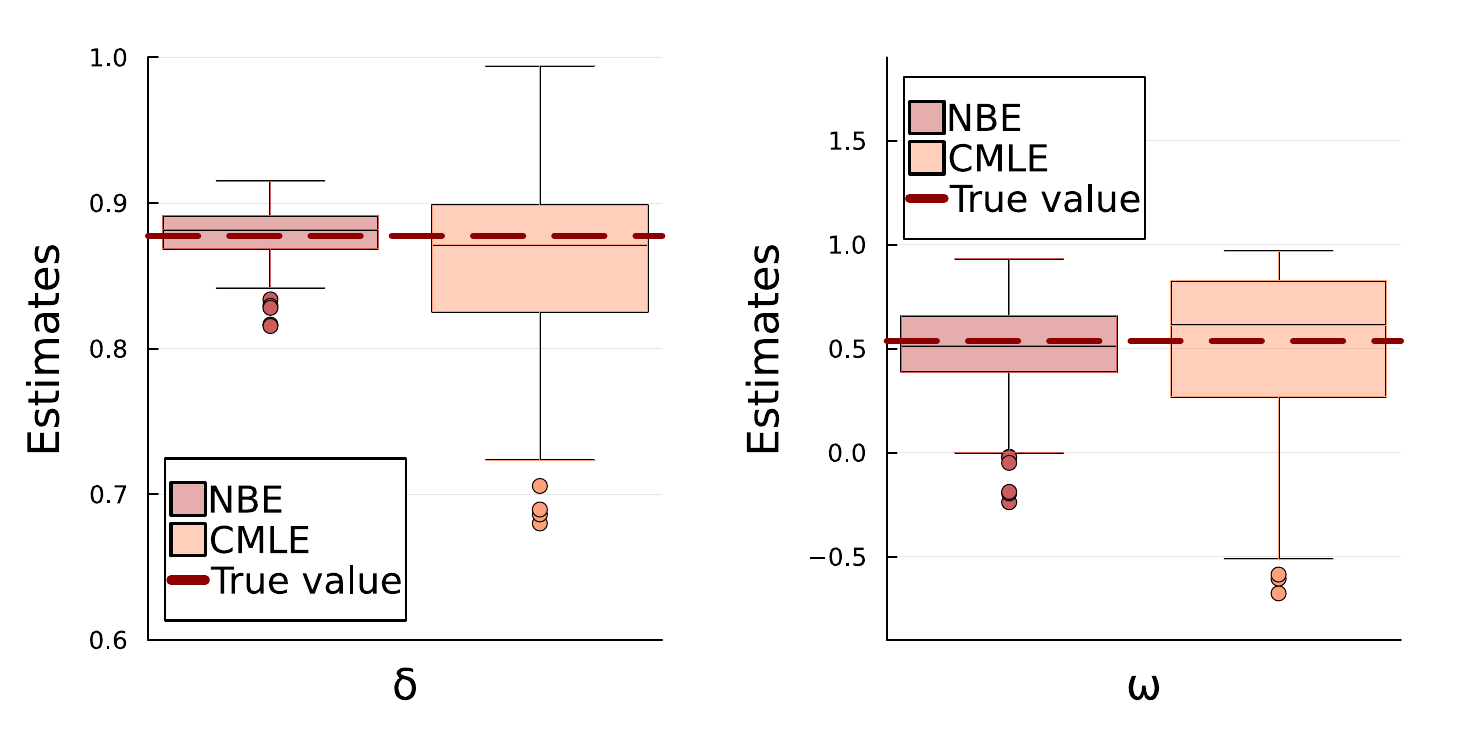}
        \caption{}
        \label{subfig:box4_hwGauss}
    \end{subfigure}
    \hfill
    \begin{subfigure}[b]{0.49\textwidth}
        \includegraphics[width=\textwidth]{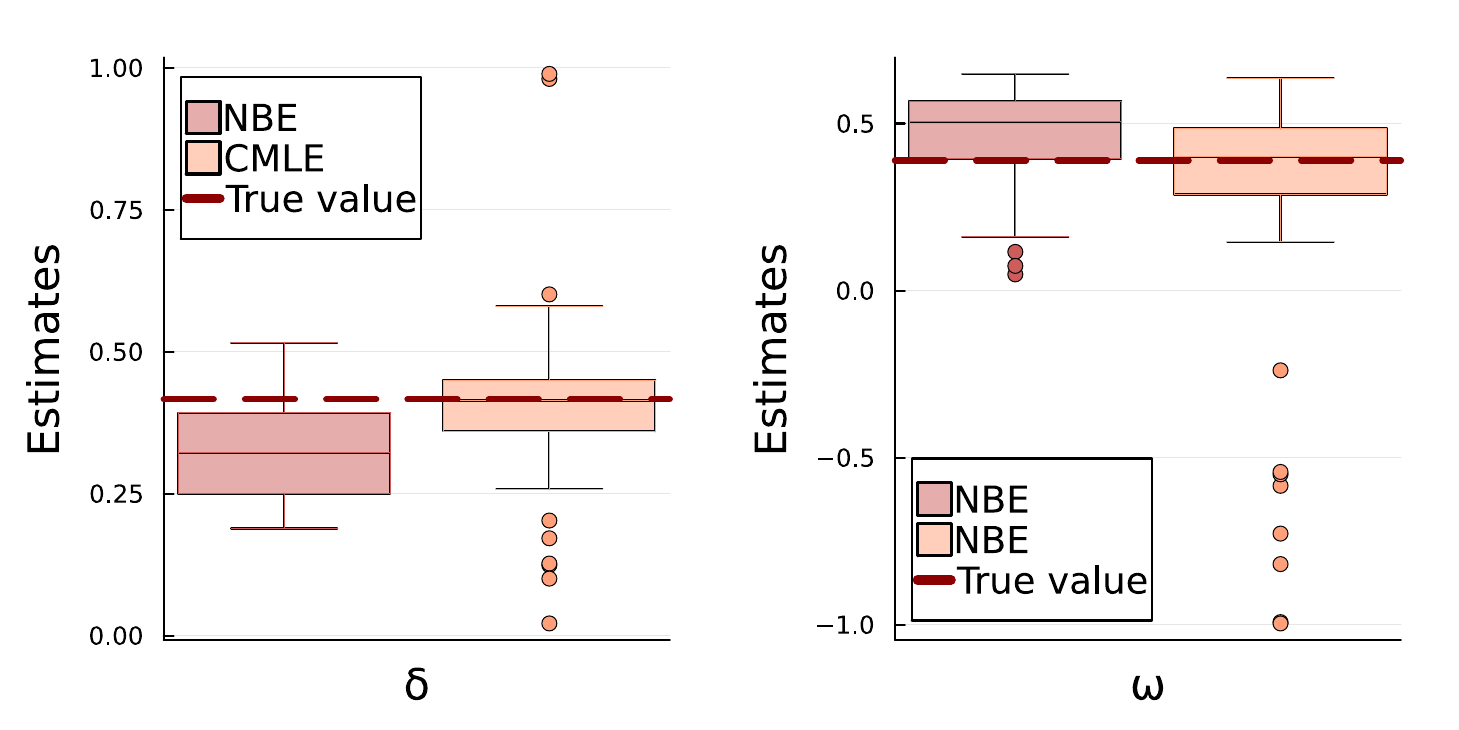}
        \caption{}
        \label{subfig:box5_hwGauss}
    \end{subfigure}
    \caption{Comparison between parameter estimates $\hat{\bm\theta}=(\hat \delta, \hat \omega)'$ given by CMLE (orange) and by NBE (red) for $100$ samples with $n=1000.$ The true parameters are given by the red line. (a) $\bm \theta = (0.20, 0.37)'$ with $\tau = 0.65,$ (b) $\bm \theta = (0.51, -0.39)'$ with $\tau = 0.76,$ (c) $\bm \theta = (0.60, -0.61)'$ with $\tau = 0.95,$ (d) $\bm \theta = (0.88, 0.54)'$ with $\tau = 0.57$ and (e) $\bm \theta = (0.42, 0.39)'$ with $\tau = 0.91.$}
    \label{fig:mlevsnbe_hwGauss}
\end{figure}

Similarly to the previous cases, we generate five parameter vectors from the priors considered in the main paper, along with the corresponding data sets of size $n=1000,$ and simulate each data set $100$ times. The comparison between the NBE and CMLE is shown in  Figure~\ref{fig:mlevsnbe_hwGauss}; the estimates given by the NBE are quite good, particularly for lower censoring levels.  As for computational times, the CMLE took 688.837 seconds on average, while the NBE was 2\,542 times faster with an average of 0.271 seconds.

\clearpage

\subsection{Model E1}  \label{supsec:eng1}

\begin{figure}[t!]
    \centering
    \includegraphics[width=\textwidth]{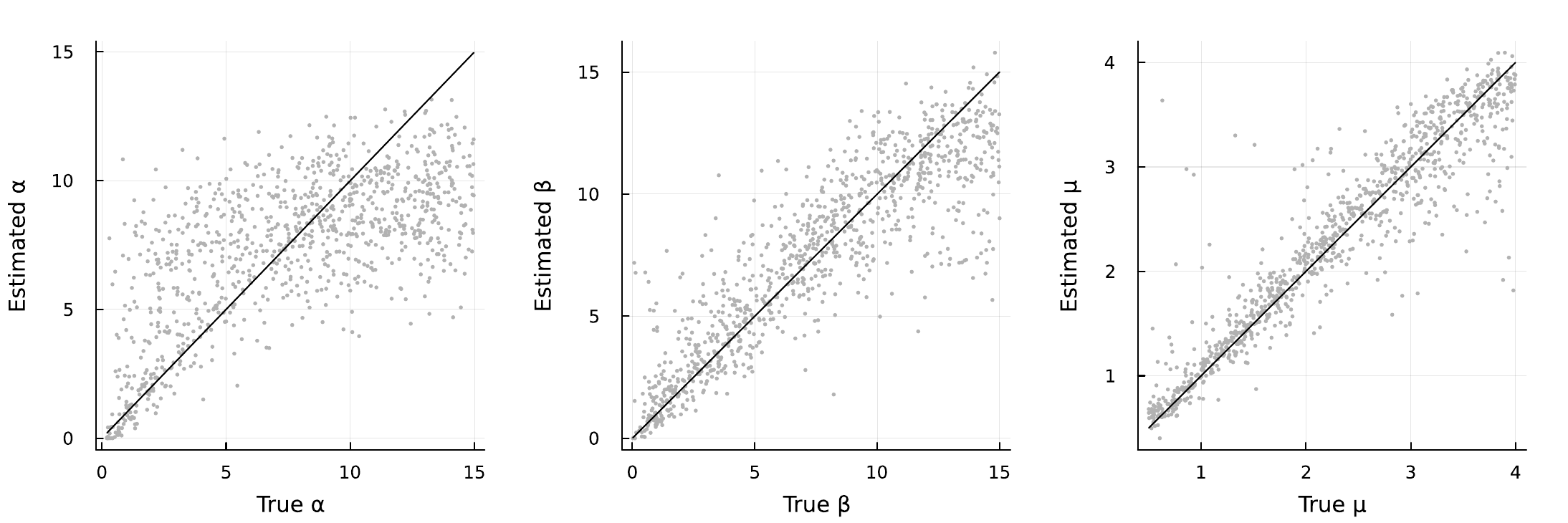}
    \caption{Assessment of the NBE for Model E1 with parameters $\bm \theta=(\alpha, \beta,\mu)'$ for a sample size of $n=1000.$}
    \label{fig:assessment_eng1}
\end{figure}

Figure~\ref{fig:assessment_eng1} and Table~\ref{tab:unc_eng1} show the performance of the NBE for Model E1. As can be seen, the parameters $\alpha$ and $\beta$ have the lowest coverage probability and highest average length of their $95\%$ uncertainty intervals; this is in agreement with the variability shown when comparing the true values with their estimated values in Figure~\ref{fig:assessment_eng1}. As before, we compute the coverage probabilities of the $95\%$ confidence intervals for $\chi(u)$ at levels $u=\{0.80, 0.95, 0.99\}$ by considering new data sets for 1000 parameter configurations, each generated with a fixed censoring level $\tau = 0.8.$ The results, given on the right of Table~\ref{tab:unc_eng1}, indicate that the bias shown by the NBE does not seem to influence the estimation of $\chi(u).$ In particular, the true value is within the confidence intervals in more than $81\%$ of the time.  

\begin{table}[!t]
    \centering
    \captionsetup{width=\textwidth}
    \caption{Coverage probability and average length of the $95\%$ uncertainty intervals for the parameters (left) and for $\chi(u)$ at levels $u=\{0.80, 0.95, 0.99\}$ (right)  obtained via a non-parametric bootstrap procedure averaged over $1000$ models fitted using a NBE (rounded to 2 decimal places).}
    \begin{tabular}{ccccccccc}
        \cmidrule[\heavyrulewidth]{1-3} \cmidrule[\heavyrulewidth]{7-9}
        Parameter & Coverage & Length & & & & $\chi(u)$ & Coverage &  Length \\
        \cmidrule{1-3} \cmidrule{7-9}
        $\alpha$ & $0.44$ & $3.60$ & & & & $\chi(0.80)$ & $0.82$ & $0.11$ \\
        $\beta$ & $0.68$ & $3.36$ & & & & $\chi(0.95)$ & $0.82$ & $0.11$ \\
        $\mu$ & $0.77$ & $0.60$ & & & & $\chi(0.99)$ & $0.81$ & $0.12$ \\
        \cmidrule[\heavyrulewidth]{1-3} \cmidrule[\heavyrulewidth]{7-9} 
    \end{tabular}
    \label{tab:unc_eng1}
\end{table}

\newpage 
\subsubsection*{Comparison with censored maximum likelihood estimation}

We compare the estimations obtained by the NBE and by the CMLE for five parameter vectors $\bm \theta = (\alpha, \beta, \mu)'$ generated from the priors considered in the main paper. Each corresponding data set has $n=1000$ and is simulated $100$ times. Similarly to the other models considered, the NBE is more biased than the CMLE and, in some cases, can be more variable than the CMLE. Despite that, on average, the CMLE took 17 minutes, whereas the NBE took 0.159 seconds, meaning that the NBE is about $6\,565$ times faster.

\begin{sidewaysfigure}[t!]
    \centering
    \begin{subfigure}[b]{0.49\textwidth}
        \includegraphics[width=\textwidth]{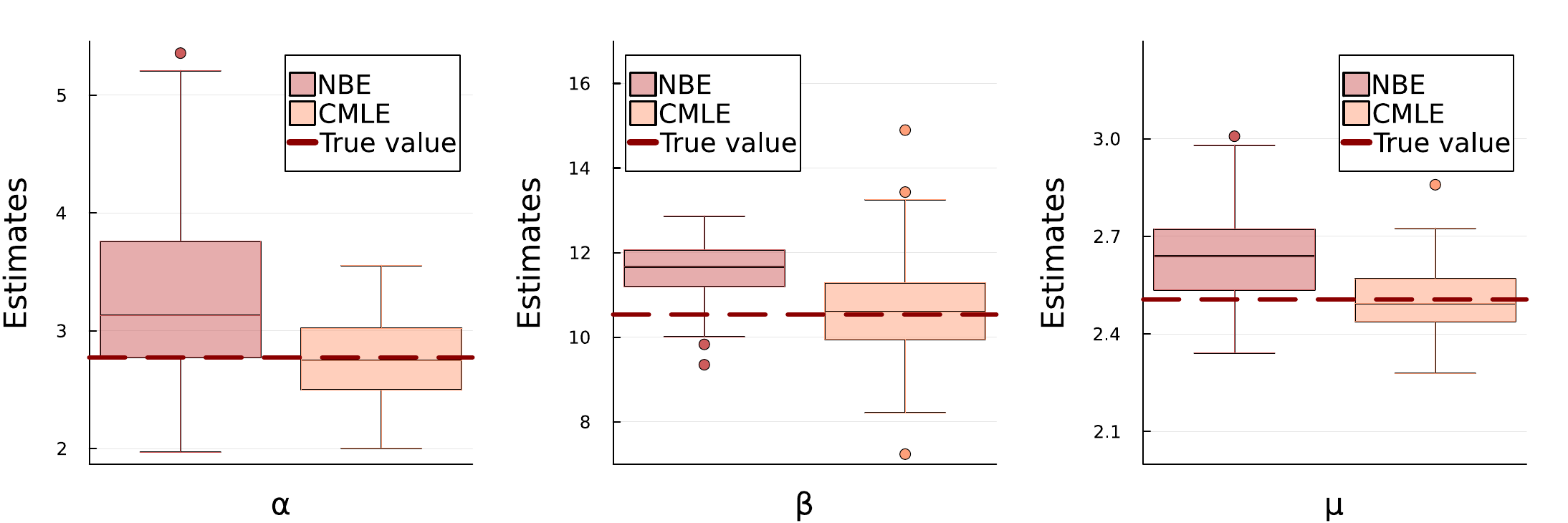}
        \caption{}
        \label{subfig:box1_eng1}
    \end{subfigure}
    \hfill
    \begin{subfigure}[b]{0.49\textwidth}
        \includegraphics[width=\textwidth]{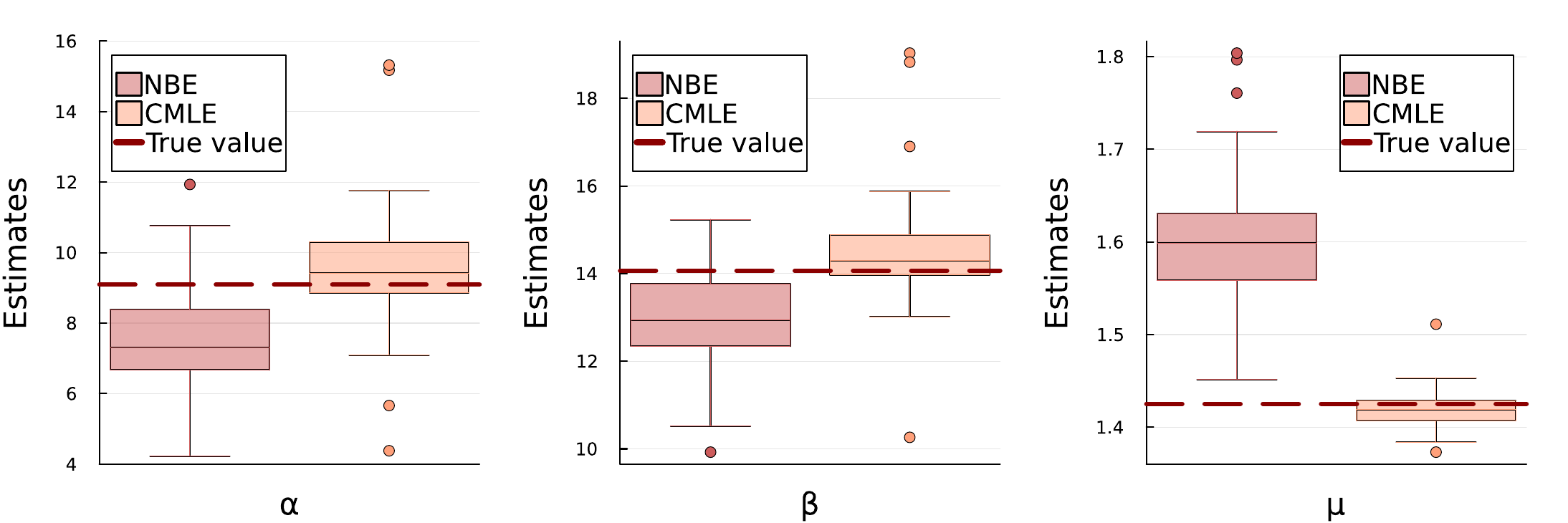}
        \caption{}
        \label{subfig:box2_eng1}
    \end{subfigure}
    \hfill
    \begin{subfigure}[b]{0.49\textwidth}
        \includegraphics[width=\textwidth]{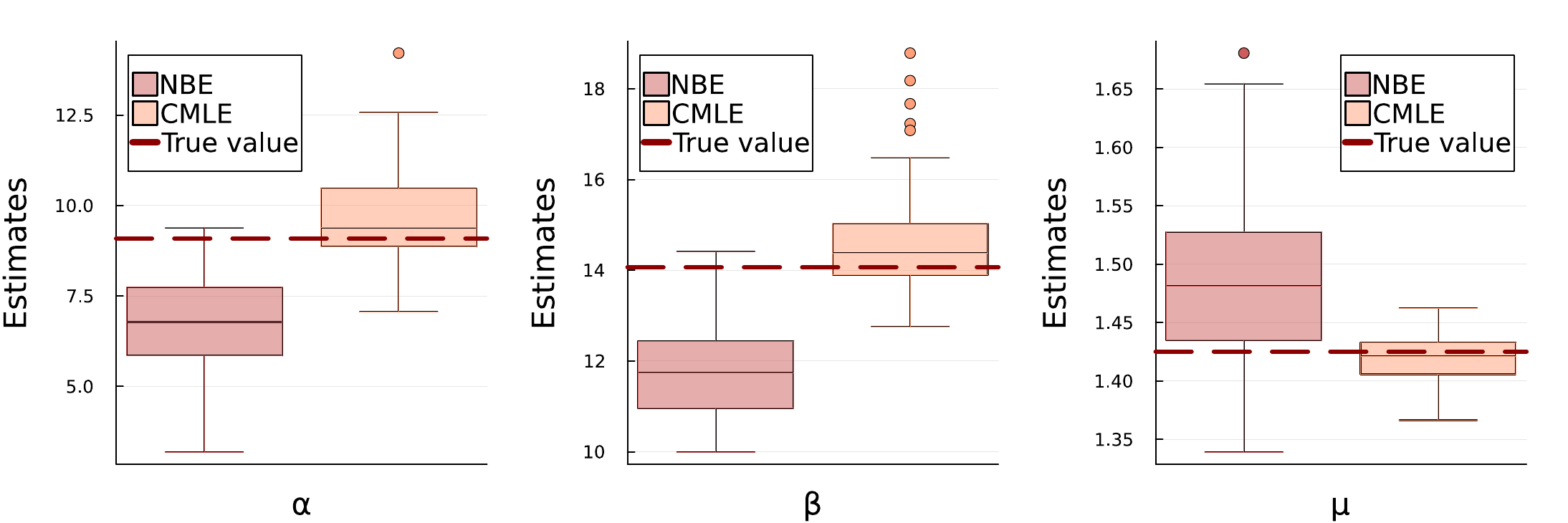}
        \caption{}
        \label{subfig:box3_eng1}
    \end{subfigure}
    \hfill
    \begin{subfigure}[b]{0.49\textwidth}
        \includegraphics[width=\textwidth]{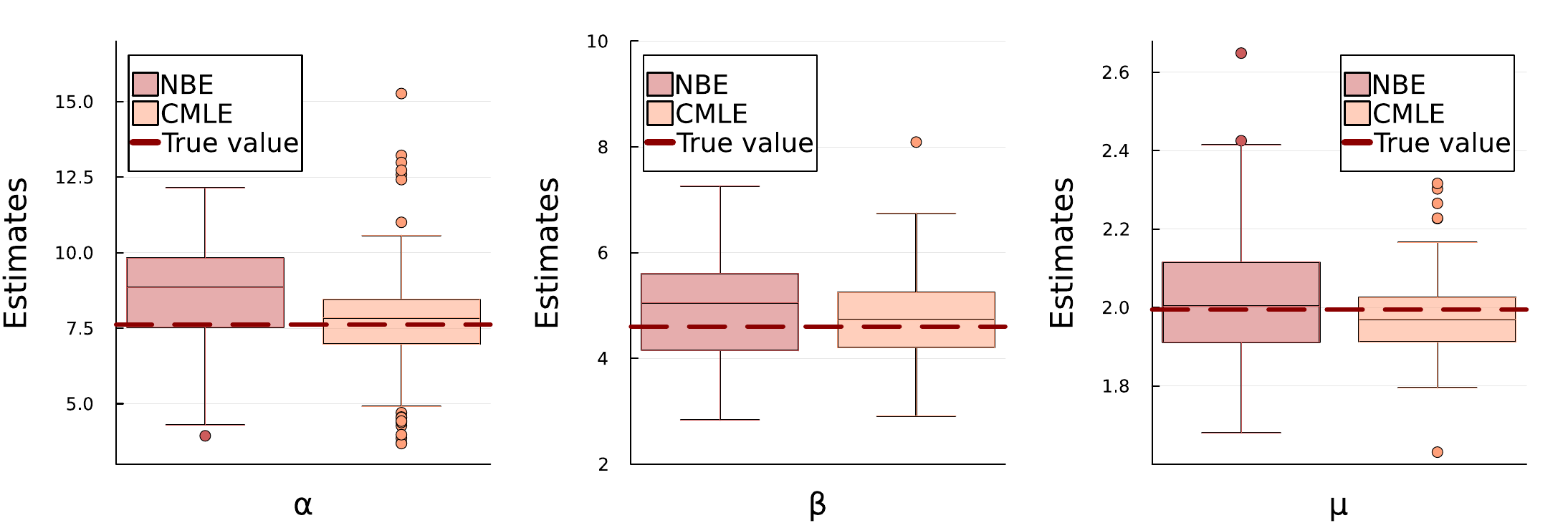}
        \caption{}
        \label{subfig:box4_eng1}
    \end{subfigure}
    \hfill
    \begin{subfigure}[b]{\textwidth}
        \includegraphics[width=\textwidth]{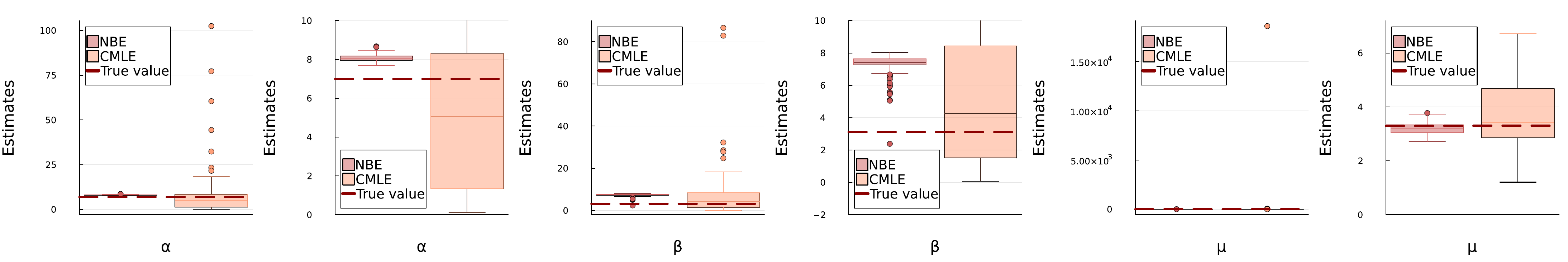}
        \caption{}
        \label{subfig:box5_eng1}
    \end{subfigure}
    \caption{Comparison between parameter estimates $\hat{\bm \theta}=(\hat\alpha, \hat\beta, \hat\mu)'$ given by CMLE (orange) and by NBE (red) for $100$ samples with $n=1000.$ The true parameter values are given by the red line. (a) $\bm \theta = (2.77, 10.54, 2.51)'$ with $\tau = 0.79,$ (b) $\bm \theta = (9.09, 14.06, 1.43)'$ with $\tau = 0.60,$ (c) $\bm \theta = (9.09, 14.06, 1.43)'$ with $\tau = 0.80,$ (d) $\bm \theta = (7.61, 4.60, 1.99)'$ with $\tau = 0.73$ and (e) $\bm \theta = (6.99, 3.12, 3.30)'$ with $\tau = 0.98.$ For better visualisation, the larger values obtained through CMLE were removed for $\bm \theta$ in (e).}
    \label{fig:mlevsnbe_eng1}
\end{sidewaysfigure}

\clearpage

\subsection{Model E2}  \label{supsec:eng2}

\begin{figure}[t!]
    \centering
    \includegraphics[width=\textwidth]{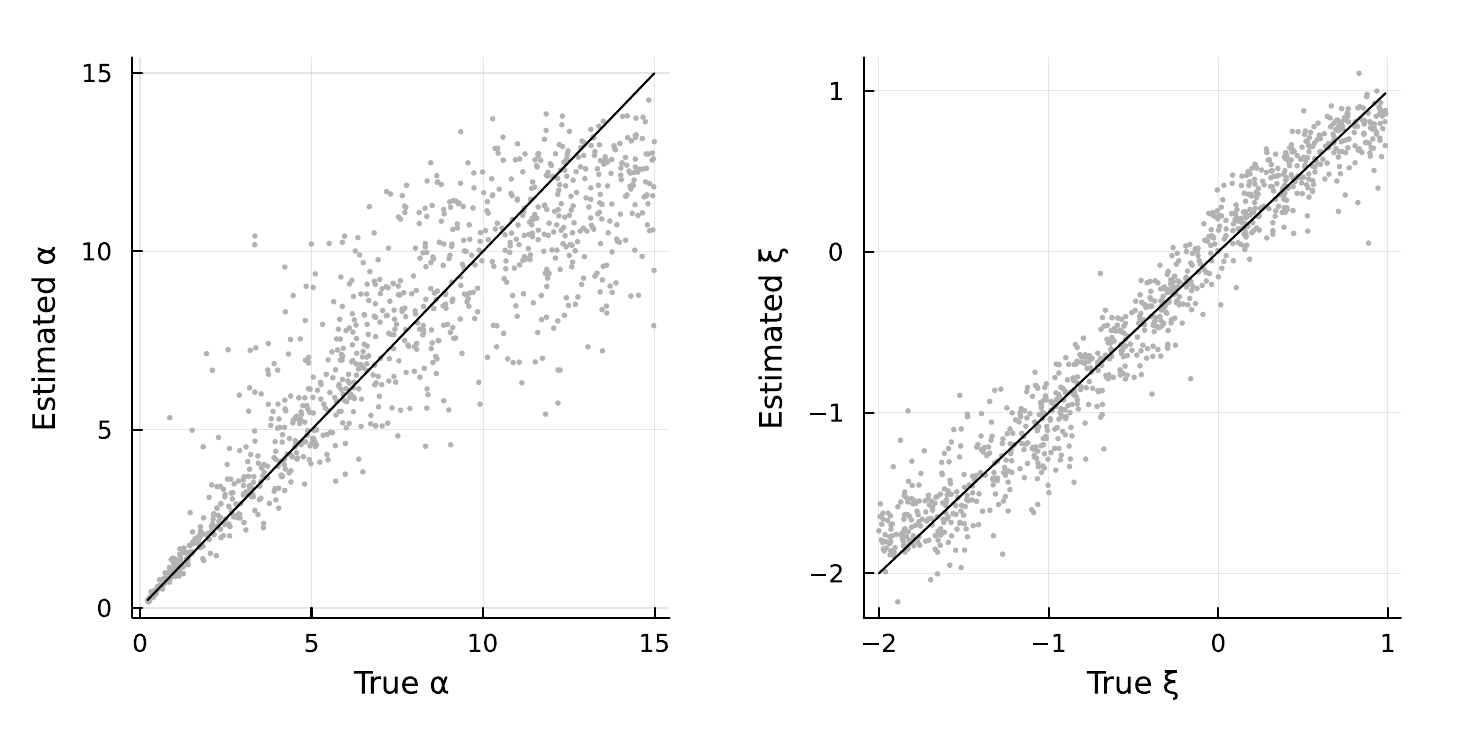}
    \caption{Assessment of the NBE for Model E2 with parameters $\bm \theta=(\alpha, \xi)'$ for a sample size of $n=1000.$}
    \label{fig:assessment_eng2}
\end{figure}

For the final model, we consider Model E2, for which the performance of the NBE is shown in Figure~\ref{fig:assessment_eng2} and Table~\ref{tab:unc_eng2}. The parameter $\alpha$ shows the highest variability, especially for larger values, with its $95\%$ uncertainty interval being wider and having lower coverage probability. The coverage probabilities of $95\%$ uncertainty intervals of $\chi(u)$ at levels $u=\{0.80, 0.95, 0.99\},$ shown on the right of Table~\ref{tab:unc_eng2}, indicate that the this measure is well calibrated, with the true $\chi(u)$ lying within the intervals at least $87\%$ of the time in spite of the bias shown by the NBE. As before, the results for $\chi(u)$ are obtained with new data sets for 1000 parameter configurations with a fixed censoring level $\tau = 0.8.$

\begin{table}[!t]
    \centering
    \captionsetup{width=\textwidth}
    \caption{Coverage probability and average length of the $95\%$ uncertainty intervals for the parameters (left) and for $\chi(u)$ at levels $u=\{0.80, 0.95, 0.99\}$ (right)  obtained via a non-parametric bootstrap procedure averaged over $1000$ models fitted using a NBE (rounded to 2 decimal places).}
    \begin{tabular}{ccccccccc}
        \cmidrule[\heavyrulewidth]{1-3} \cmidrule[\heavyrulewidth]{7-9}
        Parameter & Coverage & Length & & & & $\chi(u)$ & Coverage &  Length \\
        \cmidrule{1-3} \cmidrule{7-9}
        $\alpha$ & $0.71$ & $3.64$ & & & & $\chi(0.80)$ & $0.94$ & $0.06$ \\
        \addlinespace[1.6mm]
        $\xi$ & $0.81$ & $0.52$ & & & & $\chi(0.95)$ & $0.90$ & $0.10$ \\
        \cmidrule[\heavyrulewidth]{1-3}
        & & & & & & $\chi(0.99)$ & $0.87$ & $0.12$ \\
        \cmidrule[\heavyrulewidth]{7-9} 
    \end{tabular}
    \label{tab:unc_eng2}
\end{table}

\subsubsection*{Comparison with censored maximum likelihood estimation}

\begin{figure}[t!]
    \centering
    \begin{subfigure}[b]{0.49\textwidth}
        \includegraphics[width=\textwidth]{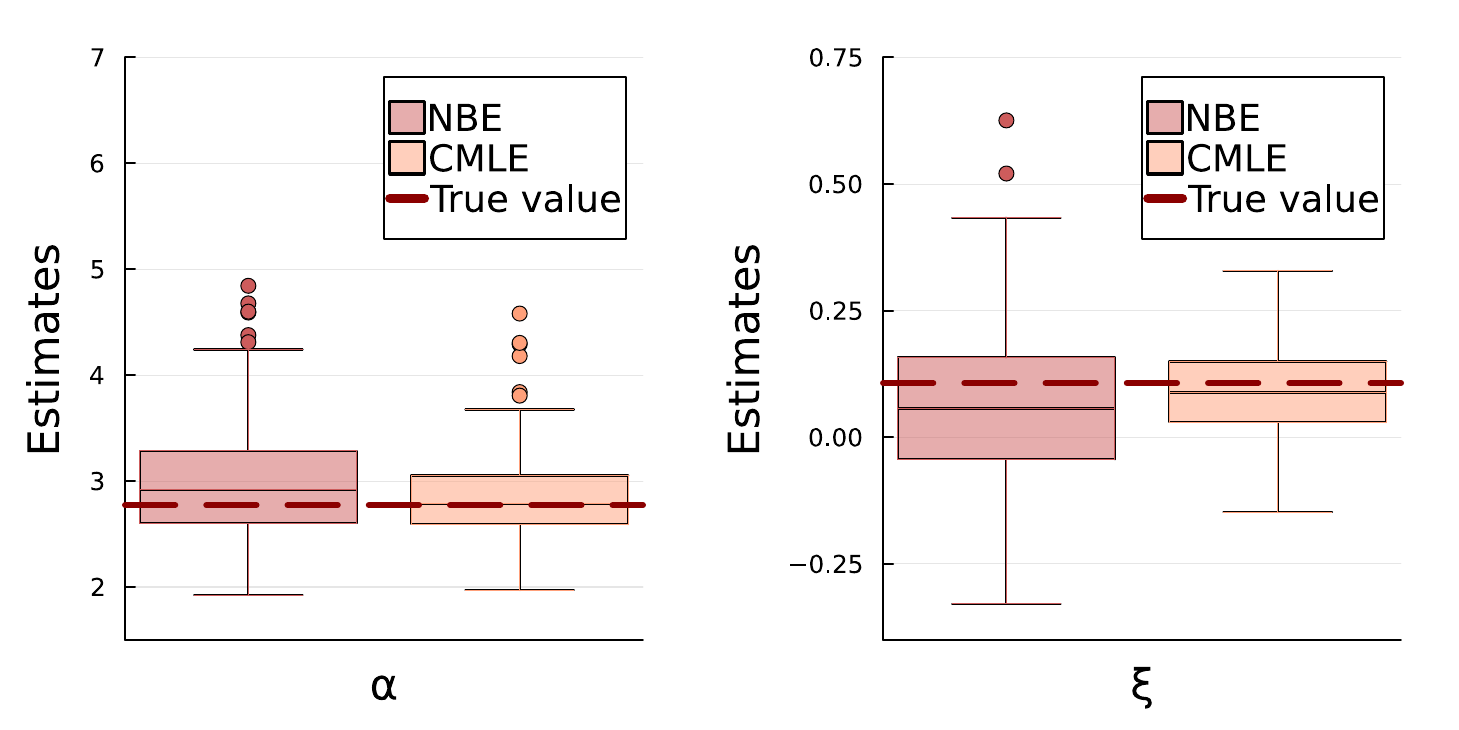}
        \caption{}
        \label{subfig:box1_eng2}
    \end{subfigure}
    \hfill
    \begin{subfigure}[b]{0.49\textwidth}
        \includegraphics[width=\textwidth]{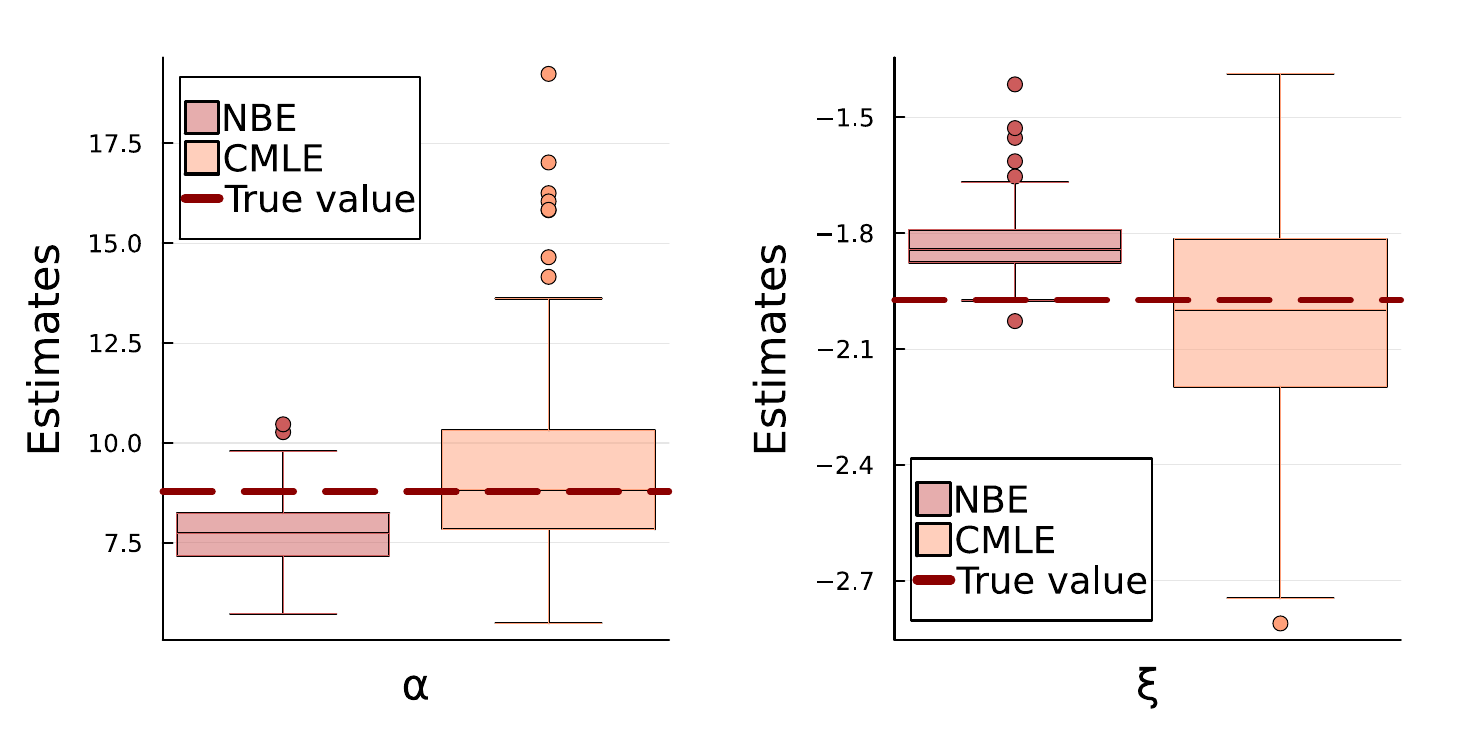}
        \caption{}
        \label{subfig:box2_eng2}
    \end{subfigure}
    \hfill
    \begin{subfigure}[b]{0.49\textwidth}
        \includegraphics[width=\textwidth]{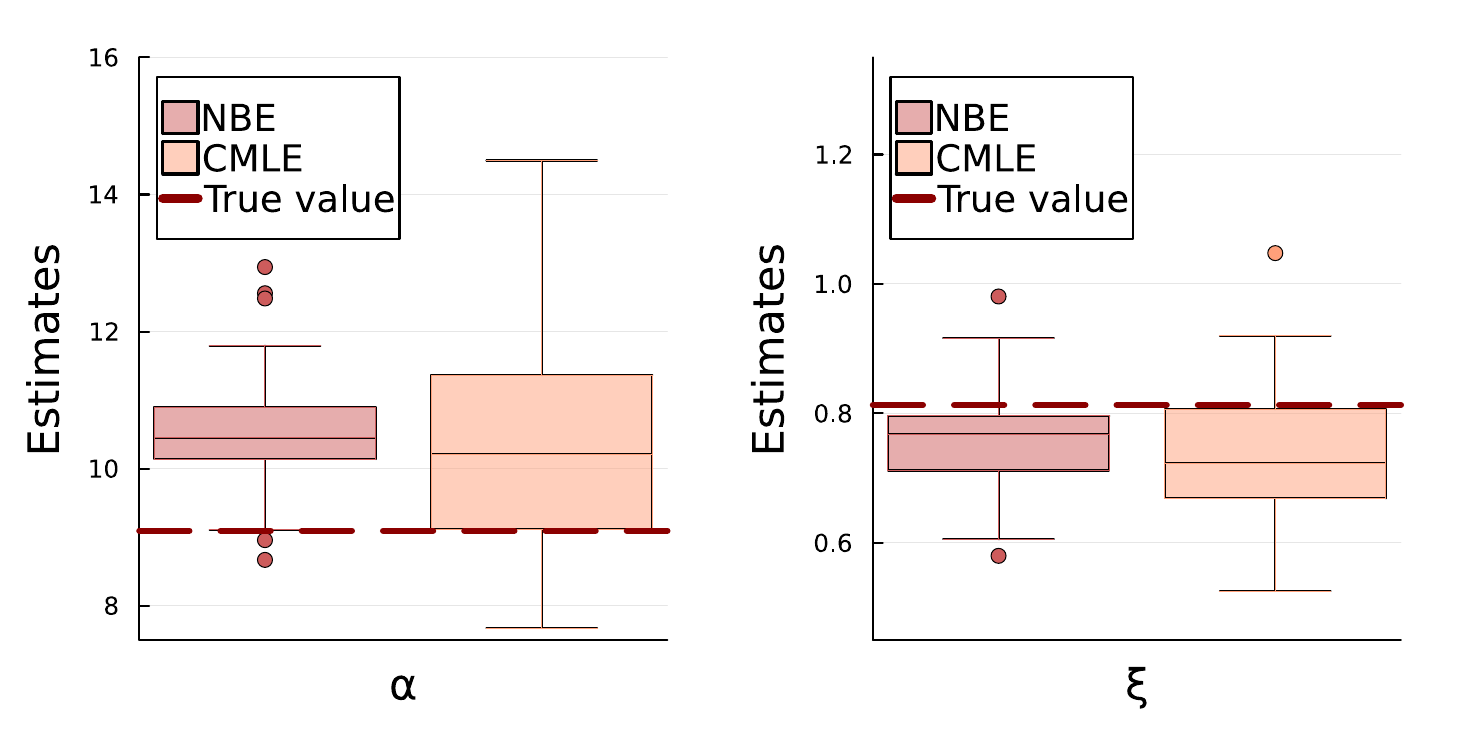}
        \caption{}
        \label{subfig:box3_eng2}
    \end{subfigure}
    \hfill
    \begin{subfigure}[b]{0.49\textwidth}
        \includegraphics[width=\textwidth]{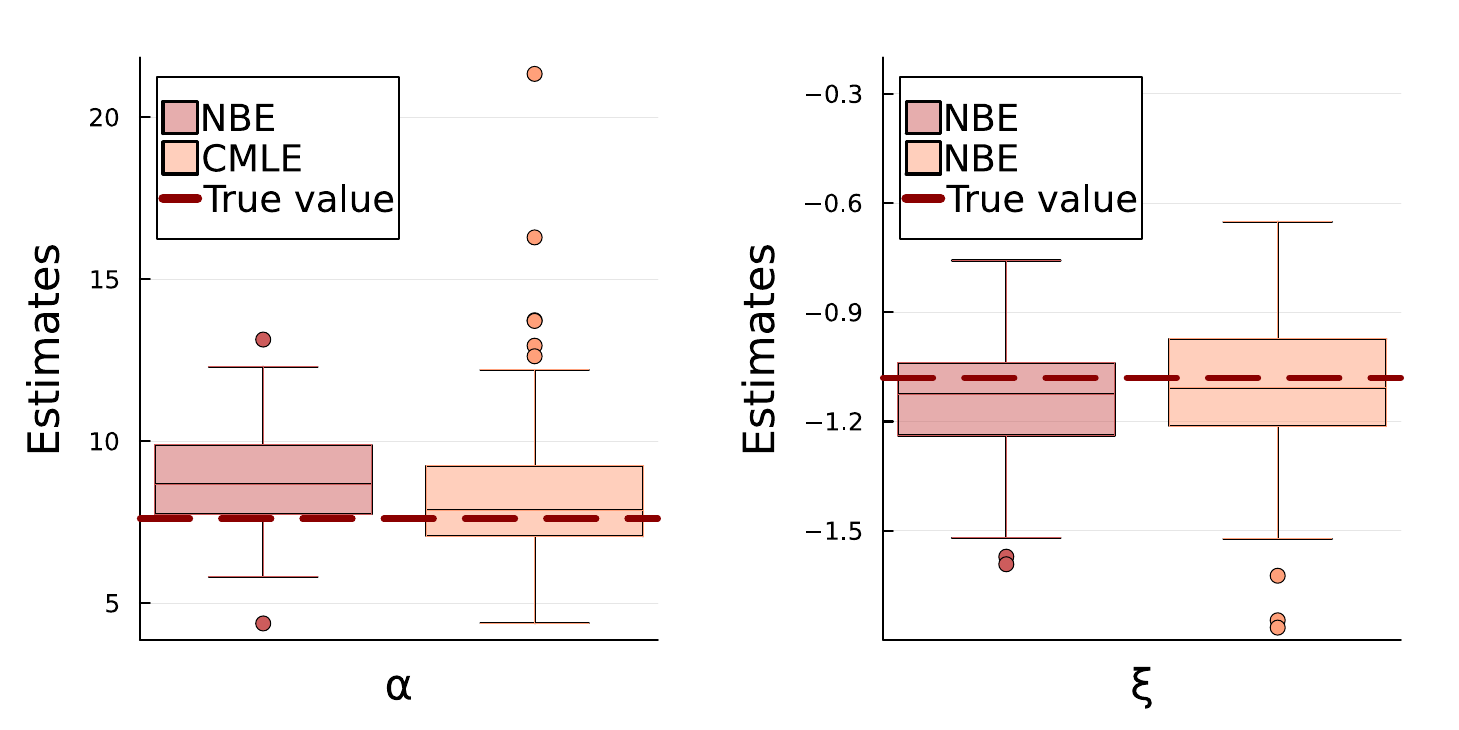}
        \caption{}
        \label{subfig:box4_eng2}
    \end{subfigure}
    \hfill
    \begin{subfigure}[b]{0.75\textwidth}
        \includegraphics[width=\textwidth]{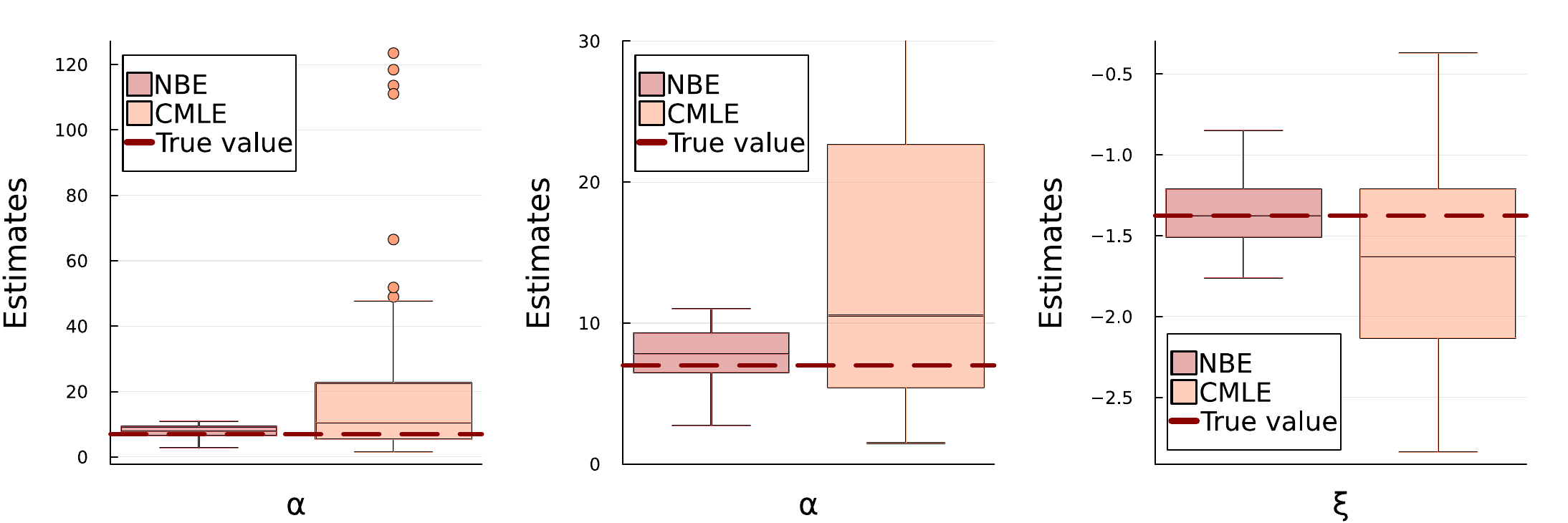}
        \caption{}
        \label{subfig:box5_eng2}
    \end{subfigure}
    \caption{Comparison between parameter estimates $\hat{\bm \theta}=(\hat\alpha, \hat\xi)'$ given by CMLE (orange) and by NBE (red) for $100$ samples with $n=1000.$ The true parameter values are given by the red line. (a) $\bm \theta = (2.77, 0.11)'$ with $\tau = 0.79,$ (b) $\bm \theta = (8.79, -1.97)'$ with $\tau = 0.60,$ (c) $\bm \theta = (9.09, 0.81)'$ with $\tau = 0.80,$ (d) $\bm \theta = (7.61, -1.08)'$ with $\tau = 0.73$ and (e) $\bm \theta = (6.99, -1.38)'$ with $\tau = 0.98.$ For better visualisation, the larger outliers obtained through CMLE were removed for $\hat\alpha$ in (e).}
    \label{fig:mlevsnbe_eng2}
\end{figure}

The estimations obtained by the NBE and the CMLE are assessed for five parameter vectors $\bm \theta=(\alpha, \xi)'$ and their corresponding data sets with $n=1000,$ each simulated $100$ times. The results, shown in Figure~\ref{fig:mlevsnbe_eng2}, indicate that the NBE is more biased than the CMLE. However, as with the previous models, the NBE is about $1\,052$ times faster than CMLE; in particular, the CMLE took $585.729$ seconds on average to compute, whilst the NBE took $0.557$ seconds.

\FloatBarrier

\newpage

\section{Model selection assessment} \label{supsec:modelsel}

The neural network architecture used for model selection (recall Section~\ref{subsec:modelselsim} of the main paper) is given in Table~\ref{tab:modelseltable}.

\begin{table}[!t]
    \centering
    \caption{Summary of the neural network architecture used for the NBC. The input array to the first layer represents the dimension $d$ of data set $\bm Z$ and the one-hot encoded vector $\bm I;$ see Section~\ref{subsec:censdata}. The output array of the last layer of neural network $\psi$ differ based on the number of models $K:$ for $K=2,$ we have $w_\psi = 128,$ while for $K=4,$ $w_\psi = 256.$ The output array $[K]$ of the last layer of neural network $\phi$ represents the output class probabilities $\bm{\hat p}.$}
    \begin{tabular}{ccc}
        \toprule
        Neural network & Input dimension & Output dimension \\
        \midrule
        \multirow{3}{*}{$\bm \psi(\cdot)$} & $[2, 2]$ & $[128]$ \\
        & $[128]$ & $[128]$ \\
        & $[128]$ & $[w_\psi]$ \\
        \midrule 
        \multirow{2}{*}{$\bm \phi(\cdot)$} & $[d_\psi + 1]$ & $[128]$ \\
        & $[128]$ & $[K]$ \\
        \bottomrule
    \end{tabular}
    \label{tab:modelseltable}
\end{table}

\subsection{Effect of sample size $N$}\label{supsec:effectn}

\begin{sidewaysfigure}[t!]
    \centering
    \begin{subfigure}[b]{0.49\textwidth}
        \includegraphics[width=\textwidth]{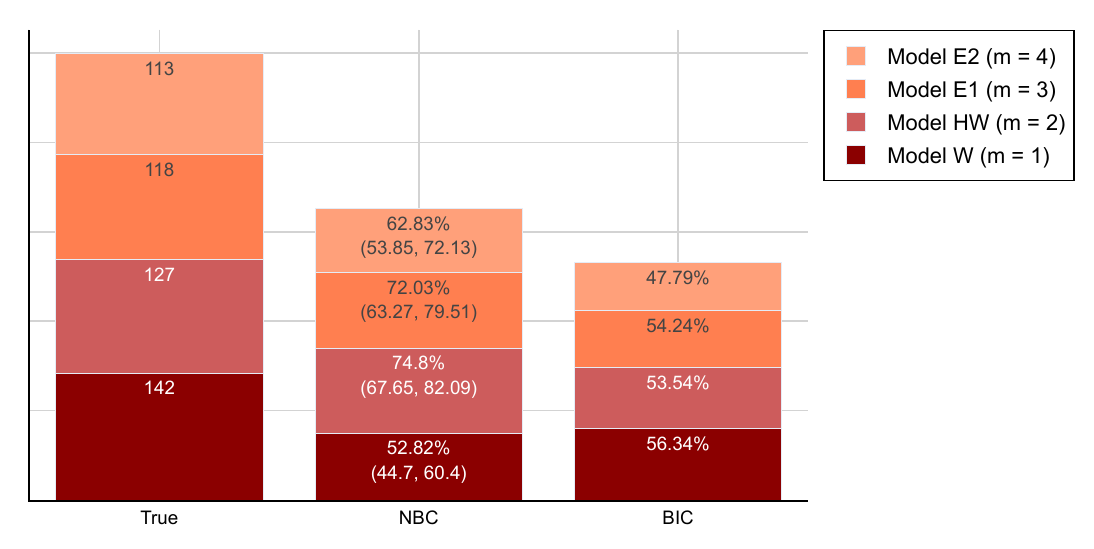}
        \caption{$n=200.$}
        \label{subfig:multi200}
    \end{subfigure}
    \hfill
    \begin{subfigure}[b]{0.49\textwidth}
        \includegraphics[width=\textwidth]{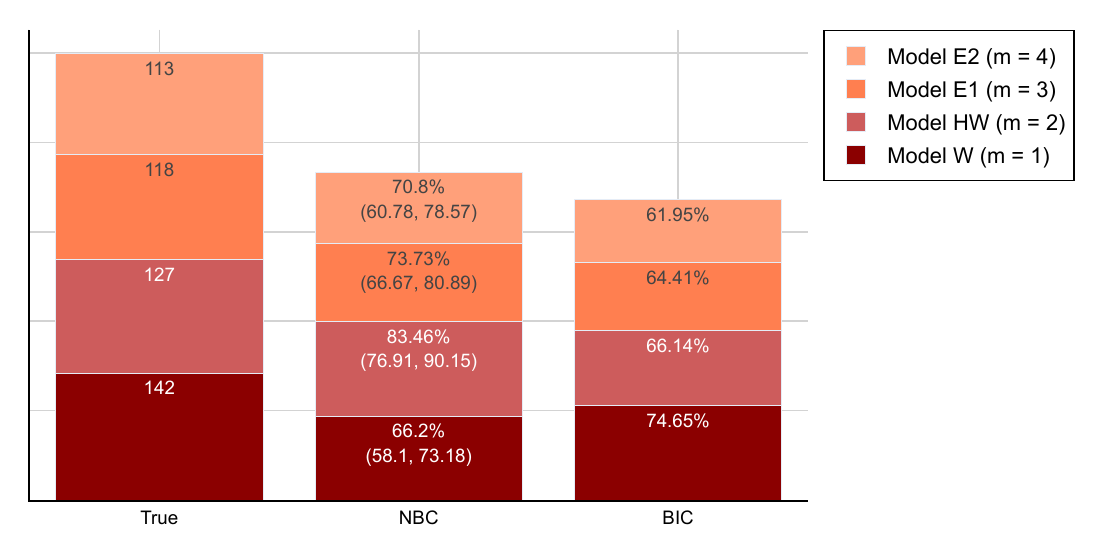}
        \caption{$n=500.$}
        \label{subfig:multi500}
    \end{subfigure}
    \hfill
    \begin{subfigure}[b]{0.49\textwidth}
        \includegraphics[width=\textwidth]{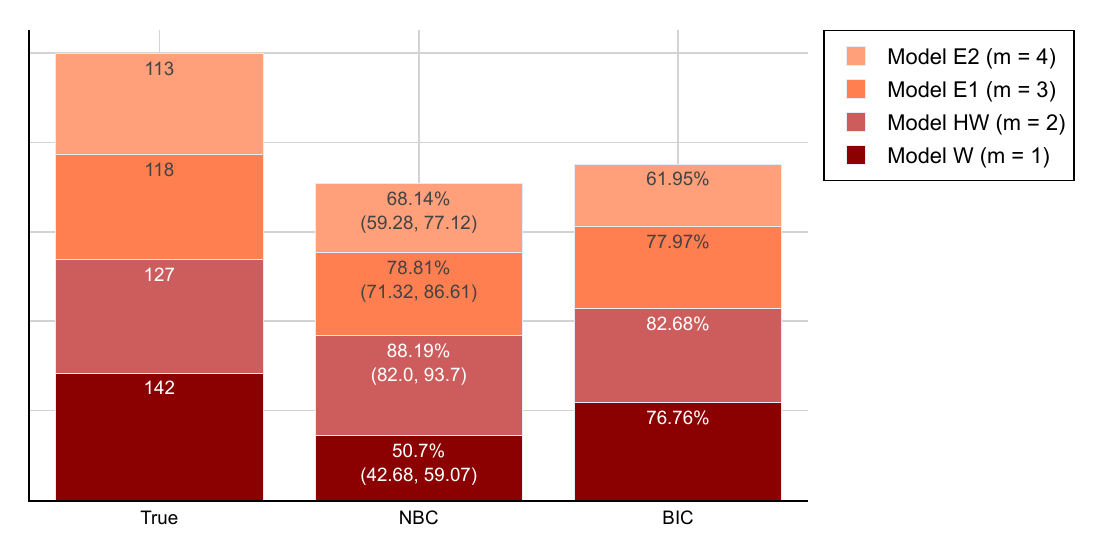}
        \caption{$n=1400.$}
        \label{subfig:multi1400}
    \end{subfigure}
    \hfill
    \begin{subfigure}[b]{0.49\textwidth}
        \includegraphics[width=\textwidth]{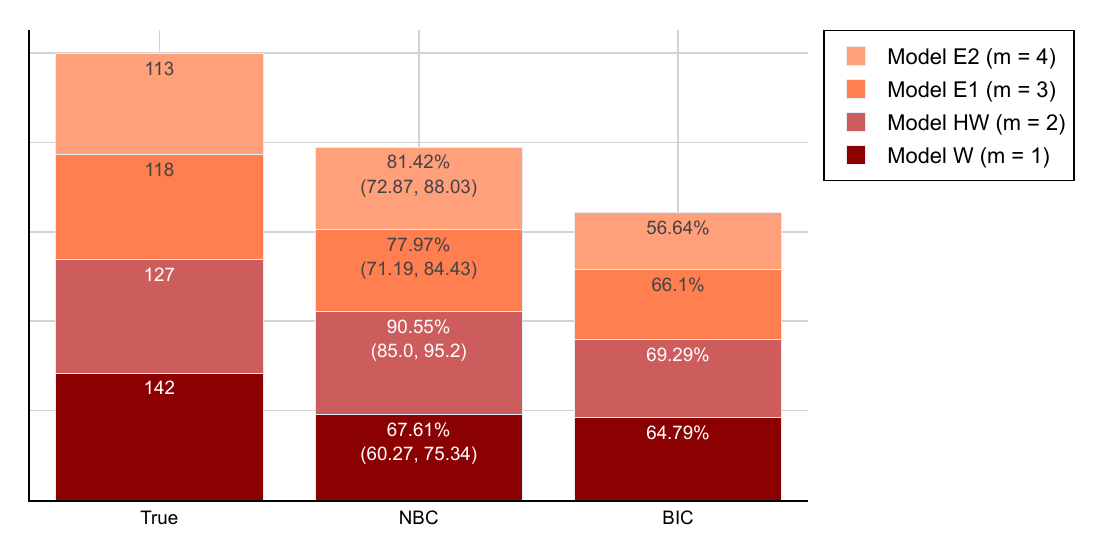}
        \caption{$n=2000.$}
        \label{subfig:multi2000}
    \end{subfigure}
    \caption{Proportion (in $\%$) of correctly identified data sets when $K=4$ through the NBCs (middle) and through BIC (right) for $n\in\{200, 500, 1400, 2000\}$. The true counts of data sets generated from models $m = 1$ (red), $m = 2$ (light red), $m = 3$ (orange) and $m = 4$ (light orange) are given in the left bar plot. The $95\%$ confidence intervals for the proportions of correctly identified data sets by the NBC are given in brackets.}
    \label{fig:samplesizenbc}
\end{sidewaysfigure}

We perform an additional study to analyse the effect of sample size $n$ on the comparative differences between the NBC and BIC. To do so, we have considered 4 cases when $K=4,$ and generated 500 models $m,$ each with samples of size $n \in N\in\{200, 500, 1400, 2000\}.$ The results are shown in Figure~\ref{fig:samplesizenbc}. As can be expected since a lower $n$ results in fewer number of exceedances for likelihood-based estimation, the BIC performs generally worse than the NBC for $n=200$ and $n=500.$ The NBC is more consistent across the different sample sizes; however, as with the study from Section~\ref{subsec:modelselsim} of the main paper, it still struggles to correctly identify samples from Model W. 

 The case of $n=2000$ presents interesting results, as well. Recall that the NBC was trained for sample sizes between 100 and 1500. By considering a sample size of $n=2000,$ we were interested in assessing the performance of the NBC for a sample size outside of the interval considered. The bottom right panel of Figure~\ref{fig:samplesizenbc} shows that the NBC is still able to correctly identify the majority of the data sets, and better than BIC.

\FloatBarrier

\section{Misspecified scenarios} \label{supsec:missscenarios}

We present now two examples, one for each study performed in Section~\ref{subsec:missscenarios} from the main paper. For each case, the best model is selected through the trained NBC, and the vector of parameters is estimated using the NBE trained for inference on the selected model. In addition, a comparison with classical model selection tools and inference is given. As a further diagnostic, we compare $\chi(u)$ for $u\in (0,1)$ obtained with the NBE for the estimated model with its empirical counterparts, the true values, and those obtained by the model selected and estimated using BIC and CMLE. 
 
\begin{figure}[t!]
    \centering
    \begin{subfigure}[b]{0.49\textwidth}
        \includegraphics[width=0.9\textwidth]{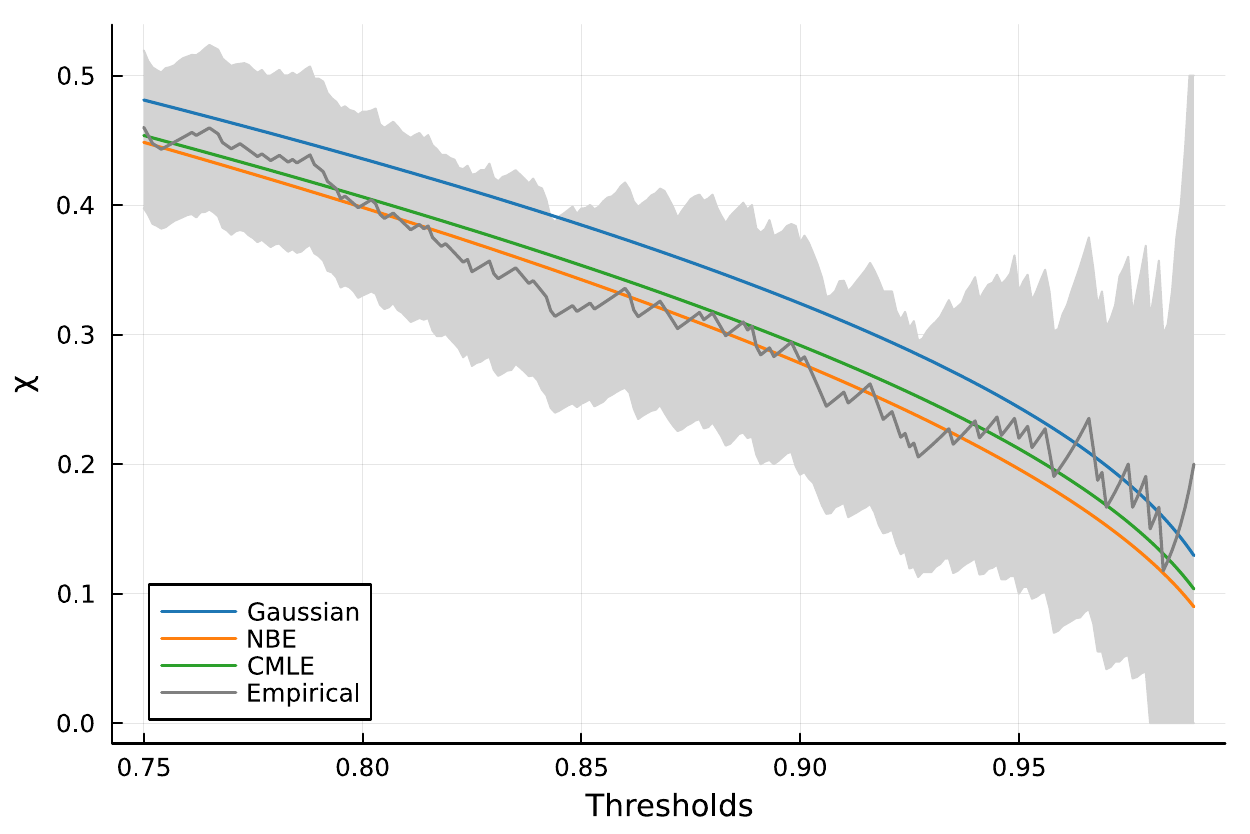}
        \caption{}
        \label{subfig:chietaNBEgaus}
    \end{subfigure}
    \hfill
    \begin{subfigure}[b]{0.49\textwidth}
        \includegraphics[width=0.9\textwidth]{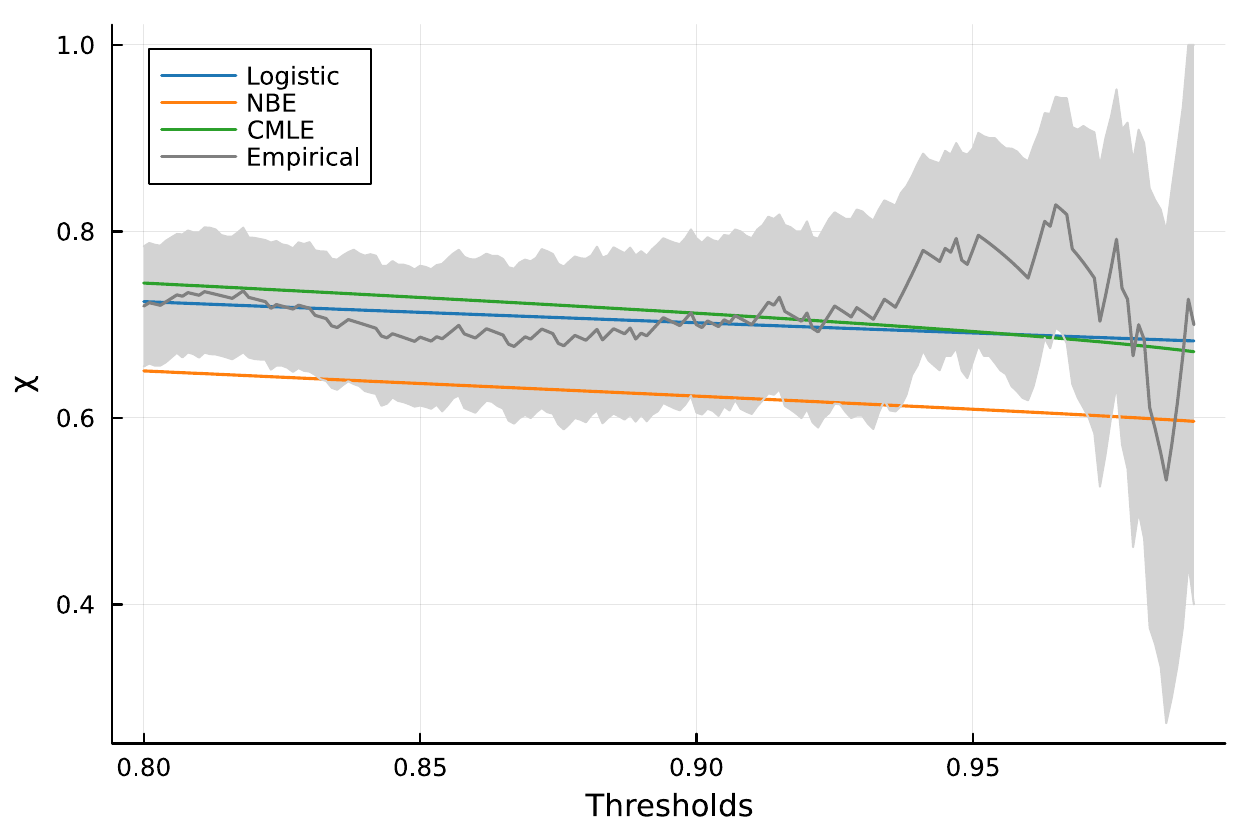}
        \caption{}
        \label{subfig:chietaNBElog}
    \end{subfigure}
    \caption{Model-based $\chi(u)$ given by the NBE (in orange) and by the CMLE (in green), and empirical $\chi(u)$ (in grey) for $u\in [\tau, 0.99].$ The 95$\%$ confidence bands were obtained by boostrapping. (a) $\chi(u)$ for a Gaussian copula with correlation parameter $\rho = 0.5$ (in blue) and censoring level $\tau = 0.75,$ and (b) $\chi(u)$ for a logistic distribution with dependence parameter $\alpha_L = 0.4$ (in blue) and censoring level $\tau = 0.8.$ Note that $\chi(u)$ for the logistic data  and for the model given by the CMLE almost overlap (right).}
    \label{fig:chimisssup}
\end{figure}

Results for model selection through the NBC and BIC for the Gaussian data case can be seen on the left of Table~\ref{tab:gausdatasup}, and the estimates for the vector of parameters obtained by the NBE and CMLE are on the right. The NBC and BIC both select Model HW as the most suitable one for the data set. Additionally,  both models indicate the presence of asymptotic independence since $\hat\delta\leq 0.5.$ This is in agreement with the underlying Gaussian data being AI. The comparison between $\chi(u)$ obtained by the models estimated through the NBE and the CMLE, with the true values of $\chi(u)$ based on the Gaussian copula, and their empirical estimates, for $u\in [0.75,0.99]$ are shown in the left panel of Figure~\ref{fig:chimisssup}. The model estimates obtained through the CMLE are slightly closer to the truth than the ones given by the NBE; however, both estimates are closer to the empirical estimates. Overall, the extremal dependence behaviour of the data is well captured by the trained NBE. 

\begin{table}[!t]
    \centering
    \captionsetup{width=\textwidth}
    \caption{Model selection procedure obtained through the probabilities given by the NBC and through BIC (left), and parameter estimates given by the NBE and by the CMLE (right) for the selected model (in bold). All the values are rounded up to 3 decimal places.} \label{tab:gausdatasup}
    \begin{tabular}{lccclc}
        \cmidrule[\heavyrulewidth]{1-3} \cmidrule[\heavyrulewidth]{5-6}
        Model & $\hat{\bm p}_{\text{NBC}}$ & BIC & & Method & Model parameters \\
        \cmidrule{1-3} \cmidrule{5-6}
        Model W & $4.609\times 10^{-5}$ & $567.348$ & & NBE (Model HW) & $(\hat\delta,\hat\omega) = (0.201, 0.400)$ \\
        \addlinespace[1.6mm]
        Model HW & $\bm{0.987}$ & $\bm{558.482}$ & & CMLE (Model HW) & $(\hat\delta,\hat\omega) = (0.107, 0.442)$  \\
        \cmidrule[\heavyrulewidth]{5-6}
        Model E1 & $2.392\times10^{-8}$ & $565.636$ & & & \\
        \addlinespace[1.6mm]
        Model E2 & $0.013$ & $564.636$ & & & \\
        \cmidrule[\heavyrulewidth]{1-3} 
    \end{tabular}
\end{table}

Table~\ref{tab:gausdatasup} gives the results for model selection and parameter estimation for the logistic data case. Again, for the model selection, the NBC and BIC agree and select Model HW as the best model to fit the data set. Looking at the parameter that indicates the extremal dependence structure, we have $\hat \delta > 0.5,$ both correctly suggesting the presence of asymptotic dependence. The comparison between $\chi(u)$ obtained by the models estimated through the NBE and the CMLE, with the true values $\chi(u)$ for the logistic data, and their empirical estimates, for $u\in [0.8,0.99]$ is shown in right panel of Figure~\ref{fig:chimisssup}. For this case, the estimated model $\chi(u)$ given by the CMLE almost overlaps with the true values for the logistic data. On the other hand, the model $\chi(u)$ estimated by the NBE seems to under-estimate the truth. However, as before, the extremal dependence structure is still reasonably well captured with the trained NBE.

\begin{table}[!t]
    \centering
    \captionsetup{width=\textwidth}
    \caption{Model selection procedure obtained through the probabilities given by the NBC and through BIC (left), and parameter estimates given by the NBE and by the CMLE (right) for the selected model (in bold). All the values are rounded to 3 decimal places.} \label{tab:logdatasup}
    \begin{tabular}{lccclc}
        \cmidrule[\heavyrulewidth]{1-3} \cmidrule[\heavyrulewidth]{5-6}
        Model & $\hat{\bm p}_{\text{NBC}}$ & BIC & & Method & Model parameters \\
        \cmidrule{1-3} \cmidrule{5-6}
        Model W & $7.774\times 10^{-5}$ & $-55.802$ & & NBE (Model HW) & $(\hat\delta,\hat\omega) = (0.640, -0.147)$ \\
        \addlinespace[1.6mm]
        Model HW & $\bm{0.999}$ & $\bm{-57.846}$ & & CMLE (Model HW) & $(\hat\delta,\hat\omega) = (0.621, 0.523)$  \\
        \cmidrule[\heavyrulewidth]{5-6}
        Model E1 & $2.104\times 10^{-7}$ & $-56.164$ & & & \\
        \addlinespace[1.6mm]
        Model E2 & $0.001$ & $-64.802$ & & & \\
        \cmidrule[\heavyrulewidth]{1-3} 
    \end{tabular}
\end{table}

\newpage

\section{Case study: changes in geomagnetic field fluctuations} \label{supsec:application}

In this section, we summarise the results for the remaining censoring levels for each pair of locations. Contrarily to the main paper, we only show the selected model for each censoring level and the type of extremal dependence estimated with it.

\subsubsection*{Pair (SCO, STF)}

\begin{table}[H]
    \centering
    \caption{Model selected by the NBC for censoring levels $\tau = \{0.60,0.65, \ldots, 0.95\}$ and parameter estimates given by the trained NBE for pair (SCO, STF). All the values are rounded up to 3 decimal places.}
    \begin{tabular}{ccccc}
        \toprule
        $\tau$ & Model & $\hat{\bm p}_{\text{NBC}}$ & $\hat{\bm \theta}_{\text{NBE}}$ & Extremal dependence \\
        \midrule
        $0.60$ & Model HW & $0.999$ & $(\hat \delta, \, \hat\omega) = (0.170,\, 0.743)$ & AI\\
        $0.65$ & Model HW & $0.998$ & $(\hat \delta, \, \hat\omega) = (0.178,\, 0.767)$ & AI\\
        $0.70$ & Model HW & $0.954$ & $(\hat \delta, \, \hat\omega) = (0.178,\, 0.800)$ & AI\\
        $0.75$ & Model HW & $0.906$ & $(\hat \delta, \, \hat\omega) = (0.195,\, 0.767)$ & AI\\
        $0.80$ & Model HW & $0.917$ & $(\hat \delta, \, \hat\omega) = (0.228,\, 0.742)$ & AI\\
        $0.85$ & Model HW & $0.922$ & $(\hat \delta, \, \hat\omega) = (0.258,\, 0.714)$ & AI\\
        $0.90$ & Model E2 & $0.935$ & $(\hat \alpha, \, \hat\xi) = (3.512,\, -0.368)$ & AI\\
        $0.95$ & Model E2 & $0.640$ & $(\hat \alpha, \, \hat\xi) = (3.616,\, -0.399)$ & AI\\
        \bottomrule
    \end{tabular}
    \label{tab:pair1sup}
\end{table}

\subsubsection*{Pair (SCO, STJ)}

\begin{table}[H]
    \centering
    \caption{Model selected by the NBC for censoring levels $\tau = \{0.60,0.65, \ldots, 0.95\}$ and parameter estimates given by the trained NBE for pair (SCO, STJ). All the values are rounded up to 3 decimal places.}
    \begin{tabular}{ccccc}
        \toprule
        $\tau$ & Model & $\hat{\bm p}_{\text{NBC}}$ & $\hat{\bm \theta}_{\text{NBE}}$ & Extremal dependence \\
        \midrule
        $0.60$ & Model HW & $1.000$ & $(\hat \delta, \, \hat\omega) = (0.085,\, 0.560)$ & AI\\
        $0.65$ & Model HW & $1.000$ & $(\hat \delta, \, \hat\omega) = (0.093,\, 0.580)$ & AI\\
        $0.70$ & Model HW & $1.000$ & $(\hat \delta, \, \hat\omega) = (0.109,\, 0.586)$ & AI\\
        $0.75$ & Model HW & $1.000$ & $(\hat \delta, \, \hat\omega) = (0.104,\, 0.591)$ & AI\\
        $0.80$ & Model HW & $0.958$ & $(\hat \delta, \, \hat\omega) = (0.105,\, 0.626)$ & AI\\
        $0.85$ & Model E2 & $0.900$ & $(\hat \alpha, \, \hat\xi) = (2.316\, -0.791)$ & AI\\
        $0.90$ & Model E2 & $0.940$ & $(\hat \alpha, \, \hat\xi) = (2.748,\, -0.834)$ & AI\\
        $0.95$ & Model E2 & $0.875$ & $(\hat \alpha, \, \hat\xi) = (3.223,\, -0.782)$ & AI\\
        \bottomrule
    \end{tabular}
    \label{tab:pair2sup}
\end{table}

\subsubsection*{Pair (STF, STJ)}

\begin{table}[H]
    \centering
    \caption{Model selected by the NBC for censoring levels $\tau = \{0.60,0.65, \ldots, 0.95\}$ and parameter estimates given by the trained NBE for pair (STF, STJ). All the values are rounded up to 3 decimal places.}
    \begin{tabular}{ccccc}
        \toprule
        $\tau$ & Model & $\hat{\bm p}_{\text{NBC}}$ & $\hat{\bm \theta}_{\text{NBE}}$ & Extremal dependence \\
        \midrule
        $0.60$ & Model HW & $1.000$ & $(\hat \delta, \, \hat\omega) = (0.106,\, 0.558)$ & AI\\
        $0.65$ & Model HW & $1.000$ & $(\hat \delta, \, \hat\omega) = (0.113,\, 0.571)$ & AI\\
        $0.70$ & Model HW & $1.000$ & $(\hat \delta, \, \hat\omega) = (0.117,\, 0.588)$ & AI\\
        $0.75$ & Model HW & $0.996$ & $(\hat \delta, \, \hat\omega) = (0.134,\, 0.585)$ & AI\\
        $0.80$ & Model HW & $0.920$ & $(\hat \delta, \, \hat\omega) = (0.125,\, 0.610)$ & AI\\
        $0.85$ & Model E2 & $0.672$ & $(\hat \alpha, \, \hat\xi) = (2.420,\, -0.849)$ & AI\\
        $0.90$ & Model E2 & $0.727$ & $(\hat \alpha, \, \hat\xi) = (2.573,\, -0.846)$ & AI\\
        $0.95$ & Model E2 & $0.832$ & $(\hat \alpha, \, \hat\xi) = (3.711,\, -0.772)$ & AI\\
        \bottomrule
    \end{tabular}
    \label{tab:pair3sup}
\end{table}



\end{document}

\end{document}